\pdfoutput=1
\documentclass[prd,aps,superscriptaddress,preprintnumbers,eqsecnum,nofootinbib,notitlepage]{revtex4-1}
\usepackage[utf8]{inputenc}
\usepackage{amsfonts,amsmath,amssymb,bm,subfigure}
\usepackage{multirow,hyperref}
\usepackage{tikz-feynhand}
\usepackage{verbatim}

\newcommand{\nn}{\nonumber \\}
\newcommand{\Mpl}{M_{\rm Pl}}
\def\dd{\mathrm{d}}

\newcommand{\Meff}{\mathcal{M} }

\begin{document}
\baselineskip=12pt

\preprint{YITP-21-70, RIKEN-iTHEMS-Report-21, IPMU21-0045}
\title{Positivity vs. Lorentz-violation: an explicit example}
\author{Katsuki Aoki}
\email{katsuki.aoki@yukawa.kyoto-u.ac.jp}
\affiliation{Center for Gravitational Physics, Yukawa Institute for Theoretical Physics, Kyoto University, 606-8502, Kyoto, Japan}

\author{Shinji Mukohyama}
\email{shinji.mukohyama@yukawa.kyoto-u.ac.jp}
\affiliation{Center for Gravitational Physics, Yukawa Institute for Theoretical Physics, Kyoto University, 606-8502, Kyoto, Japan}
\affiliation{Kavli Institute for the Physics and Mathematics of the Universe (WPI), The University of Tokyo, Kashiwa, Chiba 277-8583, Japan}

\author{Ryo Namba}
\email{ryo.namba@riken.jp}
\affiliation{RIKEN Interdisciplinary Theoretical and Mathematical Sciences (iTHEMS), Wako, Saitama 351-0198, Japan}
\affiliation{Tsung-Dao Lee Institute, Shanghai Jiao Tong University, Shanghai 200240, China}

\date{\today}

\begin{abstract}
We show how a class of multi-field scalar-field theories in a Lorentz-breaking background imposes consistency conditions on its effective theory of a single field and provides an example of order-unity violation of a naively applied positivity bound, assuming a large hierarchy between the masses of the lightest field and the others.
\end{abstract}

\maketitle

\tableofcontents

\section{Introduction}
\label{sec:intro}

Underlying assumptions on ultraviolet (UV) completion can impose constraints on its low-energy effective field theories (EFTs), meaning that not all EFTs may be consistent with the assumed UV physics even if they are ``consistent'' from low-energy perspective. One of the most well-established constraints is called positivity bounds~\cite{Adams:2006sv}, provided that EFTs admit a unitary, Poincar\'{e}-invariant, analytic, and bounded UV completion. The last two assumptions are inferred from causality and locality. Although whether the UV theory in nature indeed satisfies these assumptions is unknown, they are well-defined and self-consistent, and the positivity bounds can be used to test whether our working assumptions about UV physics are compatible with low-energy experiments/observations. The positivity bounds are often derived by means of scattering amplitudes where asymptotic states have to be well-defined. As for EFTs around a Poincar\'{e} invariant background, there would be no subtleties about the states and the bounds provide remarkably strong constraints on higher derivative operators of EFTs~\cite{Adams:2006sv,Bellazzini:2020cot,Tolley:2020gtv,Caron-Huot:2020cmc,Arkani-Hamed:2020blm}. Furthermore, the positivity bounds provide a cutoff scale of a renormalizable theory when it couples to gravity~\cite{Alberte:2020jsk,Alberte:2020bdz,Aoki:2021ckh,Noumi:2021uuv}: for example, the Standard Model of particle physics coupled to general relativity violates the positivity bound when it is extrapolated up to $10^{16}$GeV, suggesting that quantum gravity should be needed around or below $10^{16}$GeV~\cite{Aoki:2021ckh}. The framework of EFTs is robust even around a non-trivial background, which is a typical situation in realistic setups including (but not limited to) cosmology. As is well-known, the notion of ``particle'' is ambiguous in field theories in curved spacetimes and, in particular, there would be no definite notion of particles in the infrared (IR) limit. The Poincar\'{e}-invariant positivity bounds may not be directly applicable to EFTs around such non-trivial backgrounds.

We study EFTs without the Lorentz symmetry, which naturally arise in cosmology, for instance. The states may be well-defined as long as the temporal and spatial translation symmetries are preserved. Assuming dispersion relations (linear equations of motion) of particles, one may discuss positivity bounds by considering scatterings of the particles even when the Lorentz invariance is absent. This is indeed analysed in~\cite{Baumann:2015nta,Grall:2021xxm}, and the positivity bounds are discussed. However, when the Lorentz invariance is {\it spontaneously} broken by a background configuration of a field, the background serves as a source of gravity and then the temporal and/or spatial translation symmetry should be generically broken as well. The existence of gravity should impose a strong constraint on the consistency of the arguments.\footnote{Positivity bounds on gravitational EFTs are already non-trivial due to the pole associated with the graviton $t$-channel exchange even around the Minkowski background. Gravitational positivity bounds can be derived by assuming the Regge behaviour of the amplitude, which is a consistent satisfaction of the Froissart bound~\cite{Froissart:1961ux,Martin:1962rt} in gravitational theories, to cancel the pole of the graviton exchange~\cite{Hamada:2018dde,Tokuda:2020mlf,Herrero-Valea:2020wxz} (see also~\cite{Bellazzini:2019xts,Alberte:2020jsk,Alberte:2020bdz,Caron-Huot:2021rmr} for related discussions). } In fact, the paper \cite{Pajer:2020wnj} showed that the existence of graviton enforces three-particle amplitudes to be Lorentz invariant from unitarity, spacetime translation symmetries, spatial rotational symmetry, relativistic dispersion relations, and analyticity, implying that the Lorentz symmetry is an emergent symmetry from some of the assumptions. A similar observation can be found in~\cite{Khoury:2013oqa}. One should carefully examine working assumptions in the bottom-up approaches, as the input may already include the Lorentz symmetry, implicitly.

It is therefore important to consider different approaches that can be applied to systems in which gravity is essential. Some of such approaches are called swampland conjectures, ranging from the weak gravity conjecture~\cite{ArkaniHamed:2006dz} to the de Sitter conjecture~\cite{Obied:2018sgi,Garg:2018reu,Ooguri:2018wrx}. Those swampland conjectures are supposed to tell which EFTs are consistent with quantum gravity and which ones are not (see \cite{Palti:2019pca} for a review). However, none of the swampland conjectures enjoys a rigorous proof based on a fundamental theory, and indeed there can be exceptions and/or counterexamples (e.g.~the KKLT scenario~\cite{Kachru:2003aw} and Large Volume Scenario~\cite{Balasubramanian:2005zx} against the de Sitter conjecture). Therefore, the statements of the swampland conjectures should at best be considered as some properties that most (but not necessarily all) of consistent EFTs tend to possess. Another related approach is based on the generalized second law of black holes~\cite{Dubovsky:2006vk,Eling:2007qd}, aiming to test the consistency of a particular Lorentz violating theory called ghost condensate~\cite{ArkaniHamed:2003uy,ArkaniHamed:2003uz}. It was later shown that in this theory the generalized second law is actually protected in a rather non-trivial way because of the accretion of the Nambu-Goldstone mode~\cite{Mukohyama:2009rk,Mukohyama:2009um}. Yet another consistency test was proposed in the context of an accelerated expansion of the universe, leading to the so-called de Sitter entropy bound~\cite{ArkaniHamed:2007ky}. Again, the bound was recently shown to hold in the ghost condensate~\cite{Jazayeri:2016jav}. These examples clearly illustrate the importance of model-independent consistency conditions that are applicable to systems with gravity.

On the other hand, model-dependent approaches are also useful, as, at the very least, they provide smoking guns for the corresponding UV physics that can be tested by observations/experiments at low energy. They also serve as examples upon which less model-dependent bounds and/or conjectures can be built. In the present paper, we therefore take a complementary top-down approach: we assume a particular but sufficiently wide class of (partial) UV completion and discuss constraints on its low-energy EFTs. While it does not apply to other types of (partial) UV completion such as the idea in \cite{Mukohyama:2006mm}, an advantage of this approach is that, once we admit such a particular class of (partial) UV completion, it can robustly be applied to the situation where Lorentz symmetry is broken and/or gravity cannot be ignored. Also, if one finds an example with a special feature then it serves as an existence proof of an EFT with the feature.

The low-energy EFT that we are interested in is given by the action of the form $\int \dd^d x \sqrt{-g}P(\varphi, X)$, called k-essence, where $\varphi$ is a scalar field, $X=-g^{\mu\nu}\partial_{\mu}\varphi \partial_{\nu}\varphi/2$ its kinetic term, and $d$ the spacetime dimension. This action can be regarded as the leading operators in the context of EFT of single-field inflation/dark energy models~\cite{Creminelli:2006xe,Cheung:2007st,Creminelli:2008wc}, and its (partial) UV completion has recently been developed in~\cite{Babichev:2016hys,Babichev:2017lrx,Babichev:2018twg,Mizuno:2019pcm,Mukohyama:2020lsu}. The single-field k-essence theory can be obtained from a non-linear sigma model with one light direction in the field space of an arbitrary geometry by integrating out all the massive modes while keeping the light one. We mainly study the k-essence theory in the Einstein frame, that is, gravity is minimally coupled to $\varphi$. We will briefly discuss a (partial) UV completion of a subclass of degenerate higher-order scalar-tensor (DHOST) theories~\cite{Langlois:2015cwa,Crisostomi:2016czh,BenAchour:2016fzp} by considering non-minimal couplings. The single-field low-energy EFT action that we obtain from the multi-field UV model is Lorentz-invariant, and thus the Lorentz violation is spontaneous. For a trivial background $\partial_{\mu}\varphi=0$, the positivity bound would conclude that the coefficient of the four-point interactions has to be positive~\cite{Adams:2006sv}, namely $P_{XX}>0$, where a subscript $X$ denotes a derivative with respect to $X$. On the other hand, our target is the case where $\varphi$ has a non-vanishing gradient $C_{\mu}=\partial_{\mu}\varphi \neq 0$ at the background level. Then, the vector $C_{\mu}$ determines a preferred direction, and the Lorentz symmetry is spontaneously broken. The standard positivity bounds may not be applied in such situations. In particular, we show by an explicit example that $P_{XX}$ is allowed to be negative in Lorentz-violating backgrounds, without any inconsistency in UV.

We highlight the following points that are important to understand the consistency conditions around Lorentz-violating backgrounds.
\begin{itemize}
\item {\it Background dependence.} The regime of validity and the consistency conditions of the EFT are background dependent and, indeed, some of $P(\varphi,X)$ is consistent only around a Lorentz-violating background, $\partial_{\mu} \varphi \neq 0$. In such a case, there is no continuous limit to recover the Lorentz symmetry within the EFT, implying that there is no need to obey the standard positivity bounds, see also e.g.~\cite{Nicolis:2015sra}. Note that this does not mean the underlying UV theory has no Lorentz-invariant background. The UV theory may admit different backgrounds, the Lorentz-invariant one and the Lorentz-violating one, and predict separate low-energy EFT around each background. Furthermore, even if the EFT has a continuous Lorentz-invariant limit within its regime of validity, it is non-trivial how the bound $P_{XX}>0$ is extended away from the Lorentz invariant background $X=0$. We shall refer to the inequality $P_{XX}(\varphi, X)>0$ evaluated away from $X=0$ as the naively applied positivity bound and will discuss the actual sign of $P_{XX}$ derived from multi-field UV theories. The papers~\cite{Baumann:2015nta,Grall:2021xxm} suggested a different form of the bound, $P_{XX}-2XP_{XXX} + \cdots >0$, around Lorentz-violating backgrounds (see Appendix A of~\cite{Davis:2021oce} for the explicit expression, keeping in mind that their convention of $X$ and ours are different). Although in later sections we find a general tendency of the sign of $P_{XX}$ even for $X\neq 0$, we do not find any top-down support for the bound of~\cite{Baumann:2015nta,Grall:2021xxm} as we discuss in Sec.~\ref{sec:comparison}.

\item {\it Stable and unstable backgrounds.} We investigate conditions that the k-essence theory admits a healthy (partial) UV completion at least within multi-field nonlinear sigma models. Here ``healthy'' means that the UV degrees of freedom are stable, that is, the k-essence theory is assumed to be obtained by integrating out stable heavy degrees of freedom that are neither ghost nor tachyon. However, we emphasize that this assumption does not exclude the existence of unstable modes in IR. Since we are interested in the system with a non-trivial background configuration of the fields, the system can exhibit an IR instability due to an attractive force even if there is no pathological instability in UV, cf.~the Jeans instability in gravitational systems. This is a major difference from the conventional argument about the positivity bounds:~the Lorentz-invariant background is supposed to be stable in all scales in the standard positivity bounds whereas our Lorentz-violating background is not necessary to be stable in IR. We indeed find that for the particular class of (partial) UV completions, the naively applied positivity $P_{XX}>0$ holds around stable backgrounds while $P_{XX}$ can be negative around an unstable background which is consistently realized only in the Lorentz-violating setup.

\item {\it UV consistency vs.~EFT predictivity.} Let us introduce the terminology, \emph{UV consistency} and \emph{EFT predictivity}, to distinguish the consistency conditions of the EFTs especially in those systems that contain instabilities in IR.\footnote{We would appreciate the anonymous referee for proposing these terms.} 
The UV consistency refers to the conditions originating from the no-ghost and no-tachyon conditions in UV while the EFT predictivity defines the conditions under which the IR instability is either absent or resolved within the regime of validity of the EFT. The violation of the UV consistency means that the EFT cannot be UV completed in a healthy way. On the other hand, when the EFT predictivity is violated, the EFT has a limited predictive power, but the EFT can be UV-completed as long as the UV consistency condition holds. We will detail these conditions in Sec.~\ref{sec:EFTgeneral} and Sec.~\ref{sec:UV_consistency}.

\end{itemize}

The rest of the paper is organized as follows. In Sec.~\ref{sec:preparation} we consider a few simple toy examples to illustrate how the reduction to a low-energy effective theory works. In Sec.~\ref{sec:EFTreduction_der} we describe the procedure of EFT reductions for two-field models and general multi-field nonlinear sigma models. Here our analysis is done at fully nonlinear levels with minimally coupled gravity. Then in Sec.~\ref{sec:UV_consistency}, we perform perturbative analyses concentrating on nonlinear sigma models composed of two scalars without gravity. We derive the consistency conditions with clear validity ranges; when the background $C_{\mu}=\partial_{\mu}\varphi$ is spacelike, namely $X<0$, the ghost-free condition and the no-tachyon condition immediately conclude $P_X>0$ and $P_{XX}>0$, while precautions are needed in the timelike background $X>0$. In particular we show an explicit example where $P_{XX}$ can be negative in the timelike background. We also conclude that the same result holds in a general multi-field case. In Sec.~\ref{sec:discussion}, extensive discussions on several aspects of our EFT reduction are provided. We discuss higher-derivative corrections to the k-essence and also comment on the implications for screening effects and the preservation of the null energy condition. We further revisit the UV consistency of the k-essence by using the Feynman-like diagrams. We then consider effects of non-minimal matter coupling and implications to a subset of DHOST theories and make a comparison with the previously discussed bounds of bottom-up approaches. Section \ref{sec:conclusion} is devoted to the summary and conclusions. In Appendix \ref{subsec:DBI}, we perform the analysis for models with the DBI-type kinetic terms in an arbitrary field space and derive essentially the same results as in the case of the non-linear sigma models. We demonstrate the EFT reduction from the $U(1)$ scalar theory as a concrete example of our general argument in Appendix \ref{app:U(1)}, and finally in Appendix \ref{sec:ghostcondensate} we make a one-to-one comparison of our EFT to the models of ghost condensate.


\section{Preparation for EFT reduction}
\label{sec:preparation}

The purpose of this section is to demonstrate the procedure of EFT reduction using a few explicit examples, before exhibiting more general multi-field cases in Sec.~\ref{sec:EFTreduction_der}. We first consider a simple toy example of a classical-mechanical system to illustrate how the reduction to a low-energy effective theory works in Sec.~\ref{sec:oscillators} and then study two-field-scalar systems and its reduction to a single-field EFT.

\subsection{Warm-up: low-energy effective theory of coupled oscillators}
\label{sec:oscillators}

As a first concrete toy model of EFT reduction, we consider a classical system of coupled oscillators $x(t)$ and $y(t)$ described by the Lagrangian,
\begin{equation}
  L = \frac{1}{2}\left(\dot{x}^2 + \dot{y}^2 + 2\alpha\dot{x}\dot{y}\right) - \frac{1}{2}\left(\mu^2x^2 + \eta M^2 y^2 + 2\beta \mu Mxy\right) - \gamma M y\dot{x}\,, \label{eqn:toyexample-Lagrangian}
\end{equation}
where a dot represents derivative with respect to the time $t$, and $\alpha,\beta,\gamma,\eta, \mu$ and $M$ are all constants. We do not perform a diagonalization to eliminate the kinetic mixing $\alpha \dot{x} \dot{y}$ at this stage, in order to make the analogy to the cases in later sections easier. We suppose the hierarchy between the two mass scales $\mu$ and $M$ as 
\begin{equation}
 \epsilon \equiv \frac{\mu}{M} \ll 1\,, \label{eqn:toyexample-hierarchy}
\end{equation}
and ($\alpha$, $\beta$, $\gamma$, $\eta$) are dimensionless constants of at most order unity. The avoidance of a ghost mode is achieved by the condition $\vert \alpha \vert <1$. The general solution for $x$ and $y$ is a linear combination of solutions of the form,
\begin{equation}
 x = x_0 e^{-i\omega \tau} \,, \quad y = y_0 e^{-i\omega \tau} \,, \quad \tau \equiv \mu t\,,
\end{equation}
where $x_0$ and $y_0$ are constants. The equations of motion are reduced to
\begin{equation}
 (1-\alpha^2)\epsilon^2 \omega^4 - (\eta +\gamma^2-2\alpha\beta\epsilon + \epsilon^2)\omega^2 + (\eta -\beta^2) = 0\,, \label{eqn:toyexample-eq-omega}
\end{equation}
and
\begin{align}
 y_0 (\eta - \epsilon^2 \omega^2)=-\epsilon x_0 [\beta - \omega(i \gamma+\epsilon \alpha \omega)]
\,.
\end{align}
Eq.~\eqref{eqn:toyexample-eq-omega} admits a couple of ``fast'' solutions, 
\begin{equation}
 \omega^2 = \frac{\eta +\gamma^2}{1-\alpha^2}\epsilon^{-2} + \mathcal{O}(\epsilon^{-1})\,,
  \label{eqn:toyexample-fastsols}
\end{equation}
and a couple of ``slow'' solutions,
\begin{equation}
 \omega^2 = \frac{\eta - \beta^2}{\eta + \gamma^2} + 2\alpha\beta\frac{\eta -\beta^2}{(\eta +\gamma^2)^2} \, \epsilon + \frac{\eta -\beta^2}{(\eta +\gamma^2)^3} \left(5\alpha^2\beta^2 -\eta\alpha^2 - \beta^2 - \gamma^2 \right) \epsilon^2  + \mathcal{O}(\epsilon^3)\,. \label{eqn:toyexample-slowsols}
\end{equation}
The ``slow'' and ``fast'' solutions are characterized by $\omega^2=\mathcal{O}(\epsilon^0)$ and $\omega^2=\mathcal{O}(\epsilon^{-2})$, respectively, where the time scale is normalized by using the mass of the ``light'' oscillator.
The ``slow'' solutions \eqref{eqn:toyexample-slowsols} describe the low-energy physics of the ``light'' oscillator $x(t)$ under the influence of the ``heavy'' oscillator $y(t)$. On the other hand, the ``fast'' solutions describe the high-energy physics of the ``heavy'' oscillator $y(t)$ under the influence of the ``light'' oscillator $x(t)$. The ``fast'' solutions \eqref{eqn:toyexample-fastsols} are stable as long as 
\begin{equation}
 \frac{\eta +\gamma^2}{1-\alpha^2} M^2 > 0\,. \label{eqn:toyexample-UVstability}
\end{equation}
This condition can be satisfied even for $\eta \leq 0$.

Although the general solution is a linear combination of the ``fast'' solutions and the ``slow'' solutions,
\begin{align}
x=x_{\rm fast}+x_{\rm slow}\,, \quad y=y_{\rm fast}+y_{\rm slow}
\,,
\label{xy_fast+slow}
\end{align}
we may extract the ``slow'' solutions by restricting our consideration to low energy phenomena since it costs energies of order $M$ ($\gg \mu$) to excite the ``fast'' solution. Since \eqref{eqn:toyexample-Lagrangian} is a linear system, the ``fast'' solutions do not affect the ``slow'' solutions. Therefore, restricting our consideration to low energy phenomenon simply means ignoring the ``fast'' solutions,
\begin{align}
x \simeq x_{\rm slow}\,, \quad y \simeq y_{\rm slow}
\; .
\label{xy_slow}
\end{align}
This procedure is justified as long as the ``fast'' solutions do not develop instabilities, i.e.~the condition \eqref{eqn:toyexample-UVstability} is respected.

We would like to find a simple effective theory that describes the low-energy physics of the ``light'' oscillator $x(t)$ under the influence of the ``heavy'' oscillator $y(t)$. For this purpose, we first rewrite the Lagrangian \eqref{eqn:toyexample-Lagrangian} as
\begin{equation}
 L = \mu^2 \tilde{L}\,, \quad \tilde{L} = \frac{1}{2}\left[(\partial_{\tau} x)^2 + \epsilon^2 (\partial_{\tau}Y)^2 + 2 \alpha \epsilon \partial_{\tau}x \partial_{\tau}Y\right]  - \frac{1}{2}( x^2 + \eta Y^2 + 2\beta xY) - \gamma Y\partial_{\tau}x\,, \quad Y \equiv \epsilon^{-1}y\,. 
 \label{eqn:toyexample-Lagrangian2}
\end{equation}
where we have introduced the variable $Y$ of which amplitude scales as $\mathcal{O}(\epsilon^0)$ when the ``slow'' solutions are considered.
Since we are interested in the ``slow'' dynamics, the frequency in the unit of $\mu$ is of the order of $\epsilon^0$, meaning $\partial_{\tau}=\mathcal{O}(\epsilon^0)$. Therefore, the scaling of $\epsilon$ is explicit in each term of the Lagrangian \eqref{eqn:toyexample-Lagrangian2}.
The Euler-Lagrange equation for $Y(t)$ is 
\begin{equation}
\epsilon^2\partial_{\tau}^2Y + \alpha\epsilon\partial_{\tau}^2x + \eta Y + \beta x + \gamma\partial_{\tau}x = 0\,. \label{eqn:toyexample-Yeom}
\end{equation}
Considering the hierarchy \eqref{eqn:toyexample-hierarchy}, we expand $Y$ with respect to $\epsilon$ and obtain the ``slow'' solution of $Y$ as 
\begin{equation}
 Y = - \frac{\beta}{\eta} x - \frac{\gamma}{\eta}\partial_{\tau}x - \frac{\alpha}{\eta} \epsilon\partial_{\tau}^2 x + \frac{\beta}{\eta^2}\epsilon^2\partial_{\tau}^2 x + \frac{\gamma}{\eta^2} \epsilon^2\partial_{\tau}^3x + \mathcal{O}(\epsilon^3)\,,  \label{eqn:toyexample-heavymode}
\end{equation}
where we have assumed that $\eta$ is non-vanishing and of order unity. By substituting this to the Lagrangian and dropping total derivative, one obtains 
\begin{equation}
 \tilde{L} = \frac{1}{2}\left(1 + \frac{\gamma^2}{\eta} - \frac{2\alpha\beta}{\eta}\epsilon + \frac{\beta^2}{\eta^2}\epsilon^2\right)(\partial_{\tau}x)^2 - \frac{1}{2}\left(1 - \frac{\beta^2}{\eta}\right)x^2 + \frac{1}{2}\left(\frac{\alpha^2}{\eta}+\frac{\gamma^2}{\eta^2}\right)\epsilon^2(\partial_{\tau}^2x)^2 + \mathcal{O}(\epsilon^3)\,. \label{eqn:toyexample-effectivetheory}
 \end{equation}
This is the low-energy effective theory describing the ``slow'' solutions. Indeed, it is easy to show that the Euler-Lagrange equation from this effective action admits a couple of ``fast'' solutions and a couple of ``slow'' solutions. As expected, while the former do not agree with the ``fast'' solutions \eqref{eqn:toyexample-fastsols} from the original Lagrangian \eqref{eqn:toyexample-Lagrangian} (since they are outside the regime of validity of the low-energy effective theory \eqref{eqn:toyexample-effectivetheory}), the latter correctly reproduce \eqref{eqn:toyexample-eq-omega} up to $\mathcal{O}(\epsilon^2)$. If one wants, one can easily increase the precision of the ``slow'' solutions by systematically expanding $Y$ up to any order in $\epsilon$. This procedure is justified under the stability condition (\ref{eqn:toyexample-UVstability}), which may be satisfied even when $\eta<0$ (negative mass square of $y$ in the absence of $x$), as far as the properties of the ``slow'' solutions are concerned.

\subsection{General procedure}
\label{sec:EFTgeneral}

In the rest of the present paper we shall perform essentially the same analysis for several multi-field scalar systems in order to derive a single-field effective field theory that describes the low-energy/momentum physics of a ``light'' degree of freedom under the influence of ``heavy'' degrees of freedom. Before starting a concrete analysis, we find it convenient to make a general argument of the EFT reduction, extending the mechanical example in the previous subsection to field-theoretical ones.

An EFT is derived from a theory by integrating out modes of which dynamics we are not interested in. In many situations, we are interested in low-energy/momentum dynamics, so we integrate out high-energy/momentum degrees of freedom while keeping low-energy/momentum degrees of freedom. In general, the term ``integrating out'' refers to performing integrations of uninterested degrees of freedom in path integral. In the present paper, we will restrict our attention to the case when the dynamics is well-approximated by tree-level calculations, i.e.~classical dynamics. In this case, integrating out is performed by solving classical equations of motion for uninterested modes and then by substituting the solutions into the action where the initial conditions of the uninterested modes have to be uniquely determined by the modes which are kept in the EFT.

Let us denote the original action by $S_{\rm UV}[\chi,\varphi]$ and the EFT action by $S_{\rm IR}[\varphi]$ where $\chi$ and $\varphi$ are the modes which are to be integrated out and to be kept in the EFT reduction, respectively. Note that $\chi$ and $\varphi$ here are collective notations and not necessarily different fields: for instance, a field $\phi$ may be split into the high-energy/momentum, namely UV, modes $\phi_{\rm UV} \in \chi $ and the low-energy/momentum (IR) modes $\phi_{\rm IR} \in \varphi$ to derive a low-energy EFT.
The original equations of motion are
\begin{align}
\frac{\delta S_{\rm UV}}{\delta \chi(x)}=0\,, \quad \frac{\delta S_{\rm UV}}{\delta \varphi(x)}=0
\,,
\label{SUV_eom}
\end{align}
while the EFT equation of motion from $S_{\rm IR}[\varphi]=S_{\rm UV}|_{\chi=\chi(\varphi)}$ is
\begin{align}
\frac{\delta S_{\rm IR}}{\delta \varphi(x)} = \left. \frac{\delta S_{\rm UV}}{\delta \varphi(x)} \right|_{\chi=\chi[\varphi]} + \int \dd^d y \, \left. \frac{\delta \chi[\varphi](y)}{\delta \varphi(x)} \, \frac{\delta S_{\rm UV}}{\delta \chi(y)} \right|_{\chi=\chi(\varphi)} 
\,,
\label{SIR_eom}
\end{align}
where $\chi=\chi[\varphi]$ is the solution to the first equation in \eqref{SUV_eom}.
Note that we have assumed that the solution $\chi=\chi[\varphi]$ is unique at least locally so that $\delta \chi[\varphi](y)/\delta \varphi(x)$ is well-defined. Then, the second term in \eqref{SIR_eom} vanishes since $\chi[\varphi]$ is a solution to the original equation of motion, and thus \eqref{SIR_eom} correctly reproduces \eqref{SUV_eom} under the solution $\chi=\chi[\varphi]$.

Hence, a necessary step for the EFT reduction is just to solve the equations of motion for $\chi$ under an appropriate initial and boundary conditions. This is not an easy task, in general; thus, an EFT derivation is usually accompanied with derivative expansions. The derivative expansion is essentially the same as what we did in the previous subsection as the series expansion in terms of $\epsilon$. As we have seen in \eqref{eqn:toyexample-heavymode} and \eqref{eqn:toyexample-effectivetheory}, the solution of $Y$ and the low-energy effective Lagrangian can be systematically obtained as a series of $\epsilon$. The parameter $\epsilon$ in Sec.~\ref{sec:oscillators} was defined as the ratio between $\mu$ and $M$, which are the only dimensional parameters characterizing the dynamics of the ``light'' and the ``heavy'' oscillators, respectively. However, this definition cannot be directly applied to field theories since frequencies depend on momenta. Also, there is no mass parameter when $\varphi$ is massless. Instead, one can introduce $\epsilon \equiv E/M$ as an expansion parameter where $E$ is a reference scale satisfying $\partial = \mathcal{O}(E)$, where $\partial$ is a temporal and/or spatial derivative. We thus have $\partial/M = \mathcal{O}(\epsilon)$ and then the series expansion in terms of the derivative $\partial/M$ is nothing but the series expansion in terms of $\epsilon$. We will give concrete examples to explain how the derivative expansion works in the EFT reduction in the following subsubsections.

We would like to investigate situations where the Lorentz symmetry is spontaneously broken by a background, and care is needed to accommodate this nature. We for now focus on a massless field $\varphi$ for simplicity and will generalize the argument to the massive case later. Let us split the massless field $\varphi$ into a background $\bar\varphi$ and perturbation $\pi$,
\begin{align}
\varphi=\bar{\varphi}+\pi
\,.
\end{align}
For a concrete discussion, we here restrict our interest to the solutions of the form
\begin{align}
\varphi(t,x^i) \simeq C_0 \Delta t + \pi(t,x^i) \,, \quad \pi(t,x^i) = \int_{ p \ll M} \frac{\dd^d p}{(2\pi)^d} \pi(p) e^{ip_{\mu}x^{\mu}} + {\rm h.c}\,,
\quad \Delta t = t-t_0 \, ,
\label{ex:sol_phi}
\end{align}
where $t_0$ is some (arbitrary) reference time and $C_0$ is a constant.%
\footnote{The constancy of $C_0$ can be an approximate one. The following arguments hold as long as $\varphi$ can be expanded in the form \eqref{ex:sol_phi} locally in spacetime.}
The background $\bar{\varphi}=C_0 \Delta t$ has a non-vanishing gradient $C_{\mu}\equiv \partial_{\mu}\bar{\varphi}=(C_0, \bm{0})$. As a result, the dynamics of $\pi$ does not respect the Lorentz invariance due to the preferred direction $C_{\mu}$. Let us set the mass dimension of $\varphi$ to be $[\varphi]=-1$ so that the vector $C_{\mu}$ is dimensionless. Then, the amount of the symmetry breaking is of order unity if
\begin{align}
\partial_{\mu}\bar{\varphi}=\mathcal{O}(1)
\,,
\label{max_Lorentzviolation}
\end{align}
and the Lorentz symmetry is recovered in the limit $\partial_{\mu}\bar{\varphi} \to 0$.
The case \eqref{max_Lorentzviolation} may appear to spoil the convergence of the derivative expansion, but this is not the case. Using $p_{\mu}=\mathcal{O}(E)$, we find the following scaling
\begin{align}
\partial^n \varphi = \mathcal{O}(E^{n-1})
\,,
\label{del_phi_scale_Eonly}
\end{align}
that is, all the higher derivatives, $\partial^n$ with $n\geq 2$, are tiny compared with the UV physics scale, $M^{n-1} $, provided $\epsilon \ll 1$. Therefore, we can still use the derivative expansion treating the second- and higher- than second-order derivatives as perturbations while keeping first order derivatives non-perturbatively. We will indeed confirm that the scaling \eqref{del_phi_scale_Eonly} does not prevent the usage of the derivative expansion in a concrete example.

When the field $\varphi$ is massive, the background $\varphi \propto t$ cannot be an exact solution. Nonetheless, \eqref{ex:sol_phi} can be regarded as an approximate solution. Let $m$ be the mass scale of the light field $\varphi$. As long as we focus on the scales satisfying $m \lesssim E \ll M$, $C_{\mu}=\mathcal{O}(1)$ survives and we can use \eqref{ex:sol_phi} as an approximate solution during some finite time interval $\Delta t$ with $m^{-1} \gtrsim \Delta t \gg M^{-1}$. In the inflationary cosmology for instance, the timescale $m^{-1}$ is typically supposed to be the inflationary Hubble timescale or longer so that $\Delta t$ is long enough to describe inflationary observables.
In general, the background can be a function of space as well. All in all, \eqref{ex:sol_phi} can be used as an approximate solution for the physics within temporal/spatial intervals $\Delta t$ and $\Delta x$ satisfying $m^{-1} \gtrsim \Delta t \gg M^{-1}$ and $m^{-1} \gtrsim \Delta x \gg M^{-1}$, respectively, even with the spontaneous broken Lorentz symmetry by the background, $C_{\mu}=\mathcal{O}(1)$.

Before closing this subsection, let us discuss a consistency condition for the EFT reduction. The low-energy EFT is dedicated to studying long-time/large-scale dynamics of the original theory. The EFT can make robust predictions at low-energies/momenta as long as the UV modes are stable. However, when there exists an unstable UV mode, there is no proper justification to describe long-time/large-scale dynamics by ignoring short-time dynamics. Such an unstable UV mode develops the instability during a short period which the low-energy EFT cannot resolve. As a result, such an EFT is inconsistent, or at best UV sensitive, and cannot make robust predictions. Therefore, a consistency condition for the EFT reduction is the stability condition of the modes which are integrated out. We emphasize that there is no inconsistency for the IR modes to be unstable, as long as the stability of the UV modes is respected. The predictions of the low-energy EFT are trustable even if the IR degree of freedom is unstable, similarly to situations like the Jeans instability.

Let us elaborate on the classification of these situations. We consider a UV theory that contains a light field $\varphi$ and a heavy field $\chi$. In field theories, $\chi$ and $\varphi$ consist of a collection of modes characterized by their momenta, and we call the ones with high momenta $\chi_{\rm UV}$ and $\varphi_{\rm UV}$ and those with low momenta $\chi_{\rm IR}$ and $\varphi_{\rm IR}$, respectively, where UV and IR are classified with reference to the mass of the heavy field. The EFT of the IR degree of freedom $\varphi_{\rm IR}$ is obtained by integrating out $\chi_{\rm UV}, \chi_{\rm IR}$ and $\varphi_{\rm UV}$. Let us list the following four possible instabilities:
\begin{enumerate}
\item {\it Ghost instability in $\chi_{\rm UV}$ and $\varphi_{\rm UV}$.} We should prohibit the ghost instability since $S_{\rm UV}$ is supposed to be UV complete at least partially. The no-ghost conditions lead to consistency conditions of the resultant EFT.

\item {\it Tachyonic instability in $\chi_{\rm IR}$.} Even if $S_{\rm UV}$ is UV complete, there can be an unstable vacuum where the heavy modes are tachyonic, e.g.~the false vacuum of the Higgs potential. This instability is allowed from the point of view of UV, but is problematic in the EFT reduction. By definition, the EFT cannot make predictions about $\chi_{\rm IR}$. Therefore, even if $\chi_{\rm IR}$ is unstable and develops instability during a short period in $S_{\rm UV}$, the EFT $S_{\rm IR}$ cannot observe the instability of the heavy field within its own domain, and the predictions of the EFT for the light field is totally untrustable. We should impose the no-tachyon condition on $\chi_{\rm IR}$ to keep the validity of the single-field description.

\item {\it IR instability in $\varphi_{\rm IR}$.} As we explained, the IR instability in $\varphi_{\rm IR}$ is allowed from both UV and EFT perspectives. There is no consistency condition arising from this.

\item {\it Instability in low-energy part of $\varphi_{\mathrm{UV}}$}.
  In some cases, the instability of $\varphi_{\rm IR}$ continues to exist even above the cutoff scale of the EFT, i.e.,~the lower-energy part of $\varphi_{\rm UV}$ can be unstable. Then, the EFT with the cutoff being the mass of $\chi$ would be inconsistent since the unstable part of $\varphi_{\rm UV}$ may not be integrated out; if the unstable part of $\varphi_{\rm UV}$ is integrated out, predictions of the EFT are UV sensitive as we have explained above. We should impose the condition that the IR instability does not continue beyond the cutoff as a consistency condition in order that the EFT makes reasonable predictions. Note that this situation should be distinguished from the ghost instability and the tachyonic instability in $\chi_{\rm IR}$. In principle, one can extend the validity of the single-field EFT to include the unstable modes of $\varphi_{\rm UV}$, implying that this consistency condition is a matter of the cutoff.
  Note also that in principle the same argument holds in such cases where some non-trivial dispersion relations allow a bounded region of instability in $\varphi_{\mathrm{UV}}$, disconnected from the in/stabilities in $\varphi_{\mathrm{IR}}$.
\end{enumerate}
As introduced in Sec.~\ref{sec:intro}, we refer to avoidance of the the first two types of instabilities as the UV consistency conditions and avoidance of the last type of instability as the EFT predictivity condition, respectively. The UV consistency conditions should be imposed to have a healthy UV completion while the EFT predictivity condition should be imposed to heal the IR instability without relying the physics beyond the cutoff.
The purpose of the present paper is to clarify the UV consistency conditions of multi-field models and to show how such stability conditions of UV modes impose constraints on the resulting single-field EFT action especially in Lorentz-violating backgrounds.

\subsubsection{Example 1}

As a concrete example of the EFT reduction, we consider a theory
\begin{align}
S_{\rm UV}=\int \dd^4 x \mathcal{L}_{\rm UV}\,, \quad \mathcal{L}_{\rm UV} =  -\frac{1}{2}\partial_{\mu} \chi \, \partial^{\mu} \chi -\frac{1}{2}\partial_{\mu} \varphi \, \partial^{\mu} \varphi -\frac{M^2}{2}\chi^2-\frac{m^2}{2}\varphi^2 - g M \chi \varphi^2\,,
\label{ex:UV}
\end{align}
where $M$ and $m$ are mass parameters satisfying $m\ll M$ and $g$ is a coupling constant. Here, we canonically normalize the kinetic terms so that the mass dimension of $\varphi$ is $[\varphi]=1$. Although we consider a four-dimensional flat spacetime, extensions to other dimensions and to curved spacetimes are straightforward. The equation of motion for $\chi$ is
\begin{align}
(\Box -M^2) \chi = gM \varphi^2
\,,
\end{align}
of which solution is formally expressed as
\begin{align}
\chi = \frac{ gM}{\Box -M^2 }  \varphi^2
\,,
\label{chi_full}
\end{align}
provided that initial and boundary conditions are properly specified.
We split the field $\chi$ and $\varphi$ into the UV modes $\chi_{\rm UV}, \varphi_{\rm UV}$ and the IR modes $\chi_{\rm IR}, \varphi_{\rm IR}$, respectively, with respect to the heavy mass scale $M$.
As long as IR-UV mixing can be ignored, the UV modes do not affect the IR dynamics (decoupling theorem), and we can simply set $\chi=\chi_{\rm IR}$ and $\varphi=\varphi_{\rm IR}$ to analyze the low-energy behavior, similarly to \eqref{xy_fast+slow} and \eqref{xy_slow}. Then, the solution of $\chi$ is
\begin{align}
\chi = \chi_{\rm IR} = \frac{ gM}{\Box -M^2 }  \varphi^2_{\rm IR} = -\frac{g}{M} \sum_{n'=0}^{\infty} \left( \frac{\Box}{M^2} \right)^{n'} \varphi_{\rm IR}^2
\,,
\label{chi_series}
\end{align}
where $\chi$ is uniquely determined by $\varphi_{\rm IR}$ when the infinite series converges.
In this context, ``IR modes'' are defined as the modes that respect this convergence.
For later convenience, we put a prime to the index of the sum here. We can truncate the series when $\varphi_{\rm IR}$ is so low-energy/momentum that the solution $\chi=\chi_{\rm IR}$ in \eqref{chi_series} can be well-approximated by the first few terms of the infinite series. For instance, using the approximate solution $\chi_{\rm IR} \approx -\frac{g}{M}\varphi_{\rm IR}^2-\frac{g}{M^3}\Box \varphi_{\rm IR}^2$, the EFT action up to this order is
\begin{align}
S_{\rm IR}=\int \dd^4 x \left[ -\frac{1}{2}\partial_{\mu} \varphi_{\rm IR} \partial^{\mu} \varphi_{\rm IR}-\frac{m^2}{2} \varphi_{\rm IR} + g^2 \varphi^4_{\rm IR} + \frac{g^2}{M^2} \varphi_{\rm IR}^2 \Box \varphi_{\rm IR}^2 + \cdots  \right]
.
\label{actionIR_example1}
\end{align}
The higher order terms can be added to improve the accuracy of EFT so long as the infinite series \eqref{chi_series} converge. The convergence condition is formally given by $|\Box/M^2| < 1$ and thus $M$ is the cutoff of the derivative expansion. As we have explained, the expansion in terms of $\Box/M^2$ is essentially the same as the series expansion of $\epsilon=E/M$ because we have $\Box/M^2=\mathcal{O}(\epsilon^2)$.

\subsubsection{Example 2}
\label{sec:EFT_ex2}

In the previous example, we did not need to care the scaling of $\partial_\mu \varphi $ since the UV Lagrangian \eqref{ex:UV} only has one interaction that involves no derivatives. On the other hand, we should take it into account when there are other interactions, especially interactions arising from the field-space metric.

We consider the UV theory containing the interactions
\begin{align}
\mathcal{L}_{\rm UV} = &
-\frac{1}{2}\partial_{\mu} \chi \, \partial^{\mu} \chi -\frac{1}{2}\partial_{\mu} \varphi \, \partial^{\mu} \varphi -\frac{M^2}{2} \, \chi^2-\frac{m^2}{2} \, \varphi^2 - g M \chi \varphi^2
\nonumber\\ & 
-h \, \partial_{\mu} \varphi \, \partial^{\mu} \chi - h_{\varphi} \, \varphi \, \partial_{\mu} \varphi \, \partial^{\mu} \chi - \frac{f_{\chi}}{2} \, \chi \, \partial_{\mu} \varphi \, \partial^{\mu} \varphi
\,,
\label{ex:kin_int}
\end{align}
where the first line is the same as \eqref{ex:UV}, and the second line is added with $h,h_{\varphi}$ and $f_{\chi}$ taken to be constants. We note that our UV theory is still a partial one, allowing to have non-renormalizable operators such as $\chi (\partial \varphi)^2$. While the cutoff of the single-field EFT is typically given by $M$, we introduce another higher scale $M_{\mathrm{UV}}$ above which a more fundamental description needs to take over our partial UV theory, with the hierarchy $M_{\rm UV} \gg M$.

Variation of \eqref{ex:kin_int} with respect to $\chi$ leads to
\begin{align}
(\Box-M^2)\chi+h\Box \varphi  + \frac{1}{2}(2h_{\varphi}-f_{\chi}) \partial_{\mu}\varphi \partial^{\mu} \varphi + h_{\varphi} \varphi \Box \varphi -gM\varphi^2=0
,
\label{ex:eom}
\end{align}
which we now solve via the derivative expansion. Note that we only consider IR solutions and suppress the subscripts, IR and UV, hereinafter. 
The solution can be found as
\begin{align}
\chi&=\frac{1}{M^2}\sum_{n'=0}^{\infty} \left( \frac{\Box}{M^2} \right)^{n'} \left[  h\Box \varphi  + \frac{1}{2}(2h_{\varphi}-f_{\chi}) \partial_{\mu}\varphi \partial^{\mu} \varphi + h_{\varphi} \varphi \Box \varphi \right]
-\frac{g}{M} \sum_{n'=0}^{\infty} \left( \frac{\Box}{M^2} \right)^{n'} \varphi^2
\nn
&= \frac{2h_{\varphi}-f_{\chi}}{2M^2} \, \partial_{\mu}\varphi \partial^{\mu} \varphi
+\frac{1}{M^2}(h \Box \varphi + h_{\varphi} \varphi \Box \varphi ) + \frac{2h_{\varphi}-f_{\chi}}{2M^4} \Box (\partial_{\mu}\varphi \partial^{\mu} \varphi)
-\frac{g}{M}\,\varphi^2-\frac{g}{M^3}\,\Box \varphi^2
+\cdots
\,.
\label{chi_sol_full}
\end{align}
If we simply counted the number of derivatives, we would need to equally treat the interactions, the second line of \eqref{ex:kin_int}, in the derivative expansion, as all the terms inside the square brackets in the first line of \eqref{chi_sol_full} have two derivatives. However, this is not what we are interested in; rather, we aim to investigate the cases where the Lorentz symmetry is significantly violated by the gradient of $\varphi$ as in \eqref{max_Lorentzviolation}. As one can see from \eqref{chi_sol_full}, the convergence condition of the derivative expansion requires that higher derivatives of $\varphi$, $\partial^n$ with $n\geq 2$, must be tiny compared with the mass scale $M$ while it does not require the smallness of the first derivatives. The maximal amount of the Lorentz symmetry breaking that is allowed in the EFT below the large mass scales $M$ and $M_{\rm UV}$ can be characterized by
\begin{align}
\partial_{\mu}\varphi=\mathcal{O}(M M_{\rm UV})
\,.
\label{ex:gradient_phi}
\end{align}
with the scaling
\begin{align}
  \varphi = \mathcal{O}(M M_{\rm UV}m^{-1})\,, \quad
\partial^n \varphi = \mathcal{O}(M M_{\rm UV} E^{n-1})\,, \ (n=1,2,\cdots)
\, ,
\label{del_phi_scale}
\end{align}
which validates the derivative expansion for $E \ll M$.
Note that \eqref{del_phi_scale} leads to the variation $\Delta\varphi =\mathcal{O}(M M_{\rm UV} m^{-1}) \gg M_{\rm UV}$ in the interval $\Delta t, \Delta x \lesssim m^{-1}$,%
\footnote{The fact that $\Delta \varphi$ is larger than $M_{\mathrm{UV}}$, the cutoff scale of our partial UV theory, does not invalidate our analysis within its domain, provided $\vert X \vert^{1/4} = \mathcal{O}( \sqrt{M M_{\mathrm{UV}}} ) \ll M_{\mathrm{UV}}$, similarly to the occasion where the field value of inflaton is allowed to be larger than Planck mass during inflation and to another occasion where the QCD axion may take a value much higher than the QCD scale. }
implying that, in order to achieve this scaling compatible with the equations of motion, an approximate shift symmetry of $\varphi$ should be introduced. To this end, let us suppose the hierarchy of the couplings:
\begin{align}
|f_{\chi}|=\mathcal{O}(M_{\rm UV}^{-1})\,, \quad |h|=\mathcal{O}(1)\,, \quad
|h_{\varphi}| = \mathcal{O}(m M^{-1} M_{\rm UV}^{-1})\,, \quad |g| = \mathcal{O}(m^2 M^{-1} M_{\rm UV}^{-1})
\,,
\label{ex:shift}
\end{align}
where the shift symmetry becomes exact in the limit $m\to 0$ with $M$ and $M_{\rm UV}$ fixed. The scale $M_{\rm UV}$ is introduced to suppress the interactions to validate \eqref{ex:kin_int} as a partially UV complete model where $M_{\rm UV} \to \infty$ leads to a free theory.\footnote{Generically speaking, one may consider a situation where the coupling $g$ is not suppressed by $M_{\rm UV}$, since $\chi \varphi^2$ is a renormalizable term and may be present regardless of the theory beyond $M_{\mathrm{UV}}$. If it took the scaling $|g|=\mathcal{O}(m^2M^{-2})$, the renormalizable coupling would give a dominant contribution (see \eqref{ex:chi0}-\eqref{ex:chi2} below), and the effects from $f_{\chi}$ and $h_{\varphi}$ could be ignored. With the scaling in \eqref{ex:shift}, on the other hand, all the interactions can equally contribute, which is the situation of our interest in this work.
This type of scaling with a renormalizable term suppressed by $M_{\mathrm{UV}}$ can occur in a technically natural way; for example, since $M_{\mathrm{UV}} \to \infty$ recovers a free theory of $\chi$, the field $\chi$ may be understood as a perturbation around a nontrivial vacuum, and $M_{\mathrm{UV}}$ corresponds to the scale of its vev.
We explain a more systematic way of the scaling of the coupling constants in Sec.~\ref{sec:setup}. }
Also, the origin of $\varphi$ is supposed to be chosen properly so that the first condition in (\ref{del_phi_scale}) holds in the spacetime interval of interest $\lesssim m^{-1}$. 
As a result, the solution $\chi$ is organized into
\begin{align}
\chi&=\sum_{n=0}^{\infty} \chi_n =\chi_0+\chi_1+\chi_2+\cdots
\,, 
\label{ex:chi_series}\\
\chi_0&=-\frac{f_{\chi}}{2M^2} \, \partial_{\mu}\varphi \partial^{\mu} \varphi -\frac{g}{M} \, \varphi^2
\,, 
\label{ex:chi0} \\
\chi_1&=\frac{h}{M^2}\,\Box \varphi + \frac{h_{\varphi}}{M^2}(\varphi \Box \varphi + \partial_{\mu}\varphi \partial^{\mu} \varphi )
\,, \\
\chi_2&=-\frac{f_{\chi}}{2M^4}\,\Box (\partial_{\mu}\varphi \partial^{\mu} \varphi) - \frac{g}{M^3}\,\Box \varphi^2
\,,
\label{ex:chi2}
\end{align}
with 
\begin{align}
\chi_n=\mathcal{O}(\epsilon^n M_{\rm UV})
\,,
\end{align}
where $\chi_0$ and $\chi_1$ are originated from $n'=0$ terms of \eqref{chi_sol_full} and $\chi_n~(n \geq 2)$ are from $n'\geq 1$ terms. The subscript $n$ denotes the order of the solution where the scaling \eqref{del_phi_scale} is taken into account.\footnote{The equation of motion \eqref{ex:eom} may be solved by using the derivative expansion even if we do not assume the approximate shift symmetry; however, the series \eqref{ex:chi_series} is not a systematic series in this case. For instance, we have $\frac{g}{M^3}\Box \varphi^2 = g \times \mathcal{O}(M_{\rm UV}^2 M^{-1})$ which is comparable to the term $\frac{f_{\chi}}{M^2}(\partial_{\mu}\varphi \partial^{\mu}\varphi)=f_{\chi} \times \mathcal{O}(M^2_{\rm UV})$ if $|f_{\chi}|=\mathcal{O}(M^{-1}_{\rm UV}),~ |g|=\mathcal{O}(M/M_{\rm UV})$ and \eqref{del_phi_scale} are assumed. } Since the solution of $\chi$ is found as a series of $\epsilon$, the effective Lagrangian is easily computed accordingly.

\subsubsection{Bookkeeping parameter $\tilde{\epsilon}$}
\label{sec:epsilon}

In this subsubsection, we introduce a simple, systematic way to perform truncation of the series expansion that is employed in the previous subsubsections. It practically deduce the same effective description but can make it more transparent to solve the equations of motion via the derivative expansion. We again consider the example described by the action \eqref{ex:UV}. Let us replace $\partial \chi$ with $\tilde{\epsilon} \partial \chi$ to rewrite the action as
\begin{align}
S_{\rm UV}=
\int d^4 x \mathcal{L}_{\rm UV}\,, \quad \mathcal{L}_{\rm UV} = -\frac{1}{2}\,\tilde{\epsilon}^2\partial_{\mu} \chi \partial^{\mu} \chi -\frac{1}{2}\,\partial_{\mu} \varphi \partial^{\mu} \varphi-\frac{M^2}{2}\,\chi^2-\frac{m^2}{2}\,\varphi^2  - g M \chi \varphi^2\,.
\end{align}
This action shares the same concept with \eqref{eqn:toyexample-Lagrangian2}.
The equation of motion of $\chi$ is
\begin{align}
(\tilde{\epsilon}^2 \Box -M^2) \chi = gM \varphi^2
\,.
\end{align}
We solve this equation of motion order by order by using the ansatz
\begin{align}
\chi=\sum_{n=0}^{\infty} \tilde{\epsilon}^n \chi_n = \chi_0+\tilde{\epsilon}\chi_1+\tilde{\epsilon}^2 \chi_2 +\cdots
\,.
\end{align}
We can easily find the solution
\begin{align}
\chi_0=-\frac{g}{M}\varphi^2\,,\quad \chi_1=0\,, \quad \chi_2=-\frac{g}{M} \frac{\Box}{M^2}\varphi^2 \,, \cdots
\; .
\end{align}
Plugging these back into the action, we obtain the IR action
\begin{align}
S_{\rm IR}=\int d^4 x \left[ -\frac{1}{2}\partial_{\mu} \varphi \partial^{\mu} \varphi -\frac{m^2}{2}\varphi^2  + g^2 \varphi^4 + \tilde{\epsilon}^2 \frac{g^2}{M^2} \varphi^2 \Box \varphi^2 + \mathcal{O}(\tilde{\epsilon}^4) \right]
,
\end{align}
which recovers the result \eqref{actionIR_example1} by setting $\tilde{\epsilon}=1$. In this prescription, $\tilde{\epsilon}$ denotes the order of the derivative expansion, and the expansion parameter should be understood as $\tilde{\epsilon}\partial/M = \mathcal{O}(\epsilon)$ rather than $\tilde{\epsilon}$ itself.

The same procedure can be applied into general UV Lagrangians, e.g.~the theory including the interaction \eqref{ex:kin_int}. We emphasize that the replacement rule $\partial \chi \to \tilde{\epsilon} \partial \chi$ should be applied when the Lagrangian contains at most first-order derivatives of the fields.
We consider the Lagrangian
\begin{align}
\mathcal{L}_{\rm UV}=
-\frac{1}{2}\tilde{\epsilon}^2\partial_{\mu} \chi \partial^{\mu} \chi 
- \tilde{\epsilon} (h+h_{\varphi} \varphi)   \partial_{\mu} \varphi \partial^{\mu} \chi - \frac{1}{2}(1+f_{\chi}\chi) \partial_{\mu} \varphi \partial^{\mu} \varphi 
 -\frac{M^2}{2}\chi^2-\frac{m}{2}\varphi^2- g M \chi \varphi^2
 \,,
\end{align}
and find the solution order by order as 
\begin{align}
\chi=\sum_{n=0}^{\infty}\tilde{\epsilon}^n \chi_n=\chi_0+ \tilde{\epsilon}\chi_1 + \tilde{\epsilon}^2 \chi_2 +\cdots
\,.
\end{align}
One can easily confirm that the solutions $\chi_i$ agree with eqs.~\eqref{ex:chi0}-\eqref{ex:chi2} at each order. The original result is recovered by setting $\tilde{\epsilon}=1$ at the end of calculations. We treat the operators $\partial_{\mu} \chi \partial^{\mu} \chi $ and $ \partial_{\mu} \varphi \partial^{\mu} \chi $ as perturbations while we keep full non-linearity of $\partial_{\mu} \varphi \partial^{\mu} \varphi $ because we are interested in the Lorentz-violating background with a sufficiently large gradient $\partial_{\mu} \varphi \lesssim \mathcal{O}(M M_{\rm UV})$. The order of the bookkeeping parameter $\tilde{\epsilon}$ represents the order of the derivative expansion with taking the scaling \eqref{del_phi_scale} into account. Needless to say, the scaling with respect to $E$ is important for the derivative expansion while the overall normalization is irrelevant. Although we have used the canonical normalization with the mass dimension $[\varphi]=1$ to follow the convention in the above examples, it is straightforward to conduct the same analysis with other normalizations of $\varphi$.


\section{Reduction to single-field EFT from general multi-field space}
\label{sec:EFTreduction_der}

\subsection{Setup}
\label{sec:setup}

Our aim is to deduce the consistency conditions in deriving the low-energy single-field effective field theory (EFT) from classes of UV models.
The gist of the EFT reduction procedure has been demonstrated in Sec.~\ref{sec:preparation}, and we now proceed to more general setups.
A key ingredient in our study is a kinetic coupling from a field space in UV. We consider a general field-space metric
\begin{equation}
    \gamma_{AB}(\Phi) \, \dd\Phi^A \dd\Phi^B \; ,
    \label{fieldspace}
\end{equation}
where the upper-case Latin alphabets $A,B, \dots$ denote the field indices, and the metric $\gamma_{AB}$ is a function of the fields $\Phi^A$. The field space metric is supposed to be positive definite not to have ghost states. We can define the covariant derivative, the curvature tensors, and the scalar associated with $\gamma_{AB}$ in the same manner as for the spacetime metric and characterize the geometrical structure of the field space in terms of their properties. Nevertheless, our UV consistency conditions can and will be derived without specifying the structure of the field space explicitly.

The works \cite{Mukohyama:2016ipl,Mukohyama:2020lsu} have shown that the only two classes of kinetic terms that are free from the formation of caustics singularity in a planar configuration in the Minkowski spacetime are linear and DBI kinetic terms. Hence, the DBI-type kinetic term can be also regarded as a partial UV model of the single-field EFT, which we will investigate in Appendix \ref{subsec:DBI}. In this section, however, we focus on a class of models that have linear kinetic terms with the curved field space metric \eqref{fieldspace}, characterized by the action of a non-linear sigma model
\begin{align}
 S_{\rm UV} = \int d^dx\sqrt{-g}\,\mathcal{L}_{\rm UV}\,, \quad
\mathcal{L}_{\rm UV}=-\frac{1}{2} \, \gamma_{AB}(\Phi) \, g^{\mu\nu}\partial_\mu \Phi^A \partial_\nu \Phi^B - V(\Phi)
\,,
\label{Lagrangian_full}
\end{align}
as a (partial) UV completion of a single-field EFT. Here the Greek alphabets $\mu, \nu , \dots$ denote the spacetime indices, $g^{\mu\nu}$ and $g$ are the inverse and determinant of the spacetime metric $g_{\mu\nu}$. As mentioned in Introduction, we consider the theory in the Einstein frame, and the Einstein-Hilbert action is implicit in this section since the Einstein-Hilbert action is irrelevant to the EFT reduction here.

In this section, we perform the EFT reduction by using the prescription introduced in Sec.~\ref{sec:epsilon} without assuming any explicit form of either the field space metric $\gamma_{AB}(\Phi)$ or the potential $V(\Phi)$.
In order to integrate out $N-1$ heavy fields to reduce the non-linear sigma model to a single-field EFT of one light degree of freedom, our only assumption is a large hierarchy between the ``mass'' of the lightest field $\varphi$ and those of other heavy fields $\chi^a$, namely the approximate shift symmetry to the direction of $\varphi$, where the index $a$ runs through the $(N-1)$-dimensional subspace that excludes the $\varphi$ direction. Note that, due to the presence of the non-linear interactions through $\gamma_{AB}$, the ``mass'' is not simply the second derivatives of $V$ around a trivial background, and moreover, the notion of invariant mass is no longer available around a Lorentz-violating background in general. Nonetheless, this does not prevent a self-consistent EFT reduction, which we perform below.

Let $m$ be the mass scale of the lightest field $\varphi$. We suppose that the theory enjoys the exact shift symmetry $\varphi \to \varphi + c$ in the limit $m\to 0$, similarly to the scaling \eqref{ex:shift} discussed in Sec.~\ref{sec:EFT_ex2}. More precisely, the approximate shift symmetry is introduced as follows.
By setting the mass dimension of $\varphi$ to be $-1$ so that $\partial_{\mu}\varphi$ is dimensionless, we would like to study situations where the Lorentz symmetry is spontaneously broken by the background $\partial_{\mu}\varphi = \mathcal{O}(1)$. We assume that $\varphi$ respects an approximate shift symmetry, which is reflected by the $\varphi$-dependence of the potential and the field space metric that is suppressed by the small mass scale $m$, compared to other dimensional quantities that are normalized by the scale of the heavy physic. In this case, the change of the Lagrangian under the change $\Delta \varphi = \mathcal{O}(\Delta t)$ (or $\Delta \varphi = \mathcal{O}(\Delta x)$) is at most of order unity for the intervals $\Delta t, \Delta x=\mathcal{O}(E^{-1})\lesssim \mathcal{O}(m^{-1})$, and, in particular, the change is negligible during the period $\Delta t, \Delta x\ll \mathcal{O}(m^{-1})$. Hence, in this case the effect due to he violation of the shift symmetry would not be significant within the temporal/spatial intervals $\Delta t = \mathcal{O}(E^{-1})$ and $\Delta x = \mathcal{O}(E^{-1})$ that we are interested in, with $m\lesssim E \ll M$, i.e.~one can safely assume that the shift symmetry holds approximately. If preferred, one can canonically normalize $\varphi$ (or use other normalizations) after introducing the shift symmetry in this way.
Note that the approximate shift symmetry does not need to exist globally in the field space. We only assume that the theory enjoys the approximate shift symmetry at least during the intervals $\Delta t, \Delta x=\mathcal{O}(E^{-1})\lesssim \mathcal{O}(m^{-1})$ and consider the effective description at the scale  $m \lesssim E \ll M$ during this phase. This observation concludes that the resultant EFT does not necessarily have a Lorentz-invariant vacuum in the regime of validity; some EFTs are well-defined only in the Lorentz-violating phase, $\partial_{\mu}\varphi = \mathcal{O}(1)$ (see Appendix \ref{app:U(1)} for a concrete example).
We also note that the regime of our consideration, $m \lesssim E$, is demanded due to our interest in the physics around the scale $E^{-1}$ for which $\varphi$ is well approximated by the solution \eqref{ex:sol_phi} with $C_0 = \mathcal{O}(1)$. Of course our EFT itself can be also used to describe the low-energy physics $E\ll m$ if $\partial_{\mu}\varphi \ll 1 $ is well-defined for the intervals $\Delta t = \mathcal{O}(E^{-1})$ and $\Delta x = \mathcal{O}(E^{-1})$. 

In addition, we impose a hierarchy $M_{\rm UV} \gg M$, where $M_{\rm UV}$ is the cutoff of the partial UV model (not of EFT), since our UV model is a partial one and is allowed to have a generic field-space metric and a generic potential.\footnote{We assume that renormalizable interactions are well suppressed so that the system can be governed by a non-trivial field-space metric and a potential.} Although higher derivative interactions associated with the scale $M_{\rm UV}^{-1}$ should be included in the Lagrangian \eqref{Lagrangian_full}, these higher derivative terms can be ignored as long as the time/length scales are longer than $M_{\rm UV}^{-1}$. On the other hand, interactions through the potential and the field-space metric can be important when field expectation values are sufficiently large even if they are suppressed by $M_{\rm UV}$ in an appropriate way, e.g.~\eqref{ex:shift}.

Let us elaborate on how the three scales, $m,~M$ and $M_{\rm UV}$, appear in the general Lagrangian \eqref{Lagrangian_full}. As we have explained, the scale $m$ is introduced by assuming that the field-space metric and the potential are functions of the combination $m\varphi$, in the spirit of (approximate) shift symmetry. In contrast, we set $\chi^a$ to be dimensionless (denoted by $\hat{\chi}^a$) and assume that the background value of $\hat{\chi}^a$ is of order unity. This background is consistently obtained from the Lagrangian
\begin{align}
\mathcal{L}_{\rm UV}=M_{\rm UV}^{d-2}\left[ -\frac{1}{2}\left( \hat{\gamma}_{ab} \nabla_{\mu} \hat{\chi}^a \nabla^{\mu} \hat{\chi}^b+2M \hat{h}_a \nabla_{\mu} \hat{\chi}^a \nabla^{\mu} \varphi+M^2 \hat{f}\nabla_{\mu} \varphi \nabla^{\mu} \varphi  \right)- M^2 \hat{V} \right]
, \label{Lagrangian_dim_explicit}
\end{align}
where $\hat{\gamma}_{ab}, \hat{h}_a, \hat{f}$ and $\hat{V}$ are dimensionless functions of the dimensionless variables $\hat{\chi}^a$ and $m\varphi$. When we canonically normalize the fields, all the interactions are suppressed by $M_{\rm UV}$, justifying the use of \eqref{Lagrangian_dim_explicit} as the partial UV completion of the single-field EFT. In particular, the limit $M_{\rm UV} \to \infty$ (with $d>2$) leads to the weak coupling limit. In four dimensions, one can recover the scaling \eqref{ex:shift} from \eqref{Lagrangian_dim_explicit} after the normalization. On the other hand, all the nonlinear interactions of $\hat{\chi}^a$ can equally contribute around the background with $\hat{\chi}^a, \partial_{\mu}\varphi =\mathcal{O}(1)$, providing a non-trivial IR dynamics. We notice that the scale $M_{\rm UV}$ is irrelevant to the classical dynamics, so long as interactions between $\Phi^A=(\hat{\chi}^a,\varphi)$ and other fields are ignored, since $M_{\rm UV}$ is just the overall factor of the Lagrangian. After integrating out the heavy fields, the single-field EFT is schematically given by
\begin{align}
\mathcal{L}_{\rm IR} = M_{\rm UV}^{d-2}M^2 \sum_{i,j,k} c_{ijk} (m\varphi)^k \left(\frac{\nabla}{M} \right)^j (\nabla \varphi)^i
\,, \label{EFT_schematic}
\end{align}
with dimensionless constants $c_{ijk}$. In the following subsections, we will determine the concrete form of the EFT Lagrangian via the derivative expansion.

Our partial UV theory is controlled by the three independent scales, $M,~m$ and $M_{\rm UV}$, which are in a one-to-one correspondence with the scales determining the potential, the field-space metric and the cutoff of our partial UV theory, respectively. The Lagrangian  \eqref{Lagrangian_dim_explicit} is rewritten as
\begin{align}
\mathcal{L}_{\rm UV}=-\frac{1}{2}\left[ M_{\rm UV}^{d-2} \hat{\gamma}_{ab} \nabla_{\mu} \hat{\chi}^a \nabla^{\mu} \hat{\chi}^b+2 M_{\rm UV}^{d/2-1} \mathcal{M}_{\rm UV}^{d/2-1} \hat{h}_a \nabla_{\mu} \hat{\chi}^a \nabla^{\mu} \hat{\varphi}+\mathcal{M}_{\rm UV}^{d-2} \hat{f}\nabla_{\mu} \hat{\varphi} \nabla^{\mu} \hat{\varphi}  \right]- M_{\rm UV}^{d-2} M^2 \hat{V}
, \label{Lagrangian_dim_explicit2}
\end{align}
in terms of the dimensionless variables $\hat{\chi}$ and $\hat{\varphi}=m\varphi$ where $\mathcal{M}_{\rm UV}^{d/2-1}=M_{\rm UV}^{d/2-1} M/m$. Roughly speaking, the scales $M_{\rm UV}$ and $\mathcal{M}_{\rm UV}$ are associated with the curvature scales in the direction to $\hat{\chi}^a$ and $\hat{\varphi}$ (or they might be the breaking scales of an internal symmetry if our non-linear sigma model is obtained via a spontaneous symmetry breaking from a more fundamental theory), while the combination $M_{\rm UV}^{d-2} M^2$ controls the height of the potential, providing the masses of $\hat{\chi}^a$ and $\hat{\varphi}$. Our assumption is the existence of the hierarchy $\mathcal{M}_{\rm UV} \gg M_{\rm UV} ~(> M)$ in terms of this parametrization. From the EFT perspective, on the other hand, the parameters, $m,~M$ and $M_{\rm UV}$, are more useful which we shall adopt throughout the paper.

Having detailed our setup and the justification to use \eqref{Lagrangian_full} as the partial UV completion of $P(\varphi,X)$ model, we now canonically normalize $\chi^a$ so that the mass scale $M$ is expected to be evaluated by 
\begin{align}
\left| \frac{\partial^2 \mathcal{L}_{\rm UV}}{\partial \chi^a \partial \chi^b} \right| = \mathcal{O}(M^2)
\,,
\end{align}
at the leading order, where $\chi^a$ are the canonically normalized fields and their expectation values are $\mathcal{O}\big( M_{\rm UV}^{(d-2)/2} \big)$.
Shortly, we will provide the precise definition of the parameter $M^2$. In the Lagrangian \eqref{Lagrangian_dim_explicit} where the $M$-dependence is explicit, it is clear that the operators $\hat{\gamma}_{ab} \nabla_{\mu} \hat{\chi}^a \nabla^{\mu} \hat{\chi}^b = \mathcal{O}(E^2)$ and $\hat{h}_a \nabla_{\mu} \hat{\chi}^a \nabla^{\mu} \varphi = \mathcal{O}(E)$ are subdominant in the low-energy/momentum scales $E\ll M$. After canonically normalizing, we introduce the bookkeeping parameter $\tilde{\epsilon}$ to manifest the smallness of the operators as we explained in Sec.~\ref{sec:epsilon}. The formal series expansion in terms of $\tilde{\epsilon}$ agrees with the expansion in terms of $E/M$. Hereinafter, we always use the normalized heavy fields.

Our formulation in this section applies at fully nonlinear orders, without relying on explicit configurations of the fields as long as the derivative expansion converges, i.e.~$E\ll M$. We do not need to explicitly split the field into the background and perturbations since the analysis is nonlinear. This section is dedicated to providing the general relations between the UV action and the EFT action within our setup.

\subsection{Two-field UV models}
\label{sec_twofield}

We first concentrate on two-field models $\Phi^A=\{\chi,\varphi \}$ with $N=2$ and will discuss the generic multi-field models with an arbitrary $N$ in Sec.~\ref{subsec:multifield}. The components of the field space metric are 
\begin{align}
\gamma_{AB} \, \dd \Phi^A \dd \Phi^B= \gamma(\chi,\varphi) \, \dd \chi^2 + 2 h(\chi,\varphi) \, \dd \chi \dd \varphi + f(\chi,\varphi) \, \dd \varphi^2
\,,
\end{align}
where $\gamma,h$ and $f$ are in general functions of $\chi$ and $\varphi$.

As we explained in Sec.~\ref{sec:epsilon}, the equation of motion for $\chi$ can be systematically solved by using the derivative expansion when the parameter $\tilde{\epsilon}$ is introduced. We first write the UV action as
\begin{align}
S_{\rm UV}=\int d^dx \sqrt{-g}\left[ -\frac{1}{2}\tilde{\epsilon}^2 \gamma(\chi,\varphi) (\nabla \chi)^2 - \tilde{\epsilon} h(\chi,\varphi) \nabla_{\mu} \chi \nabla^{\mu} \varphi + f(\chi,\varphi) X-V\right]
,
\label{Lagrangian_twofields}
\end{align}
where $X\equiv -(\nabla \varphi)^2/2$. The equation of motion for $\chi$ is
\begin{align}
\tilde{\epsilon}^2 \left( \gamma \Box \chi + \frac{1}{2} \gamma_{\chi} (\nabla \chi )^2 +\gamma_{\varphi} \nabla_{\mu}\chi \nabla^{\mu}\varphi 
\right)
+\tilde{\epsilon} \left(h\Box \varphi-2X h_{\varphi} \right) +Xf_{\chi}-V_{\chi}=0
\,,
\label{full_eom}
\end{align}
where the subscripts $\chi$ and $\varphi$ represent derivatives with respect to the specified variables.
We can then find a solution for the heavy field $\chi$ order by order as a series in terms of $\tilde{\epsilon}$ as
\begin{align}
\chi= \chi[\varphi] = \sum_{n=0}^{\infty}\tilde{\epsilon}^n\chi_n[\varphi]\,. \label{eqn:chi-expanded}
\end{align}
For the expansion (\ref{eqn:chi-expanded}), the equation of motion \eqref{full_eom} reads
\begin{align}
f_{\chi}(\chi_0,\varphi) X-V_{\chi}(\chi_0, \varphi) &=0\,,  
&{\rm at}~\mathcal{O}(\tilde{\epsilon}^0)\,,
\label{leading_eom} \\
h(\chi_0,\varphi) \Box \varphi-2h_{\varphi}(\chi_0,\varphi) X &=M^2(\chi_0, \varphi) \chi_1 \,, 
&{\rm at}~\mathcal{O}(\tilde{\epsilon}^1)\,,
\label{next_eom} \\
\gamma(\chi_0,\varphi) \Box \chi_0 + \frac{1}{2} \gamma_{\chi}(\chi_0,\varphi) (\nabla \chi_0 )^2 +\gamma_{\varphi}(\chi_0,\varphi) \nabla_{\mu}\chi_0 \nabla^{\mu}\varphi  & \nn
+\chi_1 \left( h_{\chi}(\chi_0,\varphi) \Box \varphi-2X h_{\varphi\chi}(\chi_0,\varphi) \right)
-\frac{1}{2}\chi_1^2 \left( V_{\chi\chi\chi}(\chi_0,\varphi) - X f_{\chi\chi\chi}(\chi_0,\varphi) \right)
&=M^2(\chi_0,\varphi) \chi_2 \,, 
&{\rm at}~\mathcal{O}(\tilde{\epsilon}^2)\,,
\label{nextnext_eom}
\end{align}
and so on. Here, we define the function
\begin{align}
M^2& \equiv V_{\chi\chi}-X f_{\chi\chi}
\,,
\label{Mdef}
\end{align}
which corresponds to a squared mass scale related to the cutoff of the derivative expansion. In fact, as in \eqref{next_eom} and \eqref{nextnext_eom}, the equation of motion at $\mathcal{O}(\tilde{\epsilon}^n)$ ($n=1,2,\cdots$) generically takes the form,
\begin{align}
\mathcal{F}_n[\varphi; \chi_0, \cdots, \chi_{n-1}]=M^2 \chi_n
\,,
\end{align}
because $\chi_n$ appears only from the last two terms of \eqref{full_eom}.
Here $\mathcal{F}_n$ are those functionals of $\varphi$ and lower orders of $\chi$ which correspond to the $n$-th order of equation of motion for $\chi$.
The leading-order equation of motion \eqref{leading_eom} is a ``constraint'' equation of $\chi_0$. The implicit function theorem guarantees that at least locally there exists a function $\chi_0=\chi_0(\varphi,X)$ that satisfies the leading-order equation of motion \eqref{leading_eom} if $M^2\neq 0$. Using this leading-order solution, the solutions higher-order in $\tilde{\epsilon}$ are uniquely determined as long as $M^2\neq 0$; schematically,
\begin{align}
\chi_0=\chi_0[\varphi]\,, \quad 
\chi_1=\chi_1[\varphi,\chi_0[\varphi] ]=\chi_1[\varphi]\,, \quad
\chi_2=\chi_2[\varphi,\chi_0[\varphi],\chi_1[\varphi] ]=\chi_2[\varphi]\,, ~\cdots \,.
\label{eqn:chi_n-sol}
\end{align}
This solution describes the response of the heavy field $\chi$ to the low-energy physics of the light field $\varphi$ and thus does not allow for independent initial conditions for $\chi$ and $\dot{\chi}$ on the initial Cauchy hypersurface. It takes high energies or/and momenta of order $M$ for $\chi$ to deviate from this particular solution. Since we are interested in physics at low energies and momenta sufficiently below $M$, we employ the solution (\ref{eqn:chi-expanded}) with (\ref{eqn:chi_n-sol}).

The effective Lagrangian for $\varphi$ is then obtained by substituting the solution $\chi=\chi[\varphi]$ given by the expansion (\ref{eqn:chi-expanded}) with (\ref{eqn:chi_n-sol}) into the Lagrangian \eqref{Lagrangian_twofields}. 
The solution up to $\mathcal{O}(\tilde{\epsilon}^1)$ is needed to obtain the effective Lagrangian up to $\mathcal{O}(\tilde{\epsilon}^2)$. The EFT Lagrangian is
\begin{align}
\mathcal{L}_{\rm IR}&=fX-V- \tilde{\epsilon} h \nabla_{\mu}\chi_0 \nabla^{\mu}\varphi
\nn
&+\tilde{\epsilon}^2\left(-\frac{1}{2}\gamma (\nabla \chi_0)^2 + \frac{h^2}{2M^2}(\Box \varphi)^2-\frac{2h h_{\varphi}}{M^2} X \Box \varphi +\frac{2h_{\varphi}^2 }{M^2}X^2 \right)
+\mathcal{O}(\tilde{\epsilon}^3)
\,,
\label{EFT_Lag_gen}
\end{align}
after integration by parts and using the equations \eqref{leading_eom} and \eqref{next_eom}, where $\chi_0=\chi_0(\varphi,X)$ is understood as the solution to \eqref{leading_eom}. Here the arguments of each function are evaluated at $\chi = \chi_0(\varphi,X)$. One can confirm that the effective Lagrangian \eqref{EFT_Lag_gen} correctly reproduces the original equation of motion for $\varphi$ under the solution \eqref{eqn:chi-expanded} up to $\mathcal{O}(\tilde{\epsilon}^2)$.\footnote{The agreement is confirmed by using e.g. the Mathematica package {\sc xTras}~\cite{Nutma:2013zea}.} Also, if needed, one can systematically increase the accuracy of EFT by using a higher-order solution.

The k-essence theory is obtained as the leading-order EFT of \eqref{Lagrangian_twofields},
\begin{align}
\mathcal{L}_{\rm IR} &\, =P(\varphi,X)+\mathcal{O}(\tilde{\epsilon}) \,,
\label{eqn:LIR_twofields}
\end{align}
where
\begin{align}
P(\varphi,X)& \equiv f(\chi_0,\varphi)  X-V(\chi_0,\varphi) \, \big\vert_{\chi_0 = \chi_0(\varphi,X)}
\,.
\end{align}
We compute the first and second derivatives of $P$ with respect to $X$ by using the chain rule and the implicit function theorem:
\begin{align}
P_X&=\frac{\partial P}{\partial X}=\left. f+(f_{\chi}X-V_\chi)\frac{\partial \chi_0}{\partial X}  \right\vert_{\chi_0 = \chi_0(\varphi,X)}= f \, \big\vert_{\chi_0 = \chi_0(\varphi,X)}
\,, \label{PX_rel}
\\
P_{XX}&=\frac{\partial^2 P}{\partial X^2}=\left. f_{\chi} \frac{\partial \chi_0}{\partial X} \right\vert_{\chi_0 = \chi_0(\varphi,X)} = \left. \frac{f_{\chi}^2}{M^2} \right\vert_{\chi_0 = \chi_0(\varphi,X)}
\,, \label{PXX_rel}
\end{align}
and other derivatives are also computed accordingly. Note that, as was shown in \cite{Mukohyama:2020lsu}, the EFT derivation at the leading order in $\tilde{\epsilon}$ is the Legendre transformation especially in the shift symmetric case and then the relations \eqref{PX_rel} and \eqref{PXX_rel} are simply the properties of the Legendre transformation.

We have not specified the sign of $M^2$ so far. The necessary condition for the existence of the solution $\chi$ required $M^2\neq 0$ while it said nothing  about the sign of $M^2$. On the other hand, the positivity bounds conclude $P_{XX}>0$ around a Lorentz-invariant background. In order to see how this constraint arises in the present setup restricted to a Lorentz-invariant background, let us consider the Lorentz-invariant background, $\chi, \varphi=$ constant, where the background values of $\chi,\varphi$ are determined by $V_{\chi}=0$ and $V_{\varphi}=0$. One can easily find that the no-tachyon condition of $\chi$ requires
\begin{align}
V_{\chi\chi}|_{\chi, \varphi= {\rm constant} }>0
\,, 
\end{align}
as a necessary condition around the Lorentz-invariant background. Since the background $\varphi$ is constant, we have $M^2=V_{\chi\chi}>0$ evaluated at this background, concluding $P_{XX}(X=0)>0$ as a UV consistency condition for the Lorentz-invariant EFT reduction.

The stability conditions around generic Lorentz-violating backgrounds are not straightforwardly obtained, on the other hand. As we have mentioned, the ``mass'' of the field is not simply evaluated by the second derivative of the potential. Furthermore, the friction term may allow to have a stable UV state even for the convex shape of the potential as we have seen in Sec.~\ref{sec:oscillators}. This issue in fact contains a rich content, and we thus leave the analysis on the stability conditions around Lorentz-violating backgrounds to Sec.~\ref{sec:UV_consistency}, while in this section we only state that $M^2>0$ is not an immediate consequence of the UV stability if the Lorentz symmetry is spontaneously broken.

Before moving to the next subsection, let us discuss a freedom of field redefinitions. The description of the theory is not unique and the EFT operators can be changed via field redefinitions. We focus on the shift symmetric theory and consider transformations which preserve the shift symmetry manifestly.
First of all, transformations of $\chi$ are irrelevant for the EFT because the field $\chi$ is integrated out. 
A field transformation that preserves the shift symmetry is the change according to
\begin{align}
\varphi \to \varphi + \tilde{\epsilon} g(\chi)
\,,
\label{change:phi+g}
\end{align}
where $g$ is an arbitrary function of $\chi$. The change must be of the order of $\tilde{\epsilon}$ because this transformation leads to
\begin{align}
\dd \varphi \to \dd \varphi + \tilde{\epsilon} g_{\chi}(\chi) \dd \chi \; ,
\end{align}
and $g_{\chi}$ contributes to $\tilde{\epsilon}^2 \dd \chi^2$ and $\tilde{\epsilon} \dd \chi \dd \varphi$. Therefore, the field redefinition \eqref{change:phi+g} does not contribute to the leading operator $P(X)$ and only changes the subleading operators. One can also consider a perturbative field redefinition including derivatives after (or before) the EFT is derived, say
\begin{align}
\varphi \to \varphi'=\varphi - \tilde{\epsilon}^2 \, \frac{\Box \varphi}{M^2}
\,,
\label{pert_trans}
\end{align}
which generates the $(\Box \varphi)^2$ operator from the leading operator $P(X)$,
\begin{align}
P \to P + \tilde{\epsilon}^2 \frac{P_X}{M^2} \nabla_{\mu}\varphi \nabla^{\mu} \Box \varphi +\cdots = P  - \tilde{\epsilon}^2 \frac{P_X}{M^2} (\Box \varphi)^2 +\cdots
\label{PX_trans}
\end{align}
where $\cdots$ are terms irrelevant for our consideration here, and we take integration by parts to obtain the last expression. The inverse of the transformation is given as a series in $\tilde{\epsilon}$. We can change the coefficient of $(\Box \varphi)^2$ in \eqref{EFT_Lag_gen} by using this transformation if $P_X\ne 0$, for instance. Nonetheless, the derivative terms are higher orders in $\tilde{\epsilon}$ and change the subleading operators only. All in all, the field redefinitions can change the subleading operators whereas the leading operators described by $P(X)$ are invariant. We can use the freedom of the field redefinitions to discuss the UV consistency conditions in Sec.~\ref{sec:UV_consistency} when the bounds on the shift symmetric parts of the leading operators $P(X)$ are concerned.\footnote{On the other hand, non-shift symmetric parts of $P(\varphi,X)$ can be changed by field redefinitions. For instance, one can consider the transformation according to $\varphi \to \varphi+m\varphi^2/M^2$. }

\subsection{General multi-field UV models}
\label{subsec:multifield}

In this subsection, we continue to perform the EFT reduction by extending the two-field model to a generic multi-field model with a field space geometry of any (finite) dimension $N$.

We introduce a mass scale $m$ as a controlling parameter so that the theory enjoys the exact shift symmetry in the limit $m\to 0$. In this limit, the existence of the exact shift symmetry,
\begin{align}
\varphi \to \varphi + c\,,
\end{align}
implies that the field space metric admits a Killing vector
\begin{equation}
     \xi = \frac{\partial}{\partial \varphi} \equiv \partial_\varphi \; ,
     \label{Killing}
\end{equation}
or $\xi^A = \delta^A_\varphi$ in the component notation, and that the potential is independent of $\varphi$. Using $\varphi$ as a coordinate~\footnote{This way, $\varphi$ is prefixed to become the actual dynamical degree of freedom in the corresponding low-energy EFT. Our setup is in this sense modified and/or illustrative as compared to other scenarios that share some common philosophy, such as the completely generic multi-field setup \cite{Sasaki:1995aw}, the gelaton scenario \cite{Tolley:2009fg}, field space with sharp turns \cite{Achucarro:2010da,Shiu:2011qw}, and the geometrical destabilization \cite{Renaux-Petel:2015mga}. }, $\Phi^N=\varphi$, and denoting other coordinates by $\Phi^a=\chi^a$ ($a=1,\cdots,N-1$), so that $\Phi^A=(\chi^a, \varphi)$, the general shift-symmetric field-space metric is written in the form
\begin{align}
\gamma_{AB}\dd \Phi^A \dd \Phi^B=\gamma_{ab}(\chi) \dd \chi^a \dd \chi^b + 2 h_a(\chi) \dd \chi^a \dd \varphi + f(\chi) \dd \varphi^2
\,.
\label{multi_fieldmetric}
\end{align}
According to Frobenius's theorem, the Killing vector is hypersurface orthogonal if and only if
\begin{align}
 \xi_{[A} \mathcal{D}_B \xi_{C]}=0 \,,
 \label{hypersurface_ortho}
\end{align}
is satisfied where $\mathcal{D}_A$ is the covariant derivative compatible with the field space metric $\gamma_{AB}$ (see e.g.~\cite{Wald:1984rg}). When \eqref{hypersurface_ortho} holds, the variable $\varphi$ can be chosen to respect the reflection symmetry,
\begin{align}
\varphi \to -\varphi
\,,
\end{align}
and we can set $h_{a}(\chi)=0$ without loss of generality.
For $N=2$, the existence of the Killing vector (the shift symmetry) immediately concludes the hypersurface orthogonality \eqref{hypersurface_ortho} while it does not necessarily hold for general $N$. Note, however, that the leading shift-symmetric operators $P(X)$ enjoys the accidental reflection symmetry whether or not the full theory does. The field space metric and the potential can depend on $\varphi$ in the way that the $\varphi$-dependence is suppressed by the small mass scale $m$ associated with $\varphi$, when the shift symmetric is not exact.

We now recover a small but non-vanishing $m$. We keep generality and do not add any extra ingredient such as the reflection and exact shift symmetries. We only assume that the non-linear sigma model has one light direction described by the approximate shift symmetry at least locally.
The general Lagrangian is given by
\begin{align}
\mathcal{L}_{\rm UV}=
-\frac{1}{2}\tilde{\epsilon}^2 \gamma_{ab}(\chi,\varphi ) \nabla_{\mu} \chi^a \nabla^{\mu} \chi^b - \tilde{\epsilon} h_a(\chi,\varphi) \nabla_{\mu} \chi^a \nabla^{\mu} \varphi + f(\chi,\varphi) X- V(\chi,\varphi)
\,,
\end{align}
where the parameter $\tilde{\epsilon}$ is introduced. The solutions for the heavy fields $\chi^a$ can be found as a series in terms of $\tilde{\epsilon}$:
\begin{align}
\chi^a=\sum_{n=0}^{\infty}\tilde{\epsilon}^n \chi^a_n[\varphi]
\end{align}
where the field space indices $a,b,\cdots$ and the order of the $\tilde{\epsilon}$ expansion, $n$, should not be confused. The zeroth-order equation of motion reads
\begin{align}
\partial_a f(\chi_0,\varphi) X-\partial_a V(\chi_0,\varphi)=0
\,,
\label{constraint_multi}
\end{align}
where $\partial_a$ is the derivative with respect to the heavy fields $\chi^a$. Provided that the matrix
\begin{align}
M^2_{ab}\equiv \partial_a \partial_b V - \partial_a \partial_b f \, X\,,
\label{defMab}
\end{align}
has a non-zero determinant, we can solve \eqref{constraint_multi} for $\chi_0^a$ in favor of $\varphi$ and $X$ according to the implicit function theorem. The equation of motion at $\mathcal{O}( \tilde{\epsilon}^n )$ with $n\geq 1$ takes the form, with $\mathcal{F}_{a,n}$ abbreviating the $n$-th order equation of motion for $\chi$,
\begin{align}
\mathcal{F}_{a,n}[\varphi; \chi_0,\cdots,\chi_{n-1}] = M^2_{ab}\chi^b_n
\,,
\end{align}
implying that the solution at $\mathcal{O}(\tilde{\epsilon}^n)$ is uniquely determined by the solutions at $\mathcal{O}(\tilde{\epsilon}^m)$ with $m<n$ as long as ${\rm det}M^2_{ab}\neq 0$ is respected.
Solving the equations of motion recurrently, all $n$-th order solutions are uniquely obtained as a function of $\varphi$ and its derivatives.

Let us focus on the leading-order EFT Lagrangian,
\begin{equation}
    {\cal L}_{\rm IR} = f(\chi_0,\varphi ) \, X - V(\chi_0,\varphi) \, \vert_{\chi_0^a = \chi_0^a(\varphi, X)} +\mathcal{O}(\tilde{\epsilon}^1)= P(\varphi, X) +\mathcal{O}(\tilde{\epsilon}^1)\; ,
\end{equation}
where $\chi_0^a(\varphi, X)$ is the solution to \eqref{constraint_multi} and $P(\varphi, X)$ is a function of $\varphi$ and $X$. Now we observe
\begin{align}
    P_X & = \frac{\partial P}{\partial X} = f + \left( \partial_a f \, X - \partial_a V \right) \frac{\partial \chi^a}{\partial X} \, \bigg\vert_{\chi_0^a = \chi^a(\varphi, X)} = f \, \big\vert_{\chi_0^a = \chi^a(\varphi, X)} \; ,
    \label{PX_multi} \\
    P_{XX} & = \frac{\partial^2 P}{\partial X^2}
    = \partial_a f \, \frac{\partial \chi^a}{\partial X} \, \bigg\vert_{\chi_0^a = \chi_0^a(\varphi, X)}
    = \partial_a f \left( M^{-2} \right)^{ab} \partial_b f \, \Big\vert_{\chi_0^a = \chi_0^a(\varphi, X)} \; ,
    \label{PXPXX_multi}
\end{align}
where $\left( M^{-2} \right)^{ab} $ is the inverse of $M^2_{ab}$. In order to obtain the above relations, we have used \eqref{constraint_multi} and its variation for a fixed $\varphi$,
\begin{equation}
    M^2_{ab} \, \delta\chi^b = \partial_a f \, \delta X 
    \implies
    \frac{\partial \chi^a}{\partial X} \, \bigg\vert_{\chi_0^a = \chi^a(\varphi, X)} =(M^{-2})^{ab} \partial_b f \, \Big\vert_{\chi_0^a = \chi_0^a(\varphi, X)}
    \,,
\end{equation}
which is a direct consequence of the implicit function theorem. It is important to stress that $P_{XX}$ is given by a quadratic form; thus, the sign of $P_{XX}$ is fixed when $M^2_{ab}$ is either the positive definite or the negative definite. We discuss the signature of $M^2_{ab}$ in Sec.~\ref{sec:UV_consistency}.


\section{UV consistency and EFT predictivity of single-field EFT}
\label{sec:UV_consistency}

The EFT reduction performed in Sec.~\ref{sec:EFTreduction_der} applies at fully nonlinear orders and is valid as far as the second and higher derivatives of $\varphi$ are sufficiently small. The typical energy/momentum scale of our interest is denoted by $E$ and the derivative expansion converges for $E \ll |M|$ where $M$ is defined via \eqref{Mdef} ($|M|$ is understood as the square root of the smallest absolute value of eigenvalues of the matrix $M^2_{ab}$ in the multi-field UV models). Here, the sign of $M^2$ is crucial and $M^2$ is precisely defined by \eqref{Mdef} (the multi-field extension $M^2_{ab}$ is defined by \eqref{defMab}). We also note that we have introduced $m$ and $M_{\rm UV}$, which are supposed to satisfy $m \ll |M| \ll M_{\rm UV}$, where $m$ and $M_{\rm UV}$ are still schematic symbols denoting the inverse of temporal/spatial scales of the light field and the cutoff of the partial UV theory, respectively.

In this section, we investigate the spectra of the full theory \eqref{Lagrangian_full} to derive the UV consistency conditions and the EFT predictivity conditions of the resultant EFT. The relevant modes of our interest in this section are the modes satisfying $M_{\rm UV}\gg E_{\rm UV} \gtrsim |M|$. Therefore, the scale $m$ is irrelevant to the analysis. We thus concentrate on the shift symmetric theories, $m\to 0$, in this section. In addition, the scale $M_{\rm UV}$ is not important for the interactions between the heavy modes and the light mode as we have explained below \eqref{Lagrangian_dim_explicit}. In the presence of gravity, on the other hand, we need a care about $M_{\rm UV}$ because the background energy density of the system is of the order of $M_{\rm UV}^{d-2}|M^2|$. We should impose $M_{\rm UV} \lesssim \Mpl$ in which the curvature of the spacetime is less than or the same order of magnitude as $|M^2|$. When $M_{\rm UV} \sim \Mpl$, the curvature size is $\mathcal{O}(|M^2|)$ and the background curvature cannot be negligible for the modes $E_{\rm UV} \sim |M|$, but this situation means that the cutoff of the single-field EFT is comparable to the curvature size, say the Hubble scale in cosmology; the EFT cannot be used to describe the sub-horizon physics of the universe.
We thus assume a sufficiently large hierarchy between $M_{\rm UV}$ and $\Mpl$. Then, the background curvature can be ignored for the modes with $M_{\rm UV}\gg E_{\rm UV} \gtrsim |M|$ that we would like to investigate. Hence, we shall simply consider the flat spacetime in this section. Note, however, that the presence of gravity puts a consistency of the argument in the way of energy conditions even if we ignore the curvature of the spacetime, which will be discussed in Sec.~\ref{subsec:nullenergycondition}

We give detailed analysis about the two-field model in Sec.\ref{sec:quadratic}-\ref{subsec:timelike} and then extend the results into the multi-field (partial) UV completion in Sec.~\ref{sec:multi_pert}.

\subsection{Quadratic action around Lorentz-violating backgrounds}
\label{sec:quadratic}

We focus on the shift symmetric two-field model for this and following two subsections. The field-space metric is
\begin{align}
\gamma_{AB} \, \dd \Phi^A \dd \Phi^B= \tilde{\epsilon}^2\gamma(\chi) \, \dd \chi^2 + 2 \tilde{\epsilon} h(\chi) \, \dd \chi \dd \varphi + f(\chi) \, \dd \varphi^2
\,.
\end{align}
As we explained, the Killing vector for $N=2$ is hypersurface orthogonal and $\varphi$ can be chosen to respect the reflection symmetry. Concretely, we can use a freedom to change $\varphi$ according to
\begin{align}
\varphi \to \varphi'=\varphi+\tilde{\epsilon} g(\chi)
\label{trans_varphi}
\end{align}
where the Lagrangian is invariant under the shift of the new $\varphi$.
Using this freedom, we can eliminate the kinetic mixing between the light direction $\varphi$ and the heavy field $\chi$, i.e.~$h=0$, without loss of generality.
Then, we can conduct a transformation $\chi\to \chi'=\mathcal{X}(\chi)$ to set $\gamma=1$, as long as the kinetic term of $\chi$ is healthy, $\gamma >0$. All these transformations keep the leading shift-symmetric operator $P(X)$ invariant. As a result, the most general Lagrangian of the two-field model under the shift symmetry is given by
\begin{align}
\mathcal{L}_{\rm UV}
&=-\frac{1}{2}\tilde{\epsilon}^2 (\nabla \chi)^2+f(\chi)X-V(\chi)
\,.
\label{UV_two-field}
\end{align}
We have the same relations as before: $P=fX-V|_{\chi_0=\chi_0(X)},~P_X=f|_{\chi_0=\chi_0(X)},~P_{XX}=f_{\chi}^2/M^2|_{\chi_0=\chi_0(X)}$ where $M^2\equiv V_{\chi\chi}-f_{\chi\chi}X \neq 0$.

The Lorentz symmetry is spontaneously broken by the gradient of $\varphi$. We split the fields into the background part and perturbations,
\begin{align}
\chi=\bar{\chi}+\delta \chi\,, \quad \varphi = \bar{\varphi}+ \pi
\,,
\end{align}
where the background is supposed to provide the large gradient, $C_{\mu}\equiv \partial_{\mu}\bar{\varphi}=\mathcal{O}(1)$. In general, the background $C_{\mu}$ can be a function of time and/or space but these dependency must be tiny compared with $|M|$ so that the same configuration of the fields can be described by the EFT (recall that the second and higher than second order derivatives of $\varphi$ must be tiny). Such a change of $C_{\mu}$ is negligible for the UV modes $E_{\rm UV} \gtrsim |M|$. Therefore, for simplicity, we consider the constant backgrounds,
\begin{align}
\bar{\chi}={\rm constant}\,, \quad C_{\mu}={\rm constant}
\,,
\label{background}
\end{align}
to realize the spontaneous Lorentz symmetry breaking.\footnote{In a more general sense, the solutions \eqref{background} should be regarded as approximate background solutions in the adiabatic limit. In particular, $C_{\mu}$ and $\bar{\chi}$ should be functions of time and/or space either when the shift symmetry is not exact or gravity is turned on.} The background equation of motion for $\bar{\chi}$ reads
\begin{align}
V_{\chi}-f_{\chi}X=0
\,,
\label{constraint_background}
\end{align}
while the equation of motion for $\bar{\varphi}$ is trivially satisfied.
From here on in this section, functions such as $X$, $f$ and $V$ and their derivatives are understood to be evaluated at the background values $\chi = \bar\chi$ and $\varphi = \bar\varphi$. The background equation \eqref{constraint_background} is identical to the leading-order equation of the derivative expansion \eqref{leading_eom} because the background should be the low-energy/momentum part of the fields. The equation \eqref{constraint_background} determines $\bar{\chi}$ in terms of $X=-\frac{1}{2}C_{\mu}C^{\mu}$ as far as $M^2 \neq 0$. On the other hand, the constant vector $C_{\mu}$ is undetermined by the equation of motion and the value and direction of $C_{\mu}$ determines how the Lorentz symmetry is broken by the background \eqref{background}.

The quadratic action for the perturbations is 
\begin{align}
\mathcal{L}^{(2)}_{\rm UV}=-\frac{1}{2}\tilde{\epsilon}^2 (\partial \delta  \chi)^2 -\frac{1}{2}M^2 \delta \chi^2 -\frac{1}{2}f (\partial \pi)^2 - f_{\chi} \delta \chi (C \cdot \partial\pi)
\,,
\label{UV_L}
\end{align}
where all the coefficients are evaluated at the background \eqref{background} and thus become functions of $X=-\frac{1}{2}C_{\mu}C^{\mu}$. Here, we use the notation $C\cdot \partial \pi=C^{\mu}\partial_{\mu}\pi$. The first three terms are nothing but the Lorentz-invariant kinetic terms and the mass term, preserving the Lorentz symmetry of the perturbations. The last term, $f_{\chi} \delta \chi (C \cdot \partial\pi)$, is precisely the origin of the Lorentz-violation.
The dispersion relation of the perturbations are obtained by solving the equation
\begin{align}
{\rm det}
\begin{pmatrix}
\tilde{\epsilon}^2 p^2+M^2 & f_{\chi} (ip\cdot C) \\
-f_{\chi} (ip\cdot C) & p^2 f
\end{pmatrix}
=
p^2f(\tilde{\epsilon}^2 p^2+M^2)-f_{\chi}^2(p\cdot C)^2=0
\,,
\label{dis_two-field}
\end{align}
where $p^{\mu}=(\omega,k^i)$. The relativistic dispersion relations, $p^2=0$ and $p^2+M^2=0$, are obtained when $f_{\chi}p\cdot C=0$ while the dispersion relations are in general nonlinear due to the presence of the last term, $f_{\chi} \delta \chi (C \cdot \partial\pi)$.

Let us clarify the stability conditions of the perturbations. The ghost-free condition is the positive definiteness of the field space metric, leading to $f>0$ in the present parametrization. We also need to identify the mass of the heavy mode.
The mass of the heavy mode may be evaluated by identifying the energy at the rest frame, $p^{\mu}=(\omega,{\bm 0})$. In the Lorentz-invariant background, we can always perform the Lorentz transformation to go the rest frame of the massive particle. However, we are interested in situations $X\neq 0$ where $X<0$ may be relevant to e.g.~astrophysical environments \cite{Babichev:2009ee,Brax:2012jr} and $X>0$ has been discussed in cosmology. There is a background vector field $C_{\mu}$ that spontaneously breaks the Lorentz symmetry.
Nonetheless, since both $\bar\chi$ and $C_\mu$ are constant on the background \eqref{background}, we can change frames globally by performing Lorentz transformations but cannot in general go to the preferred frame of the background $C_\mu$ and the rest frame of $p^\mu$ simultaneously.
We should first define the preferred frame in connection with $C_\mu$ and then the mass of the particle.  In the following, we shall discuss the spacelike $C_{\mu}$ case and the timelike one in order.

\subsection{Spacelike background}
\label{subsec:spacelike}

\subsubsection{Spectra in full theory}
\label{sec:spectraS}

As for the spacelike background $X=-\frac{1}{2}C\cdot C<0$, we can perform the Lorentz transformation so that $C^{\mu}=(0,C^i)$.
We define particles at rest in this frame. This choice would be most useful because \eqref{dis_two-field} contains $\omega^2$ and $\omega^4$ only, meaning that positive and negative frequency modes obey the same dispersion relation. The mass of the heavy particle can be identified in the limit of vanishing spatial momentum in this frame, i.e.~by substituting $p^{\mu}=(\omega,{\bm 0})$ and $C^{\mu}=(0,C^i)$ into \eqref{dis_two-field} and solving the equation in terms of $\omega$. The last term in \eqref{dis_two-field} does not contribute to the equation. The mass of the heavy mode is then simply given by $M$.

\begin{figure}[t] 
        \centering
    \subfigure[~$k_*^2<0$]{
        \includegraphics[width=0.31\linewidth]{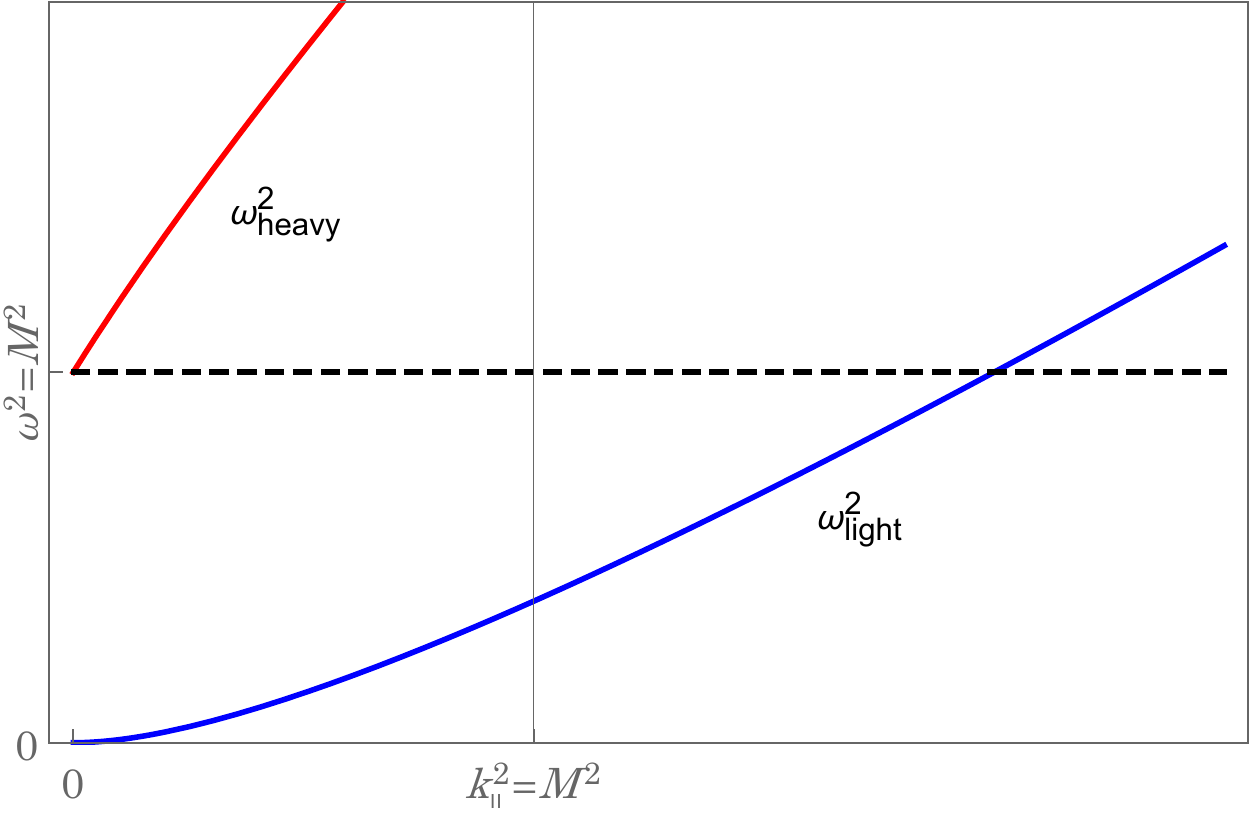}
        }
    \subfigure[~$0<k_*^2<M^2$]{
        \includegraphics[width=0.31\linewidth]{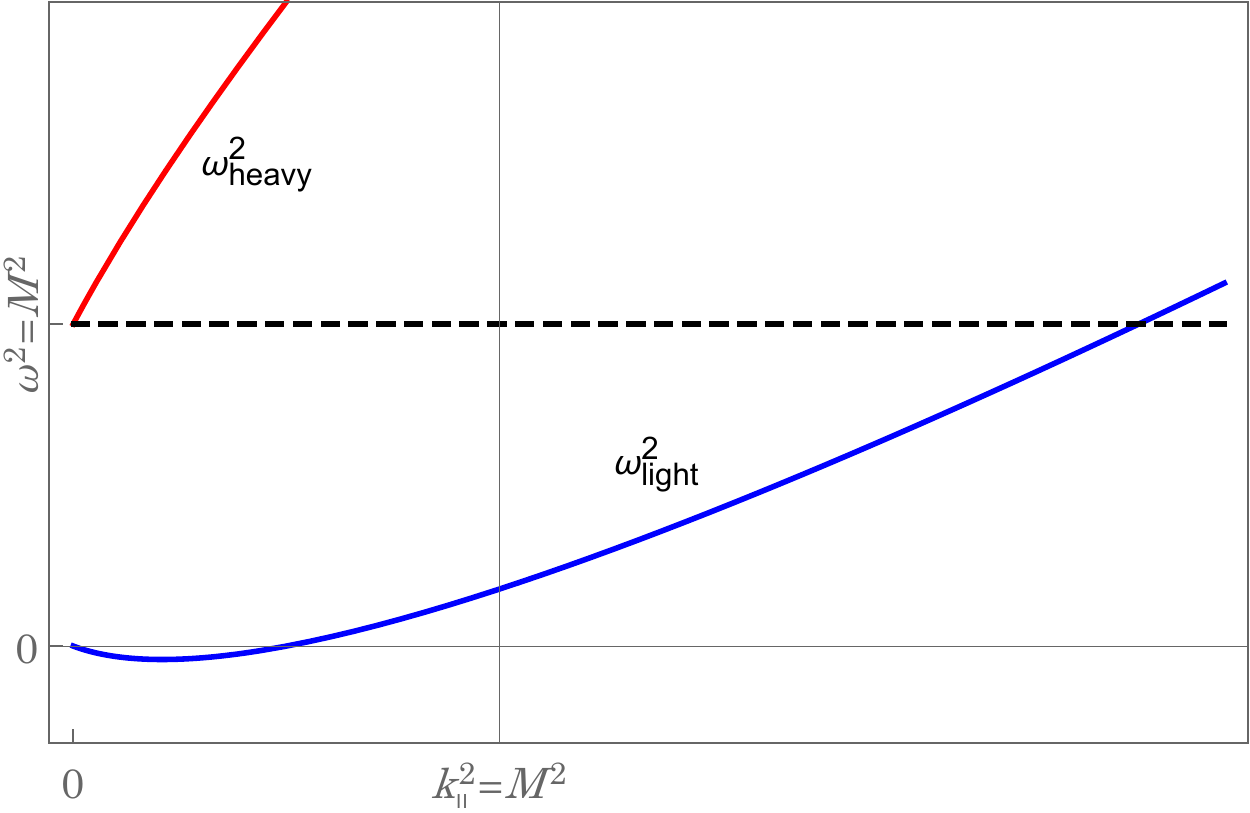} 
       }
    \subfigure[~$M^2<k_*^2$]{
        \includegraphics[width=0.31\linewidth]{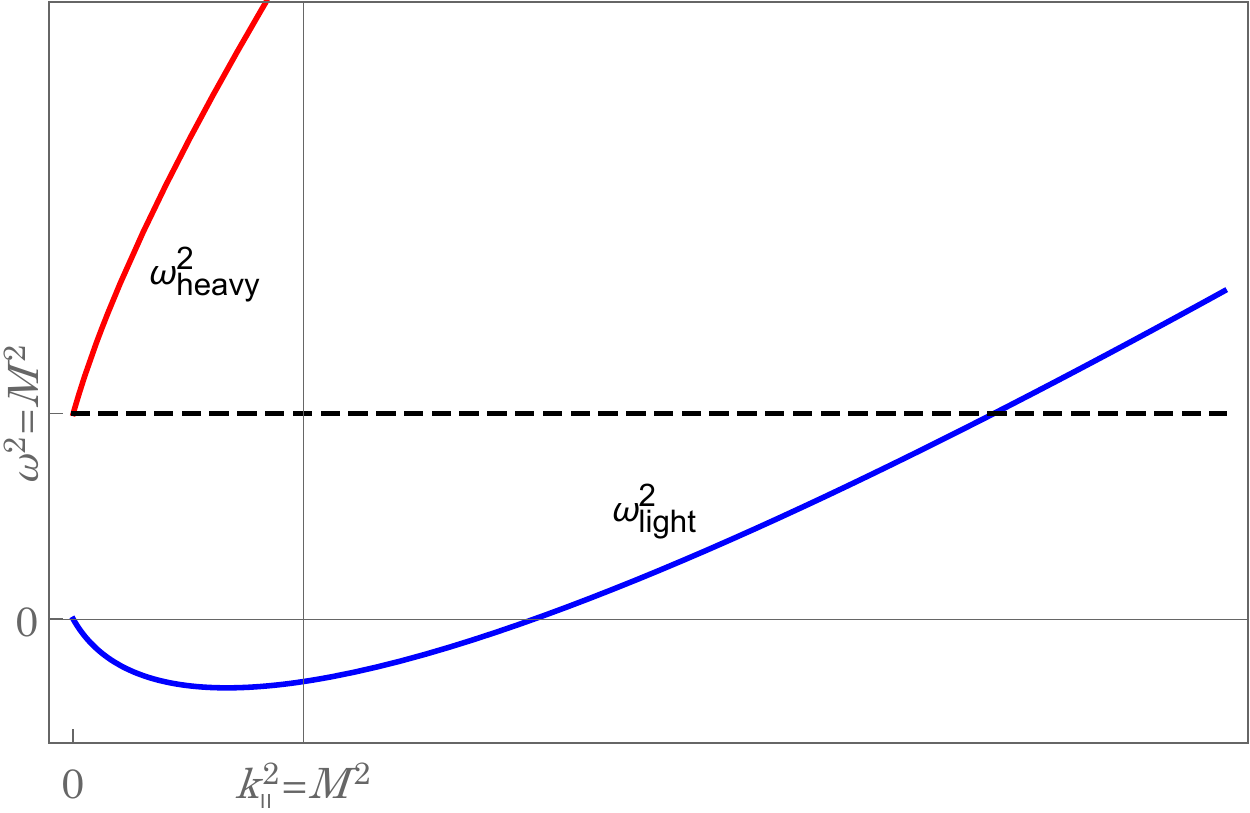} 
        }
    \caption{Illustrative figures for the dispersion relations around the spacelike background for the modes $k^i_{\perp}=0$ where the dashed lines correspond to $\omega^2=M^2$.}
    \label{fig:disS} 
\end{figure}

The dispersion relations for generic spatial momentum $k^i$ are
\begin{align}
\omega^2_{\rm heavy}&=k^2+\frac{M^2}{2\tilde{\epsilon}^2 }\left( 1 +\sqrt{1+ \tilde{\epsilon}^2\frac{4 f_{\chi}^2 (C^i k_i)^2}{f M^4}} \right)
\,,\\
\omega^2_{\rm light}&=k^2+\frac{M^2}{2\tilde{\epsilon}^2 }\left( 1 - \sqrt{1+ \tilde{\epsilon}^2 \frac{4 f_{\chi}^2 (C^i k_i)^2}{f M^4}} \right)
\nn
&=k^2-\frac{f_{\chi}^2}{f M^2} (C^i k_i)^2 +\tilde{\epsilon}^2 \frac{f_{\chi}^4}{f^2M^6}(C^i k_i)^4+ \mathcal{O}(\tilde{\epsilon}^4)
\,, \label{dis_light_spacelike}
\end{align}
where $k \equiv \vert k^i \vert$, and $f>0$ and $M^2>0$ have been used. The conditions $f>0$ and $M^2>0$ are the stability conditions for the UV modes, and we assume them throughout the analysis around the spacelike background.
We introduce  $k_{\perp}^i$ and $k_{\parallel}$ which are the perpendicular and the parallel components of $k^i$ to $C_i$, that is, $k_{\perp}^i  C_i =0$ and $C_i  k^i =\sqrt{|2X|} \, k_{\parallel}$, respectively. Since the background configuration breaks the spatial rotational symmetry, the dispersion relation is anisotropic. For the modes $k^i=k^i_{\perp}, k_{\parallel}=0$, the standard dispersion relations $\omega_{\rm heavy}^2=M^2+k^2_{\perp}, \omega_{\rm light}^2=k^2_{\perp}$ are recovered while the dispersion relations take non-linear forms for $k_{\parallel}\neq 0$. For an illustrative purpose, Fig.~\ref{fig:disS} shows some concrete forms of the dispersion relations of both heavy and light modes with $k^i_{\perp}=0$.

Let us recall that all the UV modes satisfying $\omega,k>M$ are integrated out to derive the single-field EFT via the derivative expansion. The UV modes include not only the heavy mode $\delta \chi$ but also the light mode $\pi$. The conditions $f>0$ and $M^2>0$ are the conditions for the UV consistency, but we should also discuss whether any non-trivial conditions for EFT predictivity arise by checking the full spectra of the theory.
The dispersion relations show qualitatively different features depending on the background value of $|X|$. We consider the modes $k^i_{\perp}=0$ for which the dispersion relations deviate from the standard form most significantly. We set $\tilde{\epsilon}=1$ in the rest of this subsection (see subsubsection~\ref{sec:epsilon}). The roots of $\omega_{\rm light}^2=0$ are, without expanding for small $\tilde\epsilon$,
\begin{align}
k_{\parallel}^2=0\,,~k_*^2
\,,
\end{align}
where 
\begin{align}
k_*^2 \equiv 2|X|\frac{f_{\chi}^2}{f}-M^2 \; .
\label{uns_k}
\end{align}
Therefore, all the light modes are stable if $ k_*^2 <0$ whereas there exist unstable modes at IR satisfying $k_{\parallel}^2< k_*^2$ if $k_*^2>0$. Since the modes $k_{\parallel}^2>M^2$ are integrated out, the EFT reduction would be inconsistent if the instability existed at $k_{\parallel}>M$. This requires $k_*^2<M^2$ as a consistency of the EFT predictivity. The typical behaviours are shown in Fig.~\ref{fig:disS}-(b) and \ref{fig:disS}-(c): the EFT reduction is consistent in (b) while is inconsistent in (c).\footnote{\label{footnote1}Nevertheless, there can be a consistent single-field EFT even for Fig.~\ref{fig:disS}-(c) because there is no unstable mode in the high energy limit and the heavy mode is always stable. This requires to extend the validity of the EFT into the momentum domain $k_{\parallel}^2>M^2$. We elaborate on such an extension around the timelike background in the following section. } Note that the stability of the heavy mode is guaranteed by $M^2>0$ and the instability exists only in the IR part of the light field.

We investigate the timescale of the instability in the case $M^2>k_*^2>0$. The momentum giving the minimum of $\omega^2_{\rm light}$ is determined by the condition
\begin{align}
\frac{\partial \omega_{\rm light}^2(k^i)}{\partial k_{\parallel}^2}=0
\,,
\end{align}
and then the minimum value is
\begin{align}
\omega_{\rm min}^2=k^2_{\perp}-\frac{1}{4} k_*^2 \left( 1 - \frac{M^2f}{2|X|f_{\chi}^2} \right) > -\frac{1}{4}k_*^2
\,.
\end{align}
The instability timescale is thus
\begin{align}
\Delta t_{\rm ins}\equiv \frac{1}{|\omega_{\rm min}|} > \frac{2}{k_*}
\,.
\end{align}
The timescale is long enough to be resolved by the EFT if $k_*$ is sufficiently small. In particular, $M>k_*>0$ results in $\Delta t_{\rm ins}>2/M$, that is, the instability timescale is longer than (twice) the cutoff scale $M^{-1}$. Therefore, such IR instabilities do not render the EFT reduction inconsistent, that is, the EFT predictivity condition is satisfied.

\subsubsection{UV consistency and EFT predictivity}
\label{subsubsec:consistency_spacelike}

We are ready to derive the consistency conditions on the EFT around the spacelike background, $X<0$. As observed in the previous subsubsection, the ghost-free and stability conditions are, respectively,
\begin{align}
f>0\,, \quad M^2>0
\,, \qquad \text{(UV consistency)}
\label{stabilityS}
\end{align}
and
\begin{align}
M^2 > k_*^2
\,, \qquad \text{(EFT predictivity)}
\label{IRinsS}
\end{align}
where $k_*^2$ can be either positive or negative. The conditions \eqref{stabilityS} arise from the stability conditions of the UV modes while the condition \eqref{IRinsS} is the condition that the IR instability is under control if exists.

The EFT Lagrangian up to $\mathcal{O}(\tilde{\epsilon}^2)$ is given by
\begin{align}
    \mathcal{L}_{\rm IR}=P(X)-\tilde{\epsilon}^2 \frac{P_{XX}}{2M^2}  (\partial X)^2 + \mathcal{O}\left(\tilde{\epsilon}^2 \right)
\,,
\label{EFT_simplified}
\end{align}
where the Lagrangian at subleading orders is simplified from \eqref{EFT_Lag_gen} because of the field redefinition. Using the relations $P_X=f|_{\chi_0=\chi_0(X)},~P_{XX}=f_{\chi}^2/M^2|_{\chi_0=\chi_0(X)}$, the conditions \eqref{stabilityS} immediately conclude
\begin{align}
P_X(X<0)>0\,, \quad P_{XX}(X<0)>0
\label{UV_consistencyS}
\end{align}
as the UV consistency of the EFT.\footnote{Here, we implicitly assume $f_{\chi}\neq 0$ to have a non-trivial function of $P(X)$. As seen from \eqref{leading_eom}, the solution $\chi_0$ is independent of $X$ when $f_{\chi}=0$. } We also have the relation
\begin{align}
k_*^2 = 2|X| \frac{f_{\chi}^2}{f}-M^2
=\left(2|X|\frac{P_{XX}}{P_X}-1 \right) M^2
\,.
\end{align}
Therefore, the condition \eqref{IRinsS} leads to
\begin{align}
\left| \frac{P_X+2XP_{XX}}{P_X} \right| < 1
  \quad {\rm if}\quad \frac{P_X+2XP_{XX}}{P_X} <0
 \,,
 \label{IRinsS2}
\end{align}
as the EFT predictivity, where $X<0$ is taken into account. Note that the condition \eqref{IRinsS} is trivially satisfied if $(P_X+2XP_{XX})/P_X$ is positive.

\subsubsection{Matching IR dispersion relation}

In order to verify that the procedure in the previous subsubsection correctly reproduces the IR physics, let us consider the quadratic Lagrangian of the EFT $\mathcal{L}_{\rm IR}=P(X)$ around the constant background,
\begin{align}
\mathcal{L}_{\rm IR}^{(2)}=-\frac{1}{2}\left[P_X(\partial \pi)^2-P_{XX}(C\cdot \partial \pi)^2+\tilde{\epsilon}^2  \frac{P_{XX}}{M^2} \partial_{\mu}(C\cdot\partial \pi) \partial^{\mu}(C\cdot \partial \pi) +\mathcal{O}(\tilde{\epsilon}^4) \right]
. \label{quadratic_next}
\end{align}
The EFT dispersion relation is a root of
\begin{align}
&p^2P_X-P_{XX} \left( C\cdot p \right)^2+\tilde{\epsilon}^2 \frac{P_{XX}}{M^2} \, p^2 \left( C\cdot p \right)^2 + \mathcal{O}(\tilde{\epsilon}^4)
=0
\label{dis_next_to_leading}
\end{align}
which indeed reproduces the original dispersion relation of the light mode up to $\mathcal{O}(\tilde{\epsilon}^2)$.
Using $C_{\mu}=(0,C_i)$, the EFT dispersion relation of the light mode is explicitly given by
\begin{align}
\omega^2_{\rm EFT}
&=k_{\perp}^2 + c_{\parallel}^2 k_{\parallel}^2 + \tilde{\epsilon}^2 \frac{(2XP_{XX})^2}{P_X^2 M^2} k_{\parallel}^4+\mathcal{O}(\tilde{\epsilon}^4)
\,,
\\
c_{\parallel}^2&\equiv \frac{P_X+2XP_{XX}}{P_X}
\,.
\end{align}
One can easily confirm the agreement with \eqref{dis_light_spacelike} up to $\mathcal{O}(\tilde{\epsilon}^2)$. Also, the conditions \eqref{UV_consistencyS} and \eqref{IRinsS2} lead to the bounds on the transverse part of the sound speed,
\begin{align}
-1<c_{\parallel}^2<1
\,,
\end{align}
where the upper bound and the lower bound are determined by the UV consistency condition and the EFT predictivity condition, respectively.

\subsection{Timelike background}
\label{subsec:timelike}

\subsubsection{Spectra in full theory}

\begin{figure}[t] 
        \centering
    \subfigure[~$M^2>0$]{
        \includegraphics[width=0.31\linewidth]{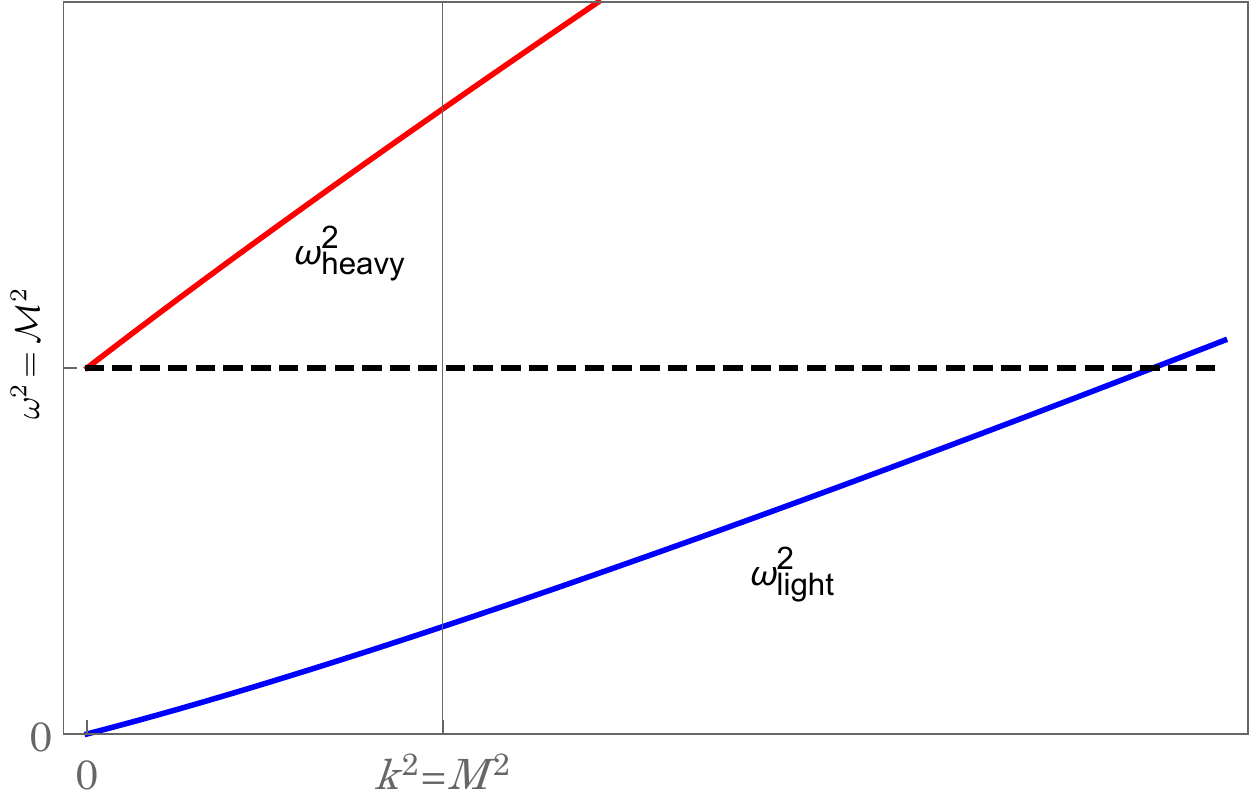}
        }
    \subfigure[~$0>M^2>-2(\sqrt{2}-1)\delta M^2$]{
        \includegraphics[width=0.31\linewidth]{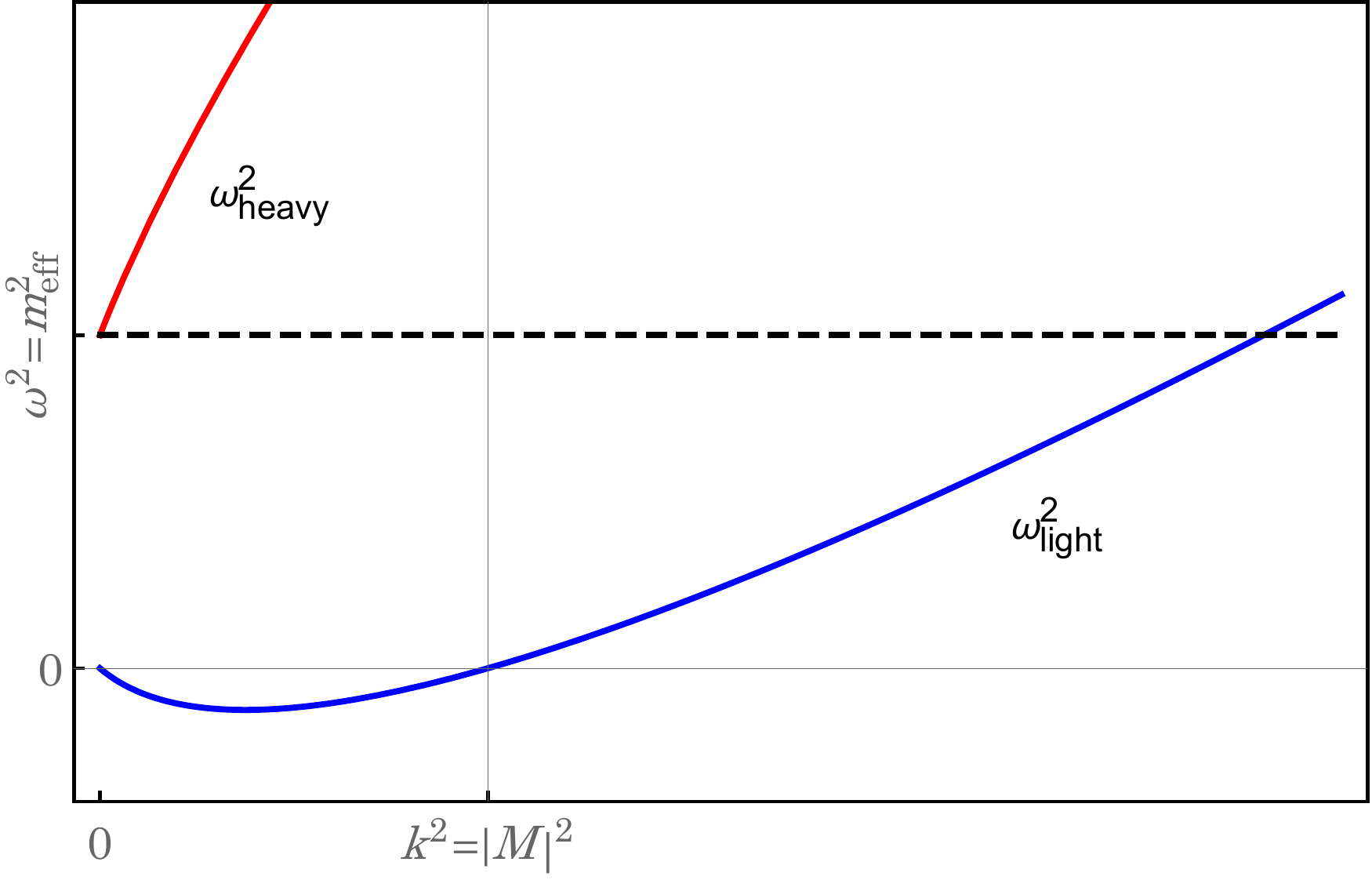} 
       }
    \subfigure[~$-2(\sqrt{2}-1)\delta M^2>M^2$]{
        \includegraphics[width=0.31\linewidth]{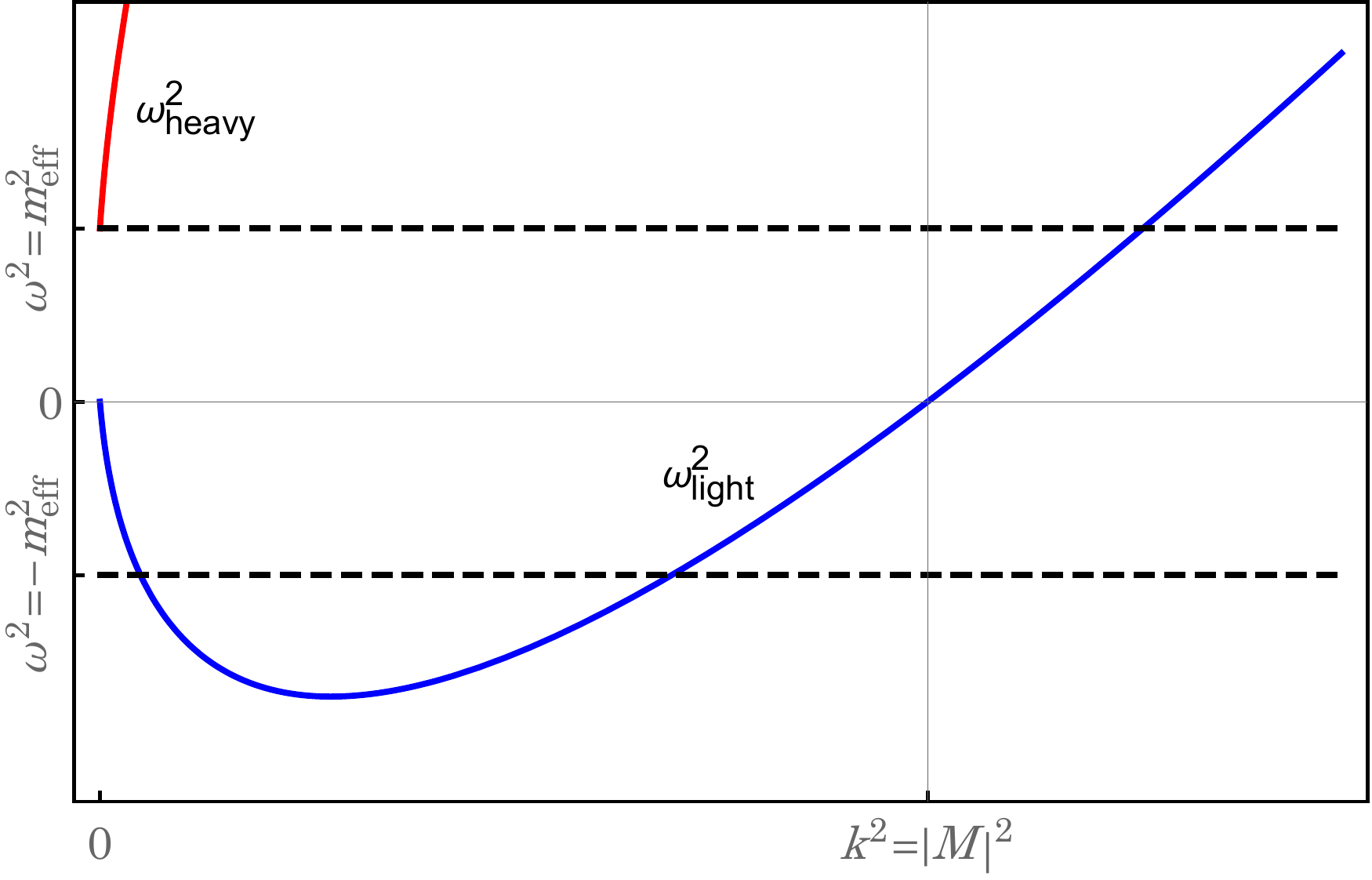} 
        }
    \caption{Dispersion relations around the timelike background where the dashed lines correspond to $|\omega^2|=\Meff^2$.}
    \label{fig:disT} 
\end{figure}

Let us now turn to the timelike case, $X>0$. We first set $C_{\mu}=(C_0, {\bm 0})$ by the use of the Lorentz transformation. The dispersion relations are then
\begin{align}
\omega_{\rm heavy}^2&=k^2+\frac{\Meff^2}{2\tilde{\epsilon}^2 }\left( 1+\sqrt{1+ \tilde{\epsilon}^2\frac{4 \, \delta M^2 k^2}{\Meff^4}} \right)
\,, \\
\omega_{\rm light}^2&=k^2+\frac{\Meff^2}{2\tilde{\epsilon}^2 }\left( 1-\sqrt{1+ \tilde{\epsilon}^2 \frac{4 \, \delta M^2 k^2}{\Meff^4}} \right)
\nn
&=c_s^2 k^2 + \tilde{\epsilon}^2 \frac{\delta M^4}{\Meff^6} k^4 + \mathcal{O}(\tilde{\epsilon}^4)
\,,
\label{dispersion_timelike}
\end{align}
where
\begin{align}
\Meff^2 =M^2 + \delta M^2 = \frac{M^2}{c_s^2}
\,, \qquad
\delta M^2 = \frac{C_0^2 f_{\chi}^2}{f}= 2X \frac{f_{\chi}^2}{f}
\,, \qquad
c_s^2 =1-\frac{\delta M^2}{\Meff^2}=\frac{M^2}{\Meff^2} \label{def_cs2}
\,,
\end{align}
and the conditions $f>0$ and $\Meff^2>0$ have been used. The mass of the heavy mode is now given by $\Meff^2$ rather than $M^2$ where the correction $\delta M^2$ comes from the friction term, $f_{\chi}\delta \chi C\cdot \partial \pi=f_{\chi}C_0 \delta \chi \dot{\pi}$.
Let us emphasize that $M^2$ can take either a positive or a negative value on the timelike background, and the heavy mode is stable so long as $\Meff^2>0$ regardless of the sign of $M^2$. The same observation was found in Sec.~\ref{sec:oscillators} (see \eqref{eqn:toyexample-UVstability} that can be satisfied even for $\eta\leq 0$.). Hence $M^2$ can be either positive or negative for the timelike $C_{\mu}=\partial_{\mu}\bar{\varphi}$.\footnote{We do not consider $M^2=0$ because the background $\bar{\chi}$ would be undetermined by $\bar{\varphi}$. On the other hand, we can discuss the limit $M^2 \to 0$.} Typical behaviours of the dispersion relations are shown in Fig.~\ref{fig:disT}.

At first glance, the dispersion relations appear to behave similarly to those around the spacelike background. IR instabilities can appear depending on the parameters. The roots of $\omega_{\rm light}^2=0$ are, absorbing $\tilde\epsilon$ into the definition of $\chi$ (equivalently setting $\tilde\epsilon = 1$),
\begin{align}
k^2=0\,, ~-M^2
\,,
\end{align}
around the timelike background. Therefore, the light modes are always stable for $M^2>0$ while the modes $k^2<|M^2|$ are unstable for $M^2<0$.
The timescale of the instability is estimated by the same way as the spacelike background. The momentum corresponding to the minimum of $\omega_{\rm light}^2$ is determined by
\begin{align}
\frac{\dd \omega_{\rm light}^2(k^2)}{\dd k^2}=0
\,,
\end{align}
yielding
\begin{align}
\omega_{\rm min}^2 = - \frac{M^4}{4\delta M^2}
\,.
\end{align}
The timescale is
\begin{align}
t_{\rm uns}\equiv \frac{1}{|\omega_{\rm min}|} = \frac{2\delta M}{|M^2|}
\,,
\end{align}
where $\delta M^2>0$ is used.

Nonetheless, there is a crucial difference from the spacelike background. In the case of the spacelike background, the boundary of the instability is determined by $k_*^2$ which can be chosen to satisfy $k_*^2< M^2$. Here, recall that the absolute value of the parameter $M$ determines the convergence radius of the derivative expansion with respect to spacetime derivatives, see \eqref{chi_series}.
On the other hand, around the timelike background with $M^2<0$, the parameter $|M^2|$ serves as not only the boundary of the convergence but also that of the instability in the EFT.
In fact all the modes with $k^2 \lesssim \vert M^2 \vert$ would be unstable if $M^2<0$. Nevertheless, this would not invoke any instabilities in the UV, and therefore the UV modes should be safely integrated out, keeping the validity of the effective description.

The time and length scales of applicability of such an unstable EFT might be rather limited in realistic setups, as the instability of the modes, especially the ones near $k^2 \sim \vert M^2 \vert$, could drive the system out of the EFT's validity range and even excite high-energy/momentum modes via nonlinear interactions.
However, the effective mass of the heavy modes is given by $\Meff^2$ rather than $M^2$, and therefore the practical cutoff scale of the single-field EFT can be raised accordingly. Indeed, we can extend the validity of the EFT into the domain
\begin{align}
|\omega^2| < \Meff^2 \,, \quad k^2 < \Lambda^2 \equiv \Meff^2+\delta M^2
\,,
\label{cutoffT2}
\end{align}
where $k^2= \Lambda^2$ corresponds to $\omega_{\rm light}^2=\Meff^2$. In this case, the single-field EFT can accommodate the stable modes above $\vert M \vert$ as well, and one can introduce hierarchy between the threshold  of the instability, $|M^2|$, and the actual cutoff $\Lambda^2$. For consistency, we need to assume the EFT predictivity condition that the timescale of the instability is sufficiently longer than the cutoff timescale, which is given by
\begin{align}
|\omega_{\rm min}^2|< \Meff^2 \implies 
 |M^2| < 2(\sqrt{2}-1) \delta M^2 \quad {\rm if} \quad M^2<0
 \,,
\end{align}
As a result, the theory corresponding to Fig.\ref{fig:disT}-(b) can be described by a single-field EFT while Fig.\ref{fig:disT}-(c) could not as instabilities would develop before the heavy modes could be stabilized.

In the following subsubsections, we separately discuss the EFTs that are applicable to the two different (but not necessarily exclusive) domains. We first consider the case in which the stability of deep IR modes is assured by the condition $M^2 >0$,
and later we study the general EFT by extending the validity of the EFT into \eqref{cutoffT2}.

\subsubsection{UV consistency without IR instability}

Let us first consider the UV consistency of the EFT \eqref{eqn:LIR_twofields} under the additional assumption of the absence of IR instabilities, i.e.~$M^2>0$. In this case we have
\begin{align}
f>0\,, \quad M^2>0
\,,
\end{align}
where the first condition is due to the UV consistency while the second one is to avoid IR instabilities. The remaining UV consistency, i.e. the no-tachyon condition $\Meff^2>0$ is automatically satisfied by these conditions. We then find
\begin{align}
P_X(X>0)>0\,, \quad P_{XX}(X>0)>0
\,, 
\label{UV_consistencyT}
\end{align}
as the UV consistency conditions of \eqref{EFT_simplified} and the above-mentioned additional assumption. The square of the sound speed of the perturbations,
\begin{align}
c_s^2=\frac{P_X}{P_X+2XP_{XX}}
\,,
\end{align}
is positive and bounded by the speed of light,
\begin{align}
0<c_s^2<1
\end{align}
as a consequence of the UV consistency \eqref{UV_consistencyT}. One can confirm that the EFT dispersion relation \eqref{dis_next_to_leading} correctly reproduces the original relation \eqref{dispersion_timelike} up to $\mathcal{O}(\tilde{\epsilon}^2)$ under $C_{\mu}=(C_0,\bm{0})$.\footnote{The dispersion relation \eqref{dis_next_to_leading} generically contains a ghost mode due to the truncation of the higher derivative operators. One has to only consider the light (physical) mode of the solution.} Since the system is stable, there is no EFT predictivity condition.

As we mentioned in the spacelike case, our consistency conditions \eqref{UV_consistencyT} hold even in the largely broken Lorentz symmetry, $C_{\mu}=\mathcal{O}(1)$, as far as we assume the class of (partial) UV completion and the absence of IR instabilities. Although we have assumed the constant $C_{\mu}$, the background can depend on time (and/or space) as long as its change is adiabatic. Furthermore, the bounds \eqref{UV_consistencyT} (and \eqref{UV_consistencyS}) are applicable to the EFT that has no consistent Lorentz-invariant background (see Appendix \ref{app:U(1)}).

\subsubsection{UV consistency and EFT predictivity with IR instability: apparent violation of positivity}
\label{sec:extendEFT}

We now discuss how we can extend the validity of EFT to go beyond the threshold of the IR instability around the timelike background, even in the case with $M^2<0$. Let us rediscuss the quadratic UV Lagrangian \eqref{UV_L},
\begin{align}
S^{(2)}_{\rm UV}=\int \dd^dx \left[-\frac{1}{2}(\partial \delta  \chi)^2 -\frac{1}{2}M^2 \delta \chi^2 -\frac{1}{2}f (\partial \pi)^2 - f_{\chi} \delta \chi (C \cdot \partial\pi) \right]
\,,
\label{UV_L2}
\end{align}
where we set $\tilde{\epsilon}=1$. The parameter $\tilde{\epsilon}$ is no longer useful because we cannot use the derivative expansion for the present purpose. In the momentum space, the action is
\begin{align}
S^{(2)}_{\rm UV}=\int \dd t \int \frac{\dd^{d-1} k}{(2\pi)^{d-1}} \left[ \frac{1}{2} \left( \left\vert \delta \dot{\chi} \right\vert^2- \left( k^2+M^2 \right) \left\vert \delta \chi \right\vert^2 \right) +\frac{f}{2} \left( \left\vert \dot{\pi} \right\vert^2-k^2 \left\vert \pi \right\vert^2 \right) + \frac{C_0 f'}{2} \left( \delta \chi^\dagger \dot{\pi} + {\rm h.c.} \right)
\right]
\,.
\label{UV_Lmom}
\end{align}

The complication arises from the fact that the variables $\delta \chi$ and $\pi$ are not eigenstates of propagations. The mass of the heavy mode is corrected by the friction term, $\Meff^2=M^2+\delta M^2$. To overcome this point, we perform field redefinition to diagonalize the fields. As for the Lorentz-invariant terms (the kinetic terms and the mass term), we can take a general linear transformation of $\delta \chi$ and $\pi$ to diagonalize them which is indeed what we did at the beginning of this section, see \eqref{trans_varphi}. On the other hand, the friction term cannot be diagonalized in this way; instead, it is more convenient to perform a canonical transformation (see e.g.~\cite{Nilles:2001fg,Gumrukcuoglu:2010yc}).

We thus consider the Hamiltonian rather than the Lagrangian. From \eqref{UV_Lmom}, defining the conjugate momenta via
\begin{align}
p&=\delta \dot{\chi} \,, \\
p_{\pi}&= f\dot{\pi}+f_{\chi} C_0 \delta \chi
\,,
\end{align}
the Hamiltonian around the timelike background is
\begin{align}
H_{\rm UV}^{(2)}&=\int \frac{\dd^{d-1}k}{(2\pi)^{d-1}} \mathcal{H}_{\rm UV}^{(2)}
\,, \\
\mathcal{H}^{(2)}_{\rm UV}&=
\frac{1}{2} \left\vert \delta \dot{\chi}  \right\vert^2+\frac{1}{2} \left( k^2+M^2 \right) \left\vert \delta \chi \right\vert^2 + \frac{f}{2} \left\vert \dot{\pi} \right\vert^2 + \frac{f}{2} \, k^2 \left\vert \pi \right\vert^2
\nn
&=\frac{1}{2} \left\vert p \right\vert^2+\frac{1}{2} \left( k^2+M^2 \right) \left\vert \delta \chi \right\vert^2 + \frac{1}{2f} \left\vert p_{\pi}-f_{\chi} C_0 \delta \chi \right\vert^2 + \frac{f}{2} \, k^2 \left\vert \pi \right\vert^2
\label{Hamiltonian_timelike}
\end{align}
in the momentum space. Here, on the first line of \eqref{Hamiltonian_timelike}, $\delta\dot{\chi}$ and $\dot{\pi}$ are understood as functions of canonical variables. We then take a canonical transformation to remove the mixing term $p_{\pi}\delta \chi$ from the Hamiltonian by considering an appropriate generating function.  One example of such a transformation is to take linear combinations to define a set of new conjugate variables by
\begin{align}
    & \tilde\pi \equiv c_1 \, \delta\chi + d_1 \, p_\pi \; , \qquad
    \tilde p_\pi \equiv c_2 \, p + d_2 \, \pi \; ,
    \\ &
    \delta \tilde\chi \equiv c_3 \, \pi + d_3 \, p \; , \qquad
    \tilde p \equiv c_4 \, p_\pi + d_4 \, \delta\chi \; ,
\end{align}
where $c_i$ and $d_i$ are constants. The exact forms of these coefficients are rather lengthy and not illuminating, so we only write the first two terms of each in the small $k$ expansion, reading
\begin{align}
    c_1 & = \frac{\delta M}{\Meff^2} \, k - \frac{3 \, \delta M^3}{2 \, \Meff^6} \, k^3
    + {\cal O}(k^5) \; , \qquad
    d_1 = \frac{1}{\sqrt{f} \, k} - \frac{\delta M^2}{2 \sqrt{f} \, \Meff^4} \, k
    + {\cal O}(k^3) \; ,
    \\
    c_2 & = \frac{\delta M}{\Meff^2} \, k - \frac{3 \, \delta M^3}{2 \, \Meff^6} \, k^3
    + {\cal O}(k^5) \; , \qquad
    d_2 = - \sqrt{f} \, k + \frac{\sqrt{f} \, \delta M^2}{2 \, \Meff^4} \, k^3
    + {\cal O}(k^5) \; ,
    \\
    c_3 & = \frac{\sqrt{f} \, \delta M}{\Meff^3} \, k^2 - \frac{\sqrt{f} \, \delta M \left( \Meff^2 + 4 \, \delta M^2 \right)}{2 \, \Meff^7} \, k^4
    + {\cal O}(k^6) \; , \qquad
    d_3 = \frac{1}{\Meff} - \frac{\Meff^2 + 2 \, \delta M^2}{2 \, \Meff^5} \, k^2
    + {\cal O}(k^4) \; ,
    \\
    c_4 & = \frac{\delta M}{\sqrt{f} \, \Meff} + \frac{\delta M \left( \Meff^2 - 2\, \delta M^2 \right)}{2 \sqrt{f} \, \Meff^5} \, k^2
    + {\cal O}(k^4) \; , \qquad
    d_4 = - \frac{k^2 + 2 \, \Meff^2}{2 \, \Meff}
    + {\cal O}(k^4) \; .
    \label{cidi_canontrans}
\end{align}
Then $\{ \tilde\pi , \tilde p_\pi \}$ and $\{ \delta\tilde\chi , \tilde p \}$ are conjugate pairs, and the Hamiltonian in terms of the new conjugate variables is diagonalized,
\begin{equation}
    {\cal H}^{(2)}_{\rm UV} = \frac{1}{2} \left( \left\vert \tilde p \right\vert^2 + \Omega_\chi^2 \left\vert \delta\tilde\chi \right\vert^2
    + \left\vert \tilde p_\pi \right\vert^2 + \Omega_\pi^2 \left\vert \tilde\pi \right\vert^2 \right) \; ,
\end{equation}
where
\begin{align}
    \Omega_\chi^2 & = \Meff^2 + k^2 \left( 1 + \frac{\delta M^2}{\Meff^2} \right)
    + {\cal O}(k^4) \; ,
    \qquad
    \Omega_\pi^2 = c_s^2 \, k^2 + \frac{\delta M^4}{\Meff^6} \, k^4 + {\cal O}(k^6) \; ,
    \label{Omega_timelike}
\end{align}
which perfectly coincide with \eqref{dispersion_timelike} up to these orders. The heavy mode variables $\{\delta\tilde{\chi}, \tilde{p} \}$ are now decoupled. Integrating the UV modes out, we find the quadratic Hamiltonian of the EFT,
\begin{align}
H^{(2)}_{\rm IR}&=\int^{\Lambda}\!\! \frac{\dd^{d-1}k}{(2\pi)^{d-1}} \mathcal{H}_{\rm UV}^{(2)}
\,, \quad
{\cal H}^{(2)}_{\rm IR} = \frac{1}{2} \left( \left\vert \tilde p_\pi \right\vert^2 + \Omega_\pi^2 \left\vert \tilde\pi \right\vert^2 \right) \; ,
\end{align}
where $\Lambda$ is added to the integral symbol in order to represent that the domain of integration is limited to $k^2<\Lambda^2 = \Meff^2+\delta M^2$ so that frequencies of $\{\tilde{p}_{\pi}, \tilde{\pi}\}$ do not exceed the mass of the heavy mode. After the Legendre transformation, the quadratic action of the EFT is
\begin{align}
S^{(2)}_{\Lambda}[\tilde{\pi}]=\int \dd t \int^{\Lambda}\!\! \frac{\dd^{d-1}k}{(2\pi)^{d-1}} \!\!
 \left( \frac{1}{2} |\dot{\tilde{\pi}}|^2 + \frac{1}{2} \Omega_\pi^2 \left\vert \tilde\pi \right\vert^2 \right) 
,
\label{L_Meff}
\end{align}
which is valid even for $M^2<0$. We also note the the Lagrangian is well-behaved even in the limit $M^2 \to 0$. Although we only compute the quadratic action, the non-linear interactions of the EFT may be computed accordingly.

It deserves care to compare the previous EFT \eqref{EFT_simplified} with the present EFT \eqref{L_Meff} because of the field transformation. Let us call the previous one the $M$-EFT and the present one the $\Lambda$-EFT, respectively, since the cutoffs of these EFTs are determined by $M^2$ and $\Lambda^2=\Meff^2+\delta M^2$. The on-shell relation between the variables is
\begin{align}
\tilde{\pi}=\frac{f(d_1d_4-c_1c_4)}{d_4+c_4 \sqrt{f} \delta M} \, \dot{\pi}
\simeq \frac{\Meff^2}{\sqrt{f}k(k^2+M^2)} \, \dot{\pi}
\,,
\label{rel_pi_tpi}
\end{align}
where we have assumed $k^2 \ll \Meff^2$ and $|M^2|\ll \Meff^2$ to get the last expression. This implies that $\pi$ and $\tilde{\pi}$ are related in a non-local way in both time and space.

The action of $\Lambda$-EFT in terms of the variable $\pi$ can be obtained by taking the canonical transformation. Here, we perform the canonical transformation in the Lagrangian level by following the technique~\cite{DeFelice:2015moy}. For simplicity, we consider the case $|M^2| \ll \Meff^2$ where $\Omega_{\pi}^2$ is approximated as
\begin{align}
\Omega_{\pi}^2 \simeq c_s^2 k^2 + \frac{1}{\Meff^2}k^4
\label{Omega_approx}
\end{align}
up to the subleading order in $k^2$.
We integrate in the variable $\pi$ and write the action \eqref{L_Meff} as
\begin{align}
S^{(2)}_{\Lambda}[\tilde{\pi},\pi]=\int \dd t \int^{\Lambda}\!\! \frac{\dd^{d-1}k}{(2\pi)^{d-1}}
 \left[ -\frac{1}{2}f k^2|\pi|^2 - \frac{k^2(k^2+M^2)}{2\Meff^2} |\tilde{\pi}|^2 -  \frac{\sqrt{f} k}{2}(\pi^\dagger \dot{\tilde{\pi}} + {\rm h.c.} ) \right]
,
\end{align}
which recovers \eqref{L_Meff} when $\pi$ is integrated out under the approximation \eqref{Omega_approx}. Instead, we integrate $\tilde{\pi}$ out. The equation of motion for $\tilde{\pi}$ yields the relation \eqref{rel_pi_tpi}. Substituting \eqref{rel_pi_tpi} into $S^{(2)}_{\Lambda}[\tilde{\pi},\pi]$, we obtain
\begin{align}
S^{(2)}_{\Lambda}[\pi]&=\int \dd t \int^{\Lambda}\!\! \frac{\dd^{d-1}k}{(2\pi)^{d-1}}
\frac{f}{2} \left[ \frac{\Meff^2}{k^2+M^2}|\dot{\pi}|^2 - k^2 |\pi|^2 \right]
. \label{L_Meff2}
\end{align}
Note that the action \eqref{L_Meff2} can be directly derived from the UV action \eqref{UV_Lmom} by integrating out $\delta \chi$ if we keep the nonlinearity of $k^2$. The case $|M^2| \ll \Meff^2$ corresponds to $|c_s^2| \ll 1 \Rightarrow |\omega^2| \ll k^2$, meaning that there is no need to equally treat the time and the space. We can perturbatively treat the time derivative $\delta \dot{\chi}$ while keeping the spacial derivative $\partial_i \delta \chi=i k_i\delta \chi$ non-perturbatively. The action \eqref{L_Meff2} is obtained by dropping $|\delta\dot{\chi}|^2$ from \eqref{UV_Lmom} and then by integrating out $\delta \chi$. In the position space, the solution of $\delta \chi$ is given by a spatial integral, yielding the non-local EFT action as one can see from the $k$-dependent denominator in \eqref{L_Meff2}. Nonetheless, such a non-locality does not provide any pathology since $\delta \chi$ can be uniquely determined by $\pi$ under an appropriate boundary condition. The action \eqref{L_Meff2} is non-local in space but local in time.

The quadratic action of the $M$-EFT around the background $C_{\mu}=(C_0,\bm{0})$ is
\begin{align}
S_{\rm IR}^{(2)}[\pi]=\int \dd t \int^{M}\!\!\! \frac{\dd^{d-1}k}{(2\pi)^{d-1}}
\frac{1}{2}\left[ (P_X+2XP_{XX})|\dot{\pi}|^2 - P_X k^2 |\pi|^2 + \frac{2XP_{XX}}{M^2}(|\ddot{\pi}|^2 - k^2 |\dot{\pi}|^2) +\cdots \right]
,
\end{align}
where we have the relations
\begin{align}
\delta M^2 = \frac{2XP_{XX}}{P_X} M^2 \,, \quad \Meff^2 = \frac{P_X+2XP_{XX}}{P_X} M^2\,, \quad c_s^2= \frac{P_X}{P_X+2XP_{XX}}
\,.
\end{align}
We find
\begin{align}
S_{\rm IR}^{(2)}[\pi]=\int \dd t \int^{M}\!\!\! \frac{\dd^{d-1}k}{(2\pi)^{d-1}}
\frac{1}{2}\left[ (P_X+2XP_{XX})|\dot{\pi}|^2 - P_X k^2 |\pi|^2 - \frac{2XP_{XX}}{M^2} k^2 |\dot{\pi}|^2 +\cdots \right]
,
\label{SIR_time}
\end{align}
when $|c_s^2| \ll 1~\Rightarrow |\ddot{\pi}|^2 \ll k^2|\dot{\pi}|^2$. This indeed agrees with the low-momentum part of the $\Lambda$-EFT,
\begin{align}
S^{(2)}_{\Lambda}[\pi]&=\int \dd t \int^{\Lambda}\!\! \frac{\dd^{d-1}k}{(2\pi)^{d-1}}
\frac{1}{2} \left[ \frac{P_X+2XP_{XX}}{k^2+M^2}|\dot{\pi}|^2 - P_Xk^2 |\pi|^2 \right]
\nn
&\simeq \int \dd t \int^{\Lambda}\!\! \frac{\dd^{d-1}k}{(2\pi)^{d-1}}
\frac{1}{2} \left[ (P_X+2XP_{XX})|\dot{\pi}|^2 - P_Xk^2 |\pi|^2 - \frac{2XP_{XX}}{M^2}k^2 |\dot{\pi}|^2 +\cdots \right]
,
\end{align}
where $|M^2|\ll \Meff^2 \Rightarrow P_X+2XP_{XX} \simeq 2XP_{XX}$ is used to make the $k^2 |\dot{\pi}|^2$ term equal to that of \eqref{SIR_time}. One should, however, notice that the integration domains are different. In the $\Lambda$-EFT, we need a resummation of the infinite number of spatial derivatives to make the EFT valid even at $|M^2|<k^2 (<\Lambda^2)$. As we mentioned, the $\Lambda$-EFT is thus spatially non-local in terms of the variable $\pi$.

\begin{figure}[t]
\centering
 \includegraphics[width=0.5\linewidth]{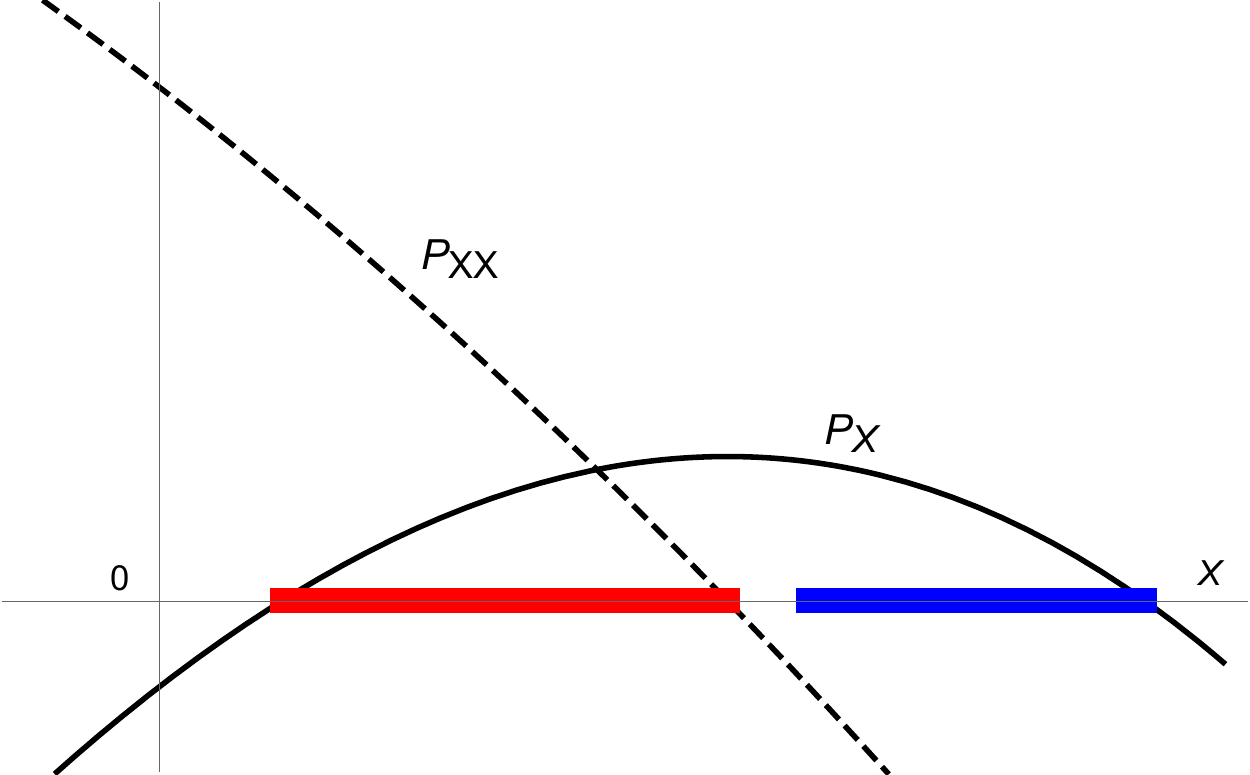}
 \caption{An example of $P_X$ and $P_{XX}$. The consistency conditions \eqref{UV_consistency_NEC} and \eqref{UVconsistency_nonlocal} conclude that the single-field EFT can only have a partial UV completion by means of a two-field model either in the red domain $(P_{XX}>0)$ or the blue domain $(P_{XX}<0)$. Note that the two consistent domains are disconnected. If $\varphi$ tends to go outside the consistent domains by considering an adiabatic motion of $X$, one should return to the UV theory before violating either \eqref{UV_consistency_NEC} or \eqref{UVconsistency_nonlocal}.
 This figure is solely for an illustrative purpose, and our choice of the shapes of $P_X$ and $P_{XX}$ has no physical consequence.}
 \label{fig_PX}
\end{figure}

We have constructed the single-field EFT that is valid in the extended domain $|M^2|<k^2<\Lambda^2$, hence the name  $\Lambda$-EFT, and that accommodates the IR instability caused by $M^2<0$. One of the advantages of the $\Lambda$-EFT over the $M$-EFT is that, due to the wider regime of validity, it can describe the truncation of the IR instabilities at $k=|M|$ without need to go back to the multi-field completion. The UV consistency conditions, combined with the condition for the existence of the $\Lambda$-EFT with this advantage, are (i) the stability conditions of the UV modes,
\begin{align}
f>0 \,, \quad \Meff^2>0\,, \qquad \text{(UV consistency)}
\end{align}
and (ii) the condition that the IR instability, if exists $(M^2<0)$, is under control within the EFT,
\begin{align}
|M^2| < 2(\sqrt{2}-1) \delta M^2
\,, \qquad \text{(EFT predictivity)}
\end{align}
where the condition (ii) does not exist if $M^2$ is positive. Using $P_X$ and $P_{XX}$, the UV consistency conditions are rewritten as
\begin{align}
P_X>0
\,, \qquad \text{(UV consistency)}
\label{UV_consistency_NEC}
\end{align}
and
\begin{align}
\begin{cases}
P_{XX}>0  &{\rm for}~ M^2>0\,,\qquad   \text{(UV consistency)} \\
|P_{XX}| > \left(\frac{1}{2}+\frac{1}{\sqrt{2}} \right)\frac{P_X}{2X} &{\rm for }~M^2<0\,,
\qquad \text{(EFT predictivity)}
\label{UVconsistency_nonlocal}
\end{cases}
\end{align}
around the timelike background, $X>0$, where the second bound is applicable only if $P_{XX}<0$. It is interesting that two consistent regions of $P_{XX}$ are disconnected, that is, $P_{XX}$ is either positive or large negative. The region $-\left(\frac{1}{2}+\frac{1}{\sqrt{2}} \right)\frac{P_X}{2X}< P_{XX}<0$ is inconsistent.\footnote{If we only impose the UV consistency conditions $f>0,~\Meff^2>0$, the consistent regions are either $P_{XX}>0$ or $P_{XX}<-\frac{P_X}{2X}<0$ where two consistent regions are still disconnected.}  We provide an illustrative figure in Fig.~\ref{fig_PX}.
The conditions \eqref{UVconsistency_nonlocal} are summarized in a simple form in terms of the sound speed, given by
\begin{align}
-0.2 \simeq  -\left(\frac{1}{\sqrt{2}}-\frac{1}{2}\right) < c_s^2 < 1
\,,
\end{align}
where the upper bound $c_s^2<1$ is the consequence from the UV consistency while lower bound $-0.2\lesssim c_s^2$ is from the EFT predictivity.
Albeit its appearance, the imaginary sound speed is in fact consistent with the UV stability, and its absolute value is bounded by a finite value of at most order unity.

The negative sign of $P_{XX}$ apparently contradicts the conventional positivity bound~\cite{Adams:2006sv}. One should recall that our setup is completely different from the argument of the conventional positivity bounds. First of all, the positivity bounds are the bounds on the EFT around the Lorentz-invariant background and are not applicable to the EFT around the Lorentz-violating background. In fact, the branch $|P_{XX}| > \left(\frac{1}{2}+\frac{1}{\sqrt{2}} \right)\frac{P_X}{2X}$ cannot have a continuous limit to the Lorentz-invariant background due to the denominator $X$. The flat limit $X\to +0$ yields $P_{XX}<-\infty$ since $P_X$ is related to the kinetic term of the UV theory and thus has to be strictly positive. This means that there is no consistent single-field EFT that can accommodate both the Lorentz-invariant background and the Lorentz-violating background with negative $P_{XX}$. (This does not exclude the possibility that a UV theory may reduce to a single-field EFT with positive $P_{XX}$ around a Lorentz-invariant background and to a different single-field EFT with negative $P_{XX}$ around a Lorentz-violating background. In order to connect the two single-field EFTs one needs to go back to the UV theory.) Furthermore, the negative $P_{XX}$ exhibits the IR instability, meaning that the scattering amplitudes may not be well-defined. Nonetheless, such an EFT is consistent and can describe transient IR instabilities (see e.g.~\cite{Garcia-Saenz:2018vqf}).

\subsection{UV consistency for multi-field UV models}
\label{sec:multi_pert}

We generalize the previous analysis to the multi-field UV models.
Assuming the shift symmetry, the UV Lagrangian is given by
\begin{align}
\mathcal{L}_{\rm UV}^{(2)}&=- \frac{1}{2}\tilde{\epsilon}^2 \gamma_{ab}(\chi) (\partial \chi^a \cdot \partial \chi^b)- \tilde{\epsilon} h_a(\chi) (\partial \chi^a \cdot \partial \varphi) - \frac{1}{2}f(\chi) (\partial \varphi)^2- V(\chi)
\; ,
\label{general_sigma}
\end{align}
where lower-case alphabets $a,b,\dots$ run for the heavy fields $\chi^a$. The metric $\gamma_{AB}$ is supposed to be positive definite which is equivalent to $\gamma_{ab}-h_a h_b/f>0$ and $f>0$. The Killing vector is $\xi^A=\{0, 1 \}$ which is hypersurface orthogonal if and only if
\begin{align}
 f\partial_{[a}h_{b]} + h_{[a} \partial_{b]} f=0\,,
 \label{Frob_multi}
\end{align}
according to Frobenius's theorem.

We consider the constant background,
\begin{align}
\bar{\chi}^a={\rm constant}\,, \quad C_{\mu}=\partial_{\mu}\bar{\varphi}={\rm constant}
\,,
\label{sol_multi}
\end{align}
being subject to the equations
\begin{align}
f_a(\bar{\chi}) X -V_a(\bar{\chi})=0
\,,
\end{align}
where the subscript indices $a,b,\cdots $ is the derivative with respect to $\chi^a$, e.g.~$f_a=\frac{\partial f}{\partial \chi^a}$.
All $\chi^a$ are determined by $X$ as long as the determinant of the matrix
\begin{align}
M^2_{ab}= -f_{ab }X+V_{ab}
\,,
\end{align}
is non-zero. The quadratic action for the perturbations,
\begin{align}
\chi^a=\bar{\chi}^a+\delta \chi^a \,, \quad
\varphi=\bar{\varphi}+\pi
\,,
\end{align}
is
\begin{align}
\mathcal{L}_{\rm UV}^{(2)}= -\frac{1}{2}\tilde{\epsilon}^2 \gamma_{ab} (\partial \delta \chi^a \cdot \partial \delta \chi^b)-\frac{1}{2}M^2_{ab} \delta \chi^a \delta \chi^b - \tilde{\epsilon}h_{[ab]}  (C \cdot \partial \delta \chi^a) \delta \chi^b  -\frac{1}{2} f (\partial \pi)^2 -\tilde{\epsilon} h_a (\partial \pi \cdot \partial \delta \chi^a) - f_a \delta \chi^a (C\cdot \partial \pi)
\,,
\end{align}
where the coefficients are evaluated at the background.

We can use the freedom of the field redefinition
\begin{align}
\varphi \to \varphi' = \varphi + \tilde{\epsilon} g(\chi^a)  \Rightarrow \pi \to \pi'=\pi + \tilde{\epsilon} g_a(\bar{\chi}) \delta \chi^a
\end{align}
so that $h_a(\bar{\chi})=0$ evaluated at the background \eqref{sol_multi}. The quadratic Lagrangian is simplified to be
\begin{align}
\mathcal{L}_{\rm UV}^{(2)}= -\frac{1}{2} \tilde{\epsilon}^2 \gamma_{ab} (\partial \delta \chi^a \cdot \partial \delta \chi^b)-\frac{1}{2}M^2_{ab} \delta \chi^a \delta \chi^b - \tilde{\epsilon} h_{[ab]}  (C \cdot \partial \delta \chi^a) \delta \chi^b  -\frac{1}{2} f (\partial \pi)^2 - f_a \delta \chi^a (C\cdot \partial \pi)
\; ,
\label{LUV_multi_nokinmixing}
\end{align}
Note that $h_{[ab]}=\frac{1}{2}(\partial_b h_a - \partial_a h_b)$ does not vanish, in general, since we can only set $h_a(\bar{\chi})=0$ via a field redefinition while we cannot set $h_{[ab]}(\bar{\chi})=0$ (i.e.~we cannot set $h_a(\chi)=0$ unless the Killing vector is hypersurface orthogonal). The third term vanishes only when \eqref{Frob_multi} holds. 
We have an additional freedom to change $\chi^a$ according to
\begin{align}
\chi^a \to \mathcal{X}^a{}=\mathcal{X}^a{}(\chi) \Rightarrow \delta \chi \to \delta \chi'=G^a{}_b \delta \chi^b
\,,
\end{align}
which will be used later.
The dispersion relations of \eqref{LUV_multi_nokinmixing} are determined by the equation
\begin{align}
D&={\rm det}
\begin{pmatrix}
D_{ab} &  f_a (i C \cdot p) \\
 - f_b (i C \cdot p)  & p^2 f
\end{pmatrix}
=0
\,,
\label{determinant}
\end{align}
where
\begin{align}
D_{ab}&\equiv \tilde{\epsilon}^2 p^2\gamma_{ab}  +2\tilde{\epsilon} h_{[ab]} (iC\cdot p) + M^2_{ab}
\,.
\end{align}
The equation \eqref{determinant} is transformed to
\begin{align}
D={\rm det}D_{ab} \times \left[ p^2f -f_a (D^{-1})^{ab} f_b (C\cdot p)^2 \right]=0,
\quad ({\rm det}D_{ab}\neq 0)\,,
\end{align}
and then the dispersion relation of the light mode is given by
\begin{align}
p^2f -f_a (D^{-1})^{ab} f_b (C\cdot p)^2=0
\,.
\end{align}
The EFT coefficients $P_X$ and $P_{XX}$ are related to the UV coefficients via
\begin{align}
P_X=f\,, \quad P_{XX}= \left. f_a (D^{-1})^{ab} f_b \right|_{p=0} = f_a (M^{-2})^{ab} f_b
\,.
\end{align}
Hereinafter, we set $\tilde{\epsilon}=1$ since it is irrelevant for the following discussions.

As in the two-field case in Sec.~\ref{subsec:spacelike}, when $C_{\mu}$ is spacelike, the terms proportional to $C\cdot p$ do not contribute to the masses of the heavy modes. The masses are determined by the eigenvalues of $M_{ab}^2$ and the no-tachyon condition is the positive definiteness of $M_{ab}^2$. We immediately obtain the same consistency conditions
\begin{align}
P_X(X<0)>0\,, \quad P_{XX}(X<0)>0
\end{align}
from the UV stability conditions $f>0$ and $M_{ab}^2>0$ around the spacelike background.

We then discuss the timelike background. It is convenient to use the Hamiltonian to find the stability conditions around the timelike background $C_\mu = (C_0 , \bm{0})$. Before discussing the multi-field system, let us revisit the Hamiltonian of the two-field system,
\begin{align}
\mathcal{H}^{(2)}_{\rm UV}
&=\frac{1}{2} \left\vert p \right\vert^2+\frac{1}{2} M^2 \left\vert \delta \chi \right\vert^2 + \frac{1}{2f} \left\vert p_{\pi}-f_{\chi} C_0 \delta \chi \right\vert^2 
\end{align}
where we have taken the limit $k \to 0$ since we are interested in the stability of the zero momentum modes. The naive boundedness of the Hamiltonian, i.e.~$M^2>0$, does not tell us the stability condition of the heavy mode since we have seen that $M^2<0$ is allowed. What we need to see is the boundedness of the Hamiltonian incorporating its dynamics. The conjugate $p_{\pi}$ is a constant of motion in $k=0$. The solution describing the coherent oscillation of the heavy mode around the constant background can always be taken to be $p_{\pi}=0$ by absorbing any non-zero constant into a shift of the background $\bar\chi$. Therefore, the stability of the heavy mode follows the boundedness of the Hamiltonian after substituting $p_{\pi}=0$,
\begin{align}
\left. \mathcal{H}^{(2)}_{\rm UV} \right|_{p_{\pi}=0}
&=\frac{1}{2} \left\vert p \right\vert^2+\frac{1}{2} \left(M^2 + \frac{f_{\chi}^2}{f}C_0^2 \right) \left\vert \delta \chi \right\vert^2 
\implies  M^2 + \frac{f_{\chi}^2}{f}C_0^2 >0
\,,
\end{align}
which correctly reproduces the condition $\Meff^2>0$.

Returning to the multi-field models, we study the Hamiltonian of this system and impose the boundedness of the Hamiltonian
(no-ghost and no-tachyon conditions) in the UV. The conjugate momenta are
\begin{align}
p_a&=\gamma_{ab}\delta \dot{\chi}^b+h_{[ab]}C_0 \delta \chi^b
\,, \\
p_{\pi}&=f\dot{\pi}+f_a C_0 \delta \chi^a
\,,
\end{align}
and the Hamiltonian is
\begin{align}
\mathcal{H}^{(2)}_{\rm UV} &=\frac{1}{2}\gamma_{ab} \delta \dot{\chi}^a{}^{\dagger}  \delta \dot{\chi}^b+\frac{1}{2}f |\dot{\pi}|^2+ \frac{1}{2}(k^2 \gamma_{ab}+M^2_{ab})\delta \chi^a{}^{\dagger} \delta \chi^b+ \frac{1}{2}k^2 f |\pi|^2
\,,
\end{align}
where $\delta \dot{\chi}^a$ and $\dot{\pi}$ are understood as the functions of the canonical variables. The boundedness of the Hamiltonian for the mode $k=0$ with $p_{\pi}=0$ requires
\begin{align}
\Meff^2_{ab}&>0
\,, \quad
\Meff^2_{ab}\equiv M^2_{ab}+\frac{f_af_b}{f}C_0^2
\; ,
\label{def_Meffab}
\end{align}
at least under the situation we are considering. The condition $M^2_{ab}>0$ is sufficient but not necessary, just like the two-field model. On the other hand, the Hamiltonian is unbounded at low $k$ if $M^2_{ab}$ is indefinite, leading to an instability at IR, whose threshold is determined by
\begin{align}
{\rm det} (k^2 \gamma_{ab}+M^2_{ab})=0
\,.
\end{align}
Since the heavy modes are stable at $k=0$, such unstable modes must be the IR modes of the light degree of freedom $\pi$.
This fact still renders the EFT reduction consistent, as the UV sector suffers no pathological behavior.

We finally show that $P_{XX}$ is negative if $M_{ab}$ contains a negative eigenvalue.
For this purpose, we use the freedom of the change of $\chi^a$ in the following sequence. Performing a general linear transformation of $\delta \chi^a$, we can diagonalize the matrix $M^2_{ab}$ and set $f_a=(f_1,f_{a'})=(f_1,0)$ simultaneously where the indices with a prime run over $a'=2,3,\cdots$. Denoting the eigenvalues of $M_{ab}^2$ by $M_1^2$ and $M^2_{a'}~(a'=2,3,\cdots)$, the stability condition $\Meff^2_{ab}>0$ is reduced to
\begin{align}
\Meff^2_1 = M_{1}^2+\frac{f_1^2}{f} C_0^2>0 \,, \quad M^2_{a'}>0
\,.
\end{align}
Hence, the number of negative eigenvalues of $M_{ab}^2$ is at most one, which is $M_1^2$ in this basis. With this choice of field-space coordinates, we simply have
\begin{align}
P_{XX}=f_a (M^{-2})^{ab} f_b = \frac{f_1^2}{M_1^2}
\,,
\end{align}
that is, the sign of $M_1^2$ is that of $P_{XX}$. The condition $\Meff_1^2>0$ yields $|P_{XX}|>P_X/2X$ as a consistency in the case of $P_{XX}<0$.

All in all, the same UV consistency conditions are obtained as in the two-field UV model. The condition $P_X>0$ is universal in both spacelike and timelike cases. In addition, the consistency condition on $P_{XX}$ is $P_{XX}>0$ around the spacelike background while the negative $P_{XX}$ is allowed around the timelike background. The positive region $P_{XX}>0$ and the negative region $P_{XX}<0$ are disconnected since the consistency condition is either $P_{XX}>0$ or $P_{XX}<-P_X/2X<0$. Note that these conditions are necessary conditions and may be sharpened by studying detailed spectra of the UV theory, i.e. investigating the EFT predictivity conditions, as we did in the case of the two-field UV model.


\section{Discussions}
\label{sec:discussion}

\subsection{Role of higher derivative operators}
We have seen that the IR instabilities around the spacelike and timelike backgrounds are qualitatively different: the threshold of the former instability can be chosen so that $k_*^2\approx M^2$, where a finite number of higher derivatives well approximate the dynamics, while the latter one requires a resummation of higher derivative operators to cure it.
In this subsection, we discuss how this distinction can be understood.

We consider the k-essence part of the quadratic Lagrangian
\begin{align}
\mathcal{L}_{\rm IR}^{(2)}=-\frac{1}{2}\left[P_X(\partial \pi)^2-P_{XX}(C\cdot \partial \pi)^2+\cdots  \right] = -\frac{1}{2}g^{\mu\nu}_{\rm eff}\partial_{\mu}\pi \partial_{\nu} \pi +\cdots
\,, 
\end{align}
where
\begin{align}
g^{\mu\nu}_{\rm eff}\equiv P_X \eta^{\mu\nu} - P_{XX} C^{\mu}C^{\nu} 
\; ,
\end{align}
and we remind $C_\mu \equiv \partial_\mu \bar\varphi$ and $\eta^{\mu\nu}$ is the Minkowski metric. We study a $1+1$ dimensional spacetime for simplicity of the explanation, but essentially the same discussions follow in general dimensions. In both spacelike and timelike cases, the IR instability exists if 
\begin{align}
P_X + 2XP_{XX}<0
\end{align}
is satisfied in terms of the k-essence coefficient. This parameter region leads to the gradient instability around the spacelike background, $C_{\mu}=(0,\sqrt{-2X})$, because the effective metric is given by
\begin{align}
g^{\mu\nu}_{\rm eff}&=
\begin{pmatrix}
-P_X & 0 \\
0 & P_X+2XP_{XX}
\end{pmatrix}
=
\begin{pmatrix}
- & 0 \\
0 & -
\end{pmatrix}
\; , \qquad \text{(spacelike background)}
 \,.
\end{align}
On the other hand, the unstable parameter corresponds to the ghost instability around the timelike background, $C_{\mu}=(\sqrt{2X},0)$,
\begin{align}
g^{\mu\nu}_{\rm eff}&=
\begin{pmatrix}
-(P_X+2XP_{XX}) & 0 \\
0 & P_X
\end{pmatrix}
=
\begin{pmatrix}
+ & 0 \\
0 & +
\end{pmatrix}
\; , \qquad \text{(timelike background)}
 \,.
\end{align}
Therefore, since the no ghost condition in the UV ensures $P_X > 0$, the IR instabilities can be distinguished by whether the instability is caused by flipping the sign of the spatial derivative (spacelike background) or the sign of the time derivative (timelike background).

The gradient instability in EFT can be cured by adding a local higher derivative operator, {\it \`a la} ghost condensate~\cite{ArkaniHamed:2003uy,ArkaniHamed:2003uz}. In the present case, such a higher derivative operator is
\begin{align}
\partial_{\mu}X\partial^{\mu}X
\end{align}
rather than $(\Box \varphi)^2$, see \eqref{EFT_simplified}. Considering the perturbations $\pi$ around the spacelike background $C_{\mu}=(0,C_i)$, the higher derivative operator $(\partial X)^2$ yields
\begin{align}
-C^i \partial_i \dot{\pi} C^j \partial_j \dot{\pi} + C^i \partial_k \partial_i \pi C^j \partial^k \partial_j \pi 
\end{align}
at the quadratic order. In general, higher derivative operators should be treated as perturbations since the higher derivative terms yield Ostrogradsky ghost state(s), which in turn indicate that the regime of validity of the EFT is restricted below the ghost state(s). On the other hand, no higher time derivative term is generated by $(\partial X)^2$ at the quadratic order around the purely spacelike background $C_0=0$ (while the higher time derivatives appear at nonlinear level of perturbations). The contribution from $(\partial X)^2$ can consistently dominate over the ``leading'' lower derivative contribution $g^{\mu\nu}_{\rm eff}\partial_{\mu}\pi \partial_{\nu} \pi$ and can change the sign of the gradient term. After adding the higher derivative operator $(\partial X)^2$, the quadratic Lagrangian \eqref{quadratic_next} around the spacelike background is explicitly given by
\begin{align}
\mathcal{L}^{(2)}_{\rm IR}=\frac{1}{2}\left[ \left( P_X + \frac{2|X|P_{XX}}{M^2}k^2 \right) |\dot{\pi}|^2 - \left( (P_X+2XP_{XX}) k^2 + \frac{2|X|P_{XX}}{M^2} k^4 \right) |\pi|^2 +\cdots \right]
\end{align}
in the momentum space where we have $k^2=k_{\parallel}^2$ since we are considering $1+1$ dimensions, for simplicity. Even if there is the IR instability, $P_{X}+2XP_{XX}<0$, the presence of $k^4$ term cures the instability within the regime of validity of the EFT.

On the other hand, the IR ghost instability can be resolved by using the idea of Jeans ghost~\cite{Gumrukcuoglu:2016jbh}. Even though the Lagrangian in terms of $\pi$ has the ghost instability below some value of $k$, we can perform the canonical transformation to write the action in terms of $\tilde{\pi}$ as in the form of \eqref{L_Meff}. The kinetic term of \eqref{L_Meff} is positive definite, and thus there is no ghost instability. The gradient term is proportional to $\Omega_{\pi}^2=c_s^2 k^2 + \frac{\delta M^4}{\Meff^6}k^4 +\cdots$ where $P_X+2XP_{XX}<0$ corresponds to $c_s^2<0$. The IR ghost instability is translated into the IR gradient instability via the canonical transformation which is as harmless as the IR instability around the spacelike background.

There is, however, a clear distinction between the remedies of the IR gradient and ghost instabilities. Let us discuss the role of the higher derivative operator $(\partial X)^2$ around the timelike background. In contrast to the spacelike background, $(\partial X)^2$ in the timelike background yields higher time derivative term $|\ddot{\pi}|^2$ in addition to $k^2|\dot{\pi}|^2$. Nonetheless, this is not a crucial difference since we may  treat $|\ddot{\pi}|^2$ as perturbations when the Lorentz symmetry is largely broken, $|c_s^2| \ll 1~\Rightarrow |\ddot{\pi}|^2 \ll k^2|\dot{\pi}|^2$, as used to derive \eqref{SIR_time}. The crucial point is the strong coupling. In the example \eqref{SIR_time}, the sign of the time derivative is negative even if $k^2|\dot{\pi}|^2$ is taken into account. Hence, we try to consider the EFT with a ``wrong'' sign of the $(\partial X)^2$ operator to change the sign of the kinetic term:
\begin{align}
\mathcal{L}^{(2)}_{\rm IR} &= \frac{1}{2}\left[ (P_X+2XP_{XX})|\dot{\pi}|^2 + \frac{2XP_{XX}}{M^2}k^2|\dot{\pi}|^2 +\cdots \right]
\nn
&\simeq \frac{1}{2} \frac{2XP_{XX}}{M^2}\left( -|M^2| + k^2 \right) |\dot{\pi}|^2 +\cdots
\end{align}
where $k^2|\dot{\pi}|^2$ is taken to be opposite in sign to \eqref{SIR_time} and we have used $P_X+2XP_{XX}\simeq 2XP_{XX}$ and $M^2<0$. Then, one realizes that the sign of the kinetic tern cannot be flipped since the light mode $\pi$ is infinitely strongly coupled at the critical point $k^2-|M^2|=0$. The EFT description breaks down before approaching $k^2-|M^2|=0$: the ghost cannot be cured by adding a local higher derivative operator, contrary to the gradient instability. On the other hand, one may consider a resummation of spatial derivative operators to construct the non-local kinetic term
\begin{align}
\mathcal{L}^{(2)}_{\rm IR} \simeq \frac{1}{2}  \frac{2XP_{XX} M^2}{k^2-|M^2|} |\dot{\pi}|^2 + \cdots
\end{align}
just like \eqref{L_Meff2}. The critical point $k^2-|M^2|=0$ now corresponds to the infinitely weakly coupling, implying that the IR ghost can be resolved be the use of a non-local operator within the EFT. In fact, this is what we have observed in the EFT from our concrete UV models (see also the theories in~\cite{Gumrukcuoglu:2016jbh}). Alternatively, one may consider a canonical transformation to translate the IR ghost instability into the IR gradient instability. Nevertheless, such a canonical transformation should be a non-local map of the on-shell variables as we have seen in \eqref{rel_pi_tpi} and the non-locality again plays an important role in this procedure.


\subsection{UV obstruction to screening}
\label{sec:screening}

Nonlinear kinetic terms have gained attention phenomenologically since it can provide a screening mechanism, in which the fifth force carried by a light scalar can be screened at a sufficiently small scale. It has been known, however, that the requirement from the positivity bounds is incompatible with a successful screening at least in some particular theories~\cite{Brax:2012jr,Burrage:2014uwa}. Here, we discuss whether or not this obstruction is a general feature under our UV consistency conditions.

We study a four-dimensional spacetime to make discussions concrete and assume that the light field $\varphi$ is canonically normalized.
We consider a situation when the light field $\varphi$ couples with a localized source via the interaction $\frac{1}{\Mpl} \varphi T$ where $T$ is the trace of the energy-momentum tensor of the source. Then, the field $\varphi$ gives rise to a force in addition to the standard gravitational force. In the asymptotic region of the spacetime outside the source, the equation of motion of $\varphi$ may be approximated by the linear equation, $\Box \varphi = 0 $, whose solution should behave as $\varphi \propto r^{-1}$ for large $r$, where $r$ is the distance from the localized source.
Both the fifth force and the gravitational force obey the Newtonian law $r^{-2}$ and are comparable in the asymptotic region. As we get closer to the source, the gradient of $\varphi$ becomes larger and larger, and at some point nonlinear kinetic terms can no longer be ignored, modifying the scaling law of $\varphi$. The expectation is that nonlinear kinetic terms may modify $\varphi$ by making its distant behavior milder than $r^{-1}$, so that the force carried by $\varphi$ is smaller than the gravitational force in the small scales. For instance, the model $P(X)=X-X^2/\Lambda_2^4$, where $\Lambda_2=(M_{\rm UV} M)^{1/2}$ is the scale determining the nonlinear kinetic term, leads to $\varphi \propto r^{1/3}$ in the nonlinear region, $\partial \varphi \gtrsim \Lambda_2^2$, which is a successful example of the screening by the nonlinear kinetic term $-X^2/\Lambda_2^4$~\cite{Brax:2012jr}. The EFT can be valid even in the nonlinear region as far as higher derivatives are small, $\partial^{n+1} \varphi \ll M^{n} \Lambda_2^2~(n\geq 1)$.

The existence of the screening can be understood by the weak coupling (see e.g.~\cite{deRham:2014zqa}). We may split the field and the source into backgrounds $\{ \bar{\varphi}, \bar{T} \}$ and perturbations $\{ \pi, \delta T \}$. Due to the presence of the nonlinear kinetic terms, the fluctuations of the scalar field propagate on the effective metric, not the spacetime metric, leading to
\begin{align}
\mathcal{L}^{(2)}_{\rm IR}=-\frac{1}{2}g_{\rm eff}^{\mu\nu}(\bar{\varphi})\partial_{\mu} \pi \partial_{\nu} \pi + \frac{1}{\Mpl } \pi \delta T\,,
\end{align}
where we ignore the higher derivative terms.
As we have seen, the k-essence theory yields $g_{\rm eff}^{00} = -P_X$ around the spacelike background. The model $P(X)=X-X^2/\Lambda_2^4$ reads $P_X \to 1 $ as $X\to 0$ (the asymptotic region) while $P_X \gg 1$ as $|X| \gg \Lambda_2^4$ (the nonlinear region). We then introduce the canonically normalized field via
\begin{align}
\hat{\pi} \equiv \sqrt{P_X} \pi
\,,
\end{align}
by which the Lagrangian is written as
\begin{align}
\mathcal{L}^{(2)}_{\rm IR}=-\frac{1}{2}\hat{g}_{\rm eff}^{\mu\nu}(\bar{\varphi})\partial_{\mu} \hat{\pi} \partial_{\nu} \hat{\pi} + \frac{1}{ \Mpl \sqrt{P_X}} \hat{\pi} \delta T+\cdots \,.
\end{align}
with $\hat{g}^{00}_{\rm eff}=-1$ where $\cdots$ are lower derivative terms of $\hat{\pi}$. One can see that the effective coupling constant is given by $(\Mpl \sqrt{P_X})^{-1}$ which is much smaller than the gravitational one $\Mpl^{-1}$ in the nonlinear region, $P_X \gg 1$. 
Indeed, $P_X$ is nothing but the coefficient in front of the kinetic term of $\varphi$ before integrating out the heavy modes in our partial UV models, cf.~\eqref{PX_rel}. This means that $P_X=f \gg 1$ is the weak coupling limit of $\varphi$.

We then use the UV consistency condition $P_{XX}(X<0)>0$ which should hold for a spacelike configuration of $\partial_{\mu} \varphi$ (recall that $P_{XX}<0$ is allowed only around timelike backgrounds, $X>0$). As we explained, the scalar field should obey the Newtonian law $\varphi \propto r^{-1}$ in the asymptotic region, implying
\begin{align}
-X|_{ \text{at a finite } r} > -X|_{r\to \infty}=0
\,. \label{localX_vs_asymptoticX}
\end{align}
Since the condition $P_{XX}>0$ implies that $P_X$ is a monotonically increasing function of $X$, the inequality \eqref{localX_vs_asymptoticX} concludes
\begin{align}
P_X|_{\text{at a finite } r} < P_X|_{r\to \infty}
\,.
\end{align}
As a result, the UV consistency condition $P_{XX}>0$ obstructs the requirement for the screening, $P_X|_{\text{at a finite } r} > P_X|_{r\to \infty}$. The fifth force at a finite $r$ ($X<0$) should be stronger than the force in the asymptotic region $X\to 0$, which is a requirement from admitting a UV completion at least by means of multi-field models.

In our analysis, the first-order derivatives are treated nonperturbatively while second- and higher-order derivatives are treated as perturbations. We have concluded that the screening via a large first-order derivative is incompatible with the UV requirement in our partial UV models.
There is another screening mechanism via nonlinear kinetic terms involving second-order derivatives of the field, called the Vainshtein screening~\cite{Vainshtein:1972sx,Nicolis:2008in}. If all derivative interactions are specified by a single UV scale, higher-derivative terms should be suppressed over lower-derivative ones within the regime of validity of EFT (this corresponds to the less hierarchical case $M_{\rm UV} \sim M$ in our partial UV model). The implicit assumption for the Vainshtein screening is, on the other hand, that the EFT contains two scales, often denoted by $\Lambda_2$ and $\Lambda_3$, and the Lagrangian is given by
\begin{align}
\mathcal{L}=X+\frac{1}{\Lambda_2^4}X^2+\frac{g_3}{\Lambda_3^3}X\Box \varphi +\frac{g_4}{\Lambda_3^6}X [(\Box \varphi)^2-(\nabla_{\mu}\nabla_{\nu}\varphi)^2] +\cdots
\,,
\label{weakly_Gal}
\end{align}
with the hierarchy $\Lambda_2\gg \Lambda_3$ and $g_3,g_4 = \mathcal{O}(1)$, where the lower-derivative term is suppressed by a larger scale $\Lambda_2$.  The terms suppressed by $\Lambda_3$ are called the Galileon terms which are invariant under the Galileon symmetry, $\varphi \to \varphi+b_{\mu}x^{\mu}+c$, in the flat spacetime. The theory \eqref{weakly_Gal} is dubbed as the weakly broken Galileon theory~\cite{Pirtskhalava:2015nla} since the operators suppressed by $\Lambda_2$ (and gravitational interactions) weakly break the Galileon symmetry. Thanks to the hierarchy $\Lambda_2 \gg \Lambda_3$, the Galileon terms can dominate over not only the linear kinetic term $X$ but also the lower-derivative interaction $X^2$ to screen the fifth force via a large second-order derivative of the field (see~\cite{Gratia:2016tgq} for a concrete analysis). On the other hand, the recent progress of the positivity bounds for the Lorentz-invariant theories concludes that such a hierarchy lowers the cutoff of the theory~\cite{Bellazzini:2020cot,Tolley:2020gtv,Caron-Huot:2020cmc,Arkani-Hamed:2020blm}, implying that the Vainshtein screening would be also incompatible with the UV requirements. In typical phenomenological situations, we suppose the scaling, $\Lambda_2=(\Mpl m)^{1/2}$ and $\Lambda_3=(\Mpl m^2)^{1/3}$, where $m$ is the scale of the IR physics in which the cutoff is of the order of $m$, contradicting the assumption of the EFT reduction, the existence of the hierarchy between the UV physics and the IR physics (see e.g.~\cite{Tolley:2020gtv}).
Indeed, our partial UV model also suggests the difficulty of the Vainshtein screening. The schematic form of the single-field EFT is given by \eqref{EFT_schematic} where the dimension of $\varphi$ is $[\varphi]=-1$. Canonically normalizing $\varphi$ and focusing on the derivative interactions in four dimensions, we obtain
\begin{align}
\mathcal{L}_{\rm IR}
&=X + M_{\rm UV}^2 M^2 \sum_{i,j} c_{ij}  \left(\frac{\nabla}{M} \right)^j \left( \frac{\nabla \varphi}{M_{\rm UV} M } \right)^i 
\\ \nonumber
&=  X + \frac{1}{\Lambda_2^4}X^2 + \frac{1}{\Lambda_3^6} X (\nabla^2 \varphi)^2 + \frac{1}{\Lambda_4^8} X (\nabla^3 \varphi)^2 + \cdots + \frac{1}{\Lambda_{n}^{2n}} X (\nabla^{n-1} \varphi)^2 +\cdots
\,,
\end{align}
where $\Lambda_{n} = (M_{\rm UV} M^{n-1})^{1/n}$. In particular, we have $\lim_{n\to \infty} \Lambda_n = M$. In the second line, we have focused on four-point interactions, omitting the dimensionless constants. If one wants to set $\Lambda_2=(\Mpl m)^{1/2},~\Lambda_3=(\Mpl m^2)^{1/3}$, one needs to assume $M_{\rm UV}= \Mpl$ and $M= m$ where the scale of the IR physic $m$ is comparable to the cutoff of the single-field EFT $M$, concluding the breakdown of the EFT in the Vainshtein regime.\footnote{As we explained at the beginning of Sec.~\ref{sec:UV_consistency}, the no-tachyon condition (the UV consistency condition) should be carefully discussed when $M_{\rm UV}= \Mpl$ since the background curvature cannot be ignored. Nonetheless, this subtlety is not important to the present argument. }

In summary, although the nonlinear kinetic terms can dominate over the linear kinetic terms without spoiling the validity of the EFT when $X\gg \Lambda_2^4$ with $\partial^{n+1} \varphi \ll M^n \Lambda_2^2= \Lambda_{n+2}^{n+2}$, the requirement from the UV consistency contradicts the requirement from exhibiting the screening. The nonlinear kinetic terms should act as enhancing the fifth force at short scales around the spacelike background. On the other hand, we cannot find any consistent Vainshtein regime where the Galileon terms dominate over the other interactions, in agreement with the recent bottom-up arguments~\cite{Tolley:2020gtv}. Note that our analysis is fully nonlinear, but is not as quantitative as the new positivity bounds~\cite{Bellazzini:2020cot,Tolley:2020gtv,Caron-Huot:2020cmc,Arkani-Hamed:2020blm}. It is an open question whether there is a fine-tuning to admit the Vainshtein screening around Lorentz-violating backgrounds, consistently with the UV requirements. See also~\cite{Kaloper:2014vqa,Keltner:2015xda,deRham:2017imi,Padilla:2017wth,Burrage:2020bxp} for related discussions.

When this manuscript was being completed, the paper~\cite{Davis:2021oce} appeared in arXiv which would contradict our discussion about the screening. The paper~\cite{Davis:2021oce} uses the result of~\cite{Grall:2021xxm}, which claims that, under the dispersion relation $\omega^2=c_s^2 k^2$, there exist positivity constraints on four-point amplitudes even around the timelike background, and applies the result into spacelike configurations of $\partial_{\mu}\varphi$. A claim of~\cite{Davis:2021oce} is that the positivity bounds can be compatible with the screening in models $P(X)\simeq X+c_n X^n/\Lambda_2^{4(n-1)}$ with an odd number $n$\footnote{The model $P(X)= X+c_n X^n/\Lambda_2^{4(n-1)}$ with $n\geq 3$ does not satisfy the strict inequality $P_{XX}>0$ at $X=0$, meaning that we need to add $X^2$ to have a smooth weak field limit $C_{\mu} \to 0$ in the asymptotic region. Here, we assume that $X^n~(n\geq 3)$ dominates over $X^2$ (and other terms) in the nonlinear kinetic regime. Note that Reference~\cite{Chandrasekaran:2018qmx} argued positivity bounds on the $n$th-order theory $P(X)=X+\sum_{i=n}^{\infty} c_i X^i/\Lambda_2^{4(i-1)}$ around the Lorentz-invariant background, where the first nonnegligible interaction is $X^n$. We will make a comparison of our results with~\cite{Chandrasekaran:2018qmx} in Sec.~\ref{sec:comparison}.} (note that our definition of $X$ is opposite in sign to their definition of $X$). 
However, as we have clarified in this paper, the dispersion relation has a rich structure which is inevitably nonlinear around Lorentz-violating backgrounds.
Careful analysis should be required to obtain generic consistency conditions which we will further discuss in Sec.~\ref{subsub:analytic_structure}. Furthermore, the constraints obtained around the timelike background cannot be applied to the spacelike configurations. For instance, as for the models $P(X)\simeq X+c_n X^n/\Lambda_2^{4(n-1)}$ with an odd number $n$, our bound for a stable EFT, $P_{XX}>0$, reads $c_n>0$ around the timelike background while reads $c_n<0$ around the spacelike background. Recall that the EFT may be consistent only around particular backgrounds, see e.g.~Fig.~\ref{fig_PX} and Appendix~\ref{app:U(1)}. If one chooses the bound obtained around the timelike background, $c_n>0$, and extrapolates the theory into the spacelike background, the model looks compatible with the requirement from the screening, $P_{XX}<0$ with $X<0$. However, the model with $c_n>0$ is inconsistent around the spacelike background at least in our consistency conditions, and cannot be used for screening phenomena. The consistent sign around the spacelike background is $c_n<0$ which is incompatible with the screening as we have discussed with a more general argument. Nonetheless, our analysis relies on a particular class of the partial UV completion and it must be interesting to investigate a theory with the screening mechanism to be consistent with the UV requirements.


\subsection{Null energy condition}
\label{subsec:nullenergycondition}

Gravity can play an important role implicitly through the energy conditions. In this subsection we thus consider the implication of the null energy condition (NEC) in our framework. Since the class of (partial) UV completion considered in the present paper does not involve any sources of NEC violation, the corresponding single-field EFT at low energy should also respect NEC.

The stress energy tensor in the UV theory and that in the single-field EFT are defined respectively by
\begin{equation}
 T_{\rm UV}^{\mu\nu} = \frac{2}{\sqrt{-g}}\frac{\delta S_{\rm UV}}{\delta g_{\mu\nu}}\,, \quad
 T_{\rm IR}^{\mu\nu} = \frac{2}{\sqrt{-g}}\frac{\delta S_{\rm IR}}{\delta g_{\mu\nu}}\,. 
\end{equation}
They should agree with each other in the regime of validity of the EFT: $T_{\rm UV}^{\mu\nu}\simeq T_{\rm IR}^{\mu\nu}$. For the UV theory (\ref{Lagrangian_full}) that we have assumed throughout the present paper, we have
\begin{equation}
 T_{\rm UV}^{\mu\nu} = \gamma_{AB}\nabla^{\mu}\Phi^A\nabla^{\nu}\Phi^B + \mathcal{L}_{UV}g^{\mu\nu}\,,
\end{equation}
and thus NEC is respected:
\begin{equation}
 T_{\rm UV}^{\mu\nu} \ell_{\mu} \ell_{\nu} = \gamma_{AB} (\ell\cdot \partial\Phi^A) (\ell\cdot \partial\Phi^B) \geq 0\,,
\end{equation}
for an arbitrary null vector $\ell^{\mu}$, provided that the matrix $\gamma_{AB}$ is positive definite. Therefore, all single-field EFTs that result from this UV theory inevitably satisfy the null energy condition. Indeed, for the EFT Lagrangian $\mathcal{L}_{\rm IR} = P(X)$ and the background $\partial_{\mu}\varphi = C_{\mu}$, the energy-momentum tensor is
\begin{align}
T_{\rm IR}^{\mu\nu} = P_X C^{\mu}C^{\nu}+g^{\mu\nu} P\,,
\end{align}
and thus 
\begin{align}
T_{\rm IR}^{\mu\nu}\ell_{\mu}\ell_{\nu}=P_X (C \cdot \ell)^2 \geq 0\,,
\end{align}
for an arbitrary null vector $\ell^{\mu}$. Here, we have used $P_X>0$.

To be more concrete, let us consider the local description of the $\Lambda$-EFT \eqref{L_Meff}. Considering the case $|M^2|\ll \Meff^2$, the quadratic Lagrangian looks similar to that of ghost condensate,
\begin{align}
\mathcal{L}_{\rm GC}=\tilde{P}(\tilde{X}) - \frac{1}{2\tilde{\Meff}^2} (\Box \tilde{\varphi})^2 +\cdots
\,, \label{ghost_condensate}
\end{align}
where $\cdots$ are even higher derivative terms.
The quadratic Lagrangian of the ghost condensate for the perturbation $\tilde{\varphi}=\bar{\varphi}+\tilde{\pi}$ is
\begin{align}
\mathcal{L}_{\rm GC}^{(2)}=\frac{1}{2}\left[ (\tilde{P}_X+2 \tilde{X}\tilde{P}_{XX}) \left\vert \dot{\tilde \pi} \right\vert^2- \left( \tilde{P}_X k^2 + \frac{1}{\tilde{\Meff}^2} k^4 \right) \left\vert \tilde{\pi} \right\vert^2 +\cdots \right] \; ,
\label{quadratic_GC}
\end{align}
where dots include higher orders in perturbations and in time derivatives. It might look plausible to identify the $\Lambda$-EFT \eqref{L_Meff} with the ghost condensate. However, this is not possible, as we shall see now. Around the timelike background, the sound speed of perturbations is given by $c_s^2\equiv P_X/(P_X+2XP_{XX})$, and from a low-energy perspective there are in principle two possible realizations of the imaginary sound speed, $P_X<0$ or $P_X+2XP_{XX}<0$. The first case, which violates NEC, is known in the context of ghost condensate~\cite{ArkaniHamed:2003uy,ArkaniHamed:2003uz} with the soft breaking of the shift symmetry~\cite{Creminelli:2006xe}~\footnote{If the shift symmetry is exact then the averaged NEC holds~\cite{Mukohyama:2009um}.}, and the latter one, which possesses the IR ghost, is discussed in~\cite{Garcia-Saenz:2018vqf}. These EFTs can be distinguished by the way how the imaginary sound speed is realized: $\tilde{P}_X>0$ and $\tilde{P}_X+2 \tilde{X}\tilde{P}_{XX}<0$ in the case of the $\Lambda$-EFT; and $\tilde{P}_X<0$ and $\tilde{P}_X+2 \tilde{X}\tilde{P}_{XX}>0$ in the case of the ghost condensate. This means that the overall sign of the quadratic action is opposite in these two EFTs. As a result, physical degrees of freedom in these two EFTs gravitate differently. (We provide a further comparison in Appendix~\ref{sec:ghostcondensate}.) Therefore, our UV models cannot deduce the ghost condensate in the low-energy limit since NEC is preserved. On the other hand, the latter one is compatible with the assumed (partial) UV completion, even though it apparently contradicts the naively applied positivity bound $P_{XX}>0$.

While our results are obtained from a particular class of partial UV completion, still they generically hold for nonlinear sigma models independently from the number of the fields, the field space metric, and the potential (see also Appendix \ref{subsec:DBI} for the DBI-type generalization in which the same consistency conditions are obtained). It would be interesting to investigate whether our UV consistency conditions have to hold even for other types of UV completion. It is also important to stress that, even if the same conditions are obtained from generic assumptions, one should not a priori exclude other possibilities like the EFT of ghost condensate~\cite{ArkaniHamed:2003uy,ArkaniHamed:2003uz}, the scordatura theory~\cite{Motohashi:2019ymr,Gorji:2020bfl}, a stable violation of NEC~\cite{Creminelli:2006xe}, and theories exhibiting the screening mechanism~\cite{Babichev:2009ee,Brax:2012jr}, as long as they are consistent on their own. Rather, any observational signatures of these theories should be understood as a smoking gun for a richer structure in the UV, for example, sources of NEC violation such as objects with negative tension motivated by orientifold constructions in string theory~\cite{Marolf:2001ne}


\subsection{Revisiting positivity}
\label{sec:diagram}
\subsubsection{Diagrammatic understanding}
We revisit our UV consistency conditions from a different perspective in this subsection. So far, we have avoided the use of amplitudes because we are interested in non-trivial backgrounds where there could in general be subtleties to define the S-matrix in Lorentz-violating systems, and we have mainly considered the linear order of perturbations. In this section, we re-discuss the two-field model and the UV consistency of the k-essence by using the Feynman-like diagrams, putting aside the subtleties about the rigorous definitions. This provides simple and intuitive explanations to the results of the previous sections and explain why we have discussed the quadratic Lagrangian rather than the four-point interaction.

The essential difference from Lorentz-symmetric cases is the existence of the mixing term $-f_{\chi} (C\cdot \partial \pi) \delta \chi$ in \eqref{UV_L}, where $C_{\mu}=\partial_{\mu}\bar{\varphi}\neq 0$ is the origin of the spontaneous Lorentz-symmetry breaking.
The dispersion relation of the two-field model around a background $C_{\mu}\neq 0$ is given by a root of
\begin{align}
{\rm det}
\begin{pmatrix}
p^2+M^2 & f_{\chi} (ip\cdot C) \\
-f_{\chi} (ip \cdot C) & p^2 f
\end{pmatrix}
&=(p^2+M^2)\left[ p^2f - \Pi(p) \right]  &(p^2+M^2 \neq 0)
\nn
&=0
\,,
\label{dis_self_energy}
\end{align}
where
\begin{align}
i\Pi(p)&=i\frac{f_{\chi}^2 (p\cdot C)^2}{p^2+M^2}
\,.
\end{align}
This result is diagrammatically reproduced by 
\begin{align}
\frac{i\Pi(p)}{|i p\cdot C|^2} &=
\begin{tikzpicture}[baseline=0]
\begin{feynhand}
\vertex [particle] (i1) at (0,0)  {};
\vertex [particle] (o1) at (3,0)  {};
\vertex [crossdot] (a1) at (1,0) {} ;
\vertex [crossdot] (a2) at (2,0) {} ;
\propagator [plain] (i1) to (a1);
\propagator [scalar] (a1) to (a2);
\propagator [plain] (a2) to (o1);
\end{feynhand}
\end{tikzpicture}
\,, \label{Pi_light} \\
\end{align}
with the rules
\begin{align}
\begin{tikzpicture}[baseline=0]
\begin{feynhand}
\vertex [dot] (a1) at (1,0)  ;
\vertex [dot] (a2) at (2,0)  ;
\propagator [scalar] (a1) to (a2);
\end{feynhand}
\end{tikzpicture}
= \frac{-i}{p^2+M^2}
\,, \qquad
\otimes = -if_{\chi}
\,,
\label{propagator}
\end{align}
where the left-hand side is divided by $|i p\cdot C|^2$ so that the vertex $\otimes$ is normalized not to include the factor $ip\cdot C$. The dashed line is interpreted as the propagator of $\delta \chi$ while the solid line is the external leg of $\pi$. We use this diagrammatic notation to compute the leading-order EFT coefficients, $P^{(n)}\equiv \dd^n P/d X^n$.

We consider the leading-order EFT, namely the k-essence theory $\mathcal{L}_{\rm IR}=P(X)$. The coefficients $P^{(n)}~(n\geq 2)$ are the coefficients of the $n$-point interactions, $\mathcal{L}_{\rm IR} \supset  (-1)^n \frac{P^{(n)}}{n!} (C\cdot \partial \pi)^n$. Since the EFT is obtained by integrating out the heavy modes, $P^{(n)}$ can be computed as the Feynman diagrams of which all $n$ external legs are connected to $C\cdot \partial \pi$, while the internal lines are the UV mode $\delta \chi$. We define
\begin{align}
\mathcal{V}(\chi,X)\equiv V(\chi)-f(\chi)X
\end{align}
and its derivatives, 
\begin{align}
\mathcal{V}_{\chi\chi}=\left. \frac{\partial^2 \mathcal{V}}{\partial \chi^2} \right|_{\chi=\chi_0(X)}\,, \quad 
\mathcal{V}_{\chi\chi\chi}=\left. \frac{\partial^3 \mathcal{V}}{\partial \chi^3} \right|_{\chi=\chi_0(X)}\,, \cdots \,,
\mathcal{V}^{(n)}=\left. \frac{\partial^n \mathcal{V}}{\partial \chi^n} \right|_{\chi=\chi_0(X)}
\,.
\end{align}
We put the superscript $(n)$ to denote the general $n$-th derivative of $f$ and $\mathcal{V}$ with respect to $\chi$ evaluated at $\chi=\chi_0(X)$, instead of writing $n$ subscripts of $\chi$. Note the relation $\mathcal{V}_{\chi\chi}=M^2|_{\chi=\chi_0(X)}$. In the two-field (partial) UV models, all the interactions of perturbations are
\begin{align}
 -\frac{f^{(n)}}{n!}(C\cdot \partial \pi) \delta \chi^n \,, \quad 
 -\frac{f^{(n)}}{2\cdot n!}(\partial \pi)^2 \delta \chi^n\,, \quad -\frac{\mathcal{V}^{(n)}}{n!}\delta \chi^n
\end{align}
around the constant background where the coefficients of  $(C\cdot \partial \pi)\delta \chi^n$ and $(\partial \pi)^2 \delta \chi^n$ are not independent due to the original implementation of the Lorentz invariance at UV. The UV coefficients, $f^{(n)}$ and $\mathcal{V}^{(n)}$, can be arbitrary in the IR perspective unless additional assumptions on UV are imposed.

For instance, $P^{(2)}=P_{XX}$ is diagrammatically computed as
\begin{align}
iP_{XX}(X\neq 0) &=
\begin{tikzpicture}[baseline=0]
\begin{feynhand}
\vertex [particle] (i1) at (0,0) { $ C \!\cdot\! \partial \pi$};
\vertex [particle] (o1) at (3,0) {$ C \!\cdot\! \partial \pi$};
\vertex [crossdot] (a1) at (1,0) {} ;
\vertex [crossdot] (a2) at (2,0) {} ;
\propagator [plain] (i1) to (a1);
\propagator [scalar] (a1) to (a2);
\propagator [plain] (a2) to (o1);
\end{feynhand}
\end{tikzpicture}
= \left|
\begin{tikzpicture} [baseline=-0.1cm]
\begin{feynhand}
\vertex [particle] (i1) at (0,0) {$C \!\cdot\! \partial \pi$};
\vertex [crossdot] (a1) at (1,0) {} ;
\vertex [particle] (a2) at (2,0) {$\delta \chi$};
\propagator [plain] (i1) to (a1);
\propagator [scalar] (a1) to (a2);
\end{feynhand}
\end{tikzpicture}
\right|^2
,
\label{diagram_PXX}
\end{align}
where the propagator of $\delta \chi$ is approximated by $-i/M^2$ since we are focusing on the leading-order EFT coefficients. We put $C\cdot \partial \pi$ to the external legs to represent that the diagram corresponds to the EFT coefficient of $(C\cdot \partial \pi)^2$. The $|{\rm diagram}|^2$ means that the diagram is factorized by the same diagrams connected by a propagator which is positive, provided the condition $M^2>0$. The positivity of $P_{XX}$ is guaranteed by the factorization property of the diagram in addition to the positivity of $M^2$. This is the same as the diagrammatic explanation of the optical theorem.

Let us investigate third and fourth derivatives of $P(X)$ with respect to $X$, namely $P_{XXX}$ and $P_{XXXX}$.
From the relation $P=fX-V$, by the use of the chain rule and the implicit function theorem, the third derivative of $P(X)$ is computed as
\begin{align}
P_{XXX} &=\frac{\dd}{\dd X}P_{XX}=\left[ \frac{\partial }{\partial X} \left( \frac{f_{\chi} ^2}{M^2} \right) + \frac{\dd \chi }{\dd X} \frac{\partial }{\partial \chi} \left( \frac{f_{\chi}^2}{M^2} \right) \right]_{\chi=\chi_0(X)}
\nn
&=\left[3 \, \frac{f_{\chi}^2f_{\chi\chi}}{M^4}-\frac{\mathcal{V}_{\chi\chi\chi} f_{\chi}^3}{M^6} \right]_{\chi=\chi_0(X)}
\,,
\label{direct_PX3}
\end{align}
and then
\begin{align}
P_{XXXX}&=
\left[ 4 \, \frac{f_{\chi}^3 f_{\chi\chi\chi}}{M^6} - \frac{f_{\chi}^4 \mathcal{V}_{\chi\chi\chi\chi}}{M^8} + \frac{3}{M^2} \left(  2\, \frac{f_{\chi} f_{\chi\chi}}{M^2} -\frac{f_{\chi}^2 \mathcal{V}_{\chi\chi\chi}}{M^4}\right)^2 \right]_{\chi=\chi_0(X)}
\,.
\label{direct_PX4}
\end{align}
The same results can be also computed by the use of Feynman-like diagrams as follows:
\begin{align}
-iP_{XXX}(X\neq 0)&=
\begin{tikzpicture}[baseline=0]
\begin{feynhand}
\vertex [particle] (i1) at (-1,1) {$C \!\cdot\! \partial \pi$};
\vertex [particle] (i2) at (-1,-1) {$C \!\cdot\! \partial \pi$};
\vertex [particle] (i3) at (1.3,0) {$C \!\cdot\! \partial \pi$};
\vertex [crossdot] (a1) at (-1/2,1/2) {};
\vertex [crossdot] (a2) at (-1/2,-1/2) {};
\vertex [dot] (b1) at (0,0)  {};
\propagator [plain] (i1) to (a1);
\propagator [plain] (i2) to (a2);
\propagator [plain] (i3) to (b1);
\propagator [scalar] (a1) to (b1);
\propagator [scalar] (a2) to (b1);
\node at (0.2,-0.3) {$f_{\chi\chi}$};
\end{feynhand}
\end{tikzpicture}
+
\begin{tikzpicture}[baseline=0]
\begin{feynhand}
\vertex [particle] (i1) at (-1,1) {$C \!\cdot\! \partial \pi$};
\vertex [particle] (i2) at (-1,-1) {$C \!\cdot\! \partial \pi$};
\vertex [particle] (i3) at (1.6,0) {$C \!\cdot\! \partial \pi$};
\vertex [crossdot] (a1) at (-1/2,1/2) {};
\vertex [crossdot] (a2) at (-1/2,-1/2) {};
\vertex [crossdot] (a3) at (1.3/2, 0) {};
\vertex [dot] (b1) at (0,0)  {};
\propagator [plain] (i1) to (a1);
\propagator [plain] (i2) to (a2);
\propagator [plain] (i3) to (a3);
\propagator [scalar] (a1) to (b1);
\propagator [scalar] (a2) to (b1);
\propagator [scalar] (a3) to (b1);
\node at (0.3,-0.3) {$\mathcal{V}_{\chi\chi\chi}$};
\end{feynhand}
\end{tikzpicture}
, \label{diagram_PX3}
\\
iP_{XXXX}(X\neq 0) &=
\begin{tikzpicture}[baseline=0]
\begin{feynhand}
\vertex [particle] (i1) at (-1,1) {$C \!\cdot\! \partial \pi$};
\vertex [particle] (i2) at (-1,-1) {$C \!\cdot\! \partial \pi$};
\vertex [particle] (i3) at (1,-1) {$C \!\cdot\! \partial \pi$};
\vertex [particle] (i4) at (1,1) {$C \!\cdot\! \partial \pi$};
\vertex [crossdot] (a1) at (-1/2,1/2) {};
\vertex [crossdot] (a2) at (-1/2,-1/2) {};
\vertex [crossdot] (a3) at (1/2, -1/2) {};
\vertex [dot] (b1) at (0,0)  {};
\propagator [plain] (i1) to (a1);
\propagator [plain] (i2) to (a2);
\propagator [plain] (i3) to (a3);
\propagator [plain] (i4) to (b1);
\propagator [scalar] (a1) to (b1);
\propagator [scalar] (a2) to (b1);
\propagator [scalar] (a3) to (b1);
\node at (0.6,0) {$f_{\chi\chi\chi}$};
\end{feynhand}
\end{tikzpicture}
+
\begin{tikzpicture}[baseline=0]
\begin{feynhand}
\vertex [particle] (i1) at (-1,1) {$C \!\cdot\! \partial \pi$};
\vertex [particle] (i2) at (-1,-1) {$C \!\cdot\! \partial \pi$};
\vertex [particle] (i3) at (1,-1) {$C \!\cdot\! \partial \pi$};
\vertex [particle] (i4) at (1,1) {$C \!\cdot\! \partial \pi$};
\vertex [crossdot] (a1) at (-1/2,1/2) {};
\vertex [crossdot] (a2) at (-1/2,-1/2) {};
\vertex [crossdot] (a3) at (1/2, -1/2) {};
\vertex [crossdot] (a4) at (1/2, 1/2) {};
\vertex [dot] (b1) at (0,0)  {};
\propagator [plain] (i1) to (a1);
\propagator [plain] (i2) to (a2);
\propagator [plain] (i3) to (a3);
\propagator [plain] (i4) to (a4);
\propagator [scalar] (a1) to (b1);
\propagator [scalar] (a2) to (b1);
\propagator [scalar] (a3) to (b1);
\propagator [scalar] (a4) to (b1);
\node at (0.7,0) {$\mathcal{V}_{\chi\chi\chi\chi}$};
\end{feynhand}
\end{tikzpicture}
+\left|
\begin{tikzpicture}[baseline=0]
\begin{feynhand}
\vertex [particle] (i1) at (-1,1) {$C \!\cdot\! \partial \pi$};
\vertex [particle] (i2) at (-1,-1) {$C \!\cdot\! \partial \pi$};
\vertex [particle] (i3) at (1.2,0) {$\delta \chi$};
\vertex [crossdot] (a1) at (-1/2,1/2) {};
\vertex [dot] (b1) at (0,0)  {};
\propagator [plain] (i1) to (a1);
\propagator [plain] (i2) to (b1);
\propagator [scalar] (i3) to (b1);
\propagator [scalar] (a1) to (b1);
\node at (0.2,-0.3) {$f_{\chi\chi}$};
\end{feynhand}
\end{tikzpicture}
+
\begin{tikzpicture}[baseline=0]
\begin{feynhand}
\vertex [particle] (i1) at (-1,1) {$C \!\cdot\! \partial \pi$};
\vertex [particle] (i2) at (-1,-1) {$C \!\cdot\! \partial \pi$};
\vertex [particle] (i3) at (1.2,0) {$\delta \chi$};
\vertex [crossdot] (a1) at (-1/2,1/2) {};
\vertex [crossdot] (a2) at (-1/2,-1/2) {};
\vertex [dot] (b1) at (0,0)  {};
\propagator [plain] (i1) to (a1);
\propagator [plain] (i2) to (a2);
\propagator [scalar] (a1) to (b1);
\propagator [scalar] (a2) to (b1);
\propagator [scalar] (b1) to (i3);
\node at (0.3,-0.3) {$\mathcal{V}_{\chi\chi\chi}$};
\end{feynhand}
\end{tikzpicture}
\right|^2
,
\label{diagram_PX4}
\end{align}
where each vertex has a $-i$ factor in addition to the specified factors. We have not shown the crossed versions of the diagrams, e.g.~there are two other geometrically equivalent diagrams corresponding to the first diagram of \eqref{diagram_PX3}.
Note that we can compute $P^{(n)}$ by using other diagrams. For example, $P_{XXX}$ is the coefficient of $-\frac{1}{2}(C\cdot \partial \pi)^2 (\partial \pi)^2$, and thus,
\begin{align}
-iP_{XXX}(X\neq 0)&=
\begin{tikzpicture}[baseline=0]
\begin{feynhand}
\vertex [particle] (i1) at (-1.1,1.1) {$C \!\cdot\! \partial \pi$};
\vertex [particle] (i2) at (-1.1,-1.1) {$ \partial \pi$};
\vertex [particle] (i3) at (1.1,-1.1) {$\partial \pi$};
\vertex [particle] (i4) at (1.1,1.1) {$C \!\cdot\! \partial \pi$};
\vertex [crossdot] (a1) at (-1/2,1/2) {};
\vertex [crossdot] (a4) at (1/2,1/2) {};
\vertex [dot] (b1) at (0,0)  {};
\propagator [plain] (i1) to (a1);
\propagator [plain] (i4) to (a4);
\propagator [scalar] (a1) to (b1);
\propagator [plain] (i2) to (b1);
\propagator [plain] (i3) to (b1);
\propagator [scalar] (a4) to (b1);
\node at (0.5,0) {$f_{\chi\chi}$};
\end{feynhand}
\end{tikzpicture}
+
\begin{tikzpicture}[baseline=0]
\begin{feynhand}
\vertex [particle] (i1) at (-1.2,1.1) {$C \!\cdot\! \partial \pi$};
\vertex [particle] (i2) at (-1.2,-1.1) {$ \partial \pi$};
\vertex [particle] (i3) at (1.2,-1.1) {$\partial \pi$};
\vertex [particle] (i4) at (1.2,1.1) {$C \!\cdot\! \partial \pi$};
\vertex [crossdot] (a1) at (-1.2*2/5, 0.3+0.8*2/5) {};
\vertex [dot] (b1) at (0,0.3)  {};
\vertex [dot] (b2) at (0,-0.3) {};
\propagator [plain] (i1) to (a1);
\propagator [plain] (i4) to (b1);
\propagator [plain] (i2) to (b2);
\propagator [plain] (i3) to (b2);
\propagator [scalar] (b1) to (b2);
\propagator [scalar] (a1) to (b1);
\node at (0.1,0.7) {$f_{\chi\chi}$};
\node at (0,-0.6) {$f_{\chi}$};
\end{feynhand}
\end{tikzpicture}
+
\begin{tikzpicture}[baseline=0]
\begin{feynhand}
\vertex [particle] (i1) at (-1.2,1.1) {$C \!\cdot\! \partial \pi$};
\vertex [particle] (i2) at (-1.2,-1.1) {$ \partial \pi$};
\vertex [particle] (i3) at (1.2,-1.1) {$\partial \pi$};
\vertex [particle] (i4) at (1.2,1.1) {$C \!\cdot\! \partial \pi$};
\vertex [crossdot] (a1) at (-1.2*2/5, 0.3+0.8*2/5) {};
\vertex [crossdot] (a4) at (1.2*2/5, 0.3+0.8*2/5) {};
\vertex [dot] (b1) at (0,0.3)  {};
\vertex [dot] (b2) at (0,-0.3) {};
\propagator [plain] (i1) to (a1);
\propagator [plain] (i4) to (a4);
\propagator [plain] (i2) to (b2);
\propagator [plain] (i3) to (b2);
\propagator [scalar] (b1) to (b2);
\propagator [scalar] (a1) to (b1);
\propagator [scalar] (a4) to (b1);
\node at (0.6,0.2) {$\mathcal{V}_{\chi\chi\chi}$};
\node at (0,-0.6) {$f_{\chi}$};
\end{feynhand}
\end{tikzpicture}
\,.
\end{align}
All the diagrammatic computations agree with the direct calculations \eqref{direct_PX3} and \eqref{direct_PX4} as they should be.

The essential point is the presence of the mixing term, $\otimes$. The coefficients $P_{XXX},P_{XXXX}$ are generated by not only factorized exchanging diagrams but also contact diagrams since $C\cdot \partial \pi$ can be transformed into $\delta \chi$ via the mixing. The value and sign of $P^{(n)}$ with $n\geq 3$ could be arbitrary unless there are additional constraints on both field space metric and potential. It would be interesting to compare the $(\partial \pi)^4$ diagram,
\begin{align}
iP_{XX}=
\left|
\begin{tikzpicture}[baseline=0]
\begin{feynhand}
\vertex [particle] (i1) at (-0.7,0.7) {$\partial \pi$};
\vertex [particle] (i2) at (-0.7,-0.7) {$\partial \pi$};
\vertex [particle] (i3) at (1,0) {$\delta \chi$};
\vertex [dot] (b1) at (0,0)  {};
\propagator [plain] (i1) to (b1);
\propagator [plain] (i2) to (b1);
\propagator [scalar] (b1) to (i3);
\node at (0.2,-0.3) {$f_{\chi}$};
\end{feynhand}
\end{tikzpicture}
\right|^2
\,.
\label{PXX_diagram2}
\end{align}
The derivative interaction $(\partial \pi)^4$ is generated only through the $\delta \chi$ exchange which is factorized and guaranteed to be positive provided the positive mass square of $\delta \chi$, $M^2>0$. 
In the Lorentz-invariant background, \eqref{PXX_diagram2} is the only diagram of the four-point function of $\pi$ because there is no mixing (and no potential here). On the other hand, when $\varphi$ has a non-vanishing gradient at the background, other diagrams contribute to the four-point function. The only factorized diagram is the two-point diagram \eqref{diagram_PXX}, which indeed yields the bound $P_{XX}>0$ provided $M^2>0$. The same is true for the general multi-field UV models: the difference is that each vertex has field space indices while the geometrical structure of the diagrams is the same. This consideration suggests that the generic UV consistency conditions around Lorentz-violating backgrounds may be obtained by two-point functions, not four-point functions. This should be regarded physically reasonable, since the linear perturbations can be scattered by the background field. We will further discuss this point in the next subsubsection.
Note that, as shown in Sec.~\ref{sec:extendEFT}, the EFT may be well defined even in the case of $M^2 < 0$ with a timelike Lorentz violation, and the domain $P_{XX} < 0$ does not necessarily conflict with the UV physics, as seen in \eqref{UVconsistency_nonlocal}. We have the relation $P_{XX}=\frac{f_{\chi}^2}{M^2}$, and the factorization property guarantees the positivity of the numerator only.

Our UV theories are just partial ones, in the sense that they still admit a large freedom for model parameters. When one considers the fundamental theory, say quantum gravity, there could be non-trivial constraints on the structure of the field space and the form of the potential. For instance, the $U(1)$ scalar $\Phi=\chi e^{i\varphi}$, which corresponds to a flat field space, with the Higgs-type potential $V=-\frac{M_{\Phi}^2}{2}|\Phi|^2+\frac{\lambda}{4} |\Phi|^4$ yields $P^{(n)}=0$ for $n>2$. In general, the multi-field theory should be allowed to have more complicated field space and potential, and the swampland conjectures aim to find general constraints on them. We may thus be able to obtain constraints on higher derivative coefficients, $P^{(n)}$, thanks to the consistency with UV, at least in principle. Conversely, bottom-up constraints on $P^{(n)}$ with $n>2$ (either theoretically or observationally) may be translated into constraints on the field space metric and the potential of the UV theory. See e.g.~\cite{Kim:2021pbr} for implications from the perturbative unitarity and \cite{Mizuno:2019pcm,Solomon:2020viz} for those from the de Sitter swampland conjecture.

\subsubsection{Toward bottom-up derivation}
\label{subsub:analytic_structure}
Let us make a similar argument to the derivation of the Lorentz-invariant positivity bounds to clarify similarities and differences in the Lorentz-violating case. As we explained, we consider the two-point function rather than the four-point function.\footnote{One might consider the four-point function by setting special configurations of momenta and/or taking appropriate subtractions so that the non-factorized diagrams do not contribute.} In general, the dispersion relation of the light mode around Lorentz-violating backgrounds may be represented by
\begin{align}
p^2-\Pi(p^{\mu})=0
\,,
\label{dispersion_gen}
\end{align}
in the adiabatic limit of the background where we have set $f=1$ by canonically normalizing the field $\pi$. We assume that $\Pi$ is a scalar function of $p^{\mu}$ and $C_{\mu}$ the latter of which determines the preferred direction. Therefore, the ``self-energy'' $\Pi$ must be a function of two variables defined by
\begin{align}
s\equiv -p_{\mu}p^{\mu}\,, \quad q \equiv p^{\mu}C_{\mu}
\,.
\end{align}
Then, $P_{XX}$ can be defined as 
\begin{align}
P_{XX}\equiv \left. \frac{1}{2} \frac{\partial^2}{\partial q^2}\Pi(s,q) \right|_{s=0,q=0}
\,,
\end{align} 
since $P_{XX}$ is the coefficient of $q^2$ in the dispersion relation \eqref{dispersion_gen}.

Let us see properties of $\Pi(s,q)$ in our partial UV models. In the case of the general multi-field UV completion, $\Pi$ is given by
\begin{align}
\Pi(s,q)=q^2 f_a (D^{-1})^{ab}f_{b}
\end{align}
where $(D^{-1})^{ab}$ is the inverse of 
\begin{align}
D_{ab}(s,q)=-s\gamma_{ab}+2i q h_{[ab]}+M^2_{ab}
\,.
\end{align}
We can perform the field redefinition so that
\begin{align}
\gamma_{ab}=\delta_{ab}={\rm diag}[1,1,\cdots]\,, \quad M^2_{ab}={\rm diag}[M_1^2,M_2^2,\cdots]
\,,
\label{diagonal_gamma_M2}
\end{align}
that is, the fields $\delta \chi^a$ are canonically normalized and diagonalized when $q=0$.
We would like to clarify the poles and the residues of $\Pi$ as a function of $s$ for a fixed $q$.\footnote{If $q$ is also regarded as a (complex) variable, the analytic structure of $\Pi(s,q)$ would be more complicated. See, e.g.~\eqref{Pi_analytic} below.}
With the diagonalized $\gamma$ and $M^2$ as in \eqref{diagonal_gamma_M2}, the matrix $D_{ab}$ is diagonalizable by using a unitary matrix $U$ since it is hermitian (as far as $q$ is real), to have
\begin{align}
\tilde{D}_{ab}(s,q)=(U^{\dagger}DU)_{ab}=-s\delta_{ab}+\tilde{M}_{ab}(q)\,, \quad \tilde{M}^2_{ab}(q)={\rm diag}[\tilde{M}_1^2(q),\tilde{M}_2^2(q),\cdots]
\end{align}
where the components $\tilde{M}^2_a~(a=1,2,\cdots)$ are generically functions of $q$ when $h_{[ab]}\neq 0$, namely in the absence of the reflection symmetry on $\varphi$. The values of $\tilde{M}^2_a$ are real for a real $q$ since they are the eigenvalues of the hermitian matrix. Note also that the unitary matrix $U(q)$ becomes the identity matrix when $q=0$ because $D_{ab}$ is already diagonal at $q=0$. Defining $\tilde{f}_a=(Uf)_a$, $\Pi(s,q)$ is expressed as
\begin{align}
\Pi(s,q)= \sum_{a}\frac{|q\tilde{f}_a(q)|^2}{\tilde{M}_a^2(q) - s }
\,.
\label{Pi_analytic}
\end{align}
This implies
\begin{align}
{\rm Im}\, \Pi(s,q) =  \sum_{a} \pi \delta(\tilde{M}_a^2(q) - s) |q\tilde{f}_a(q)|^2 \geq 0
\end{align}
when we add $+i\epsilon$ to the denominator as usual. These properties are indeed expected properties: the singularities of $\Pi$ should arise from the ``particle'' exchanges and the imaginary part of singularity should be positive. However, we recall again that ${\rm det}D_{ab}=0\Leftrightarrow s-\tilde{M}_a^2=0$ is not an on-shell state of the heavy mode because of the coupling $\delta \chi^a (C\cdot \partial \pi)$, that is, ${\rm det}D_{ab}=0$ is not a solution to the original dispersion relation \eqref{determinant}, in general. On the other hand, $s-M_a^2=0$ with $q=0$ is a solution to the original dispersion relation \eqref{determinant} of the heavy mode.

In practice, the necessary properties for our purpose are
\begin{align}
\Pi^{(2)}(s) &= \sum_{a}\frac{|f_a|^2}{M_a^2 - s }
\,, 
\label{Pi=poles} \\
 {\rm Im}\,  \Pi^{(2)} (s) & =\sum_{a} \pi \delta( M_a^2 - s) |f_a|^2 \geq 0
\,,
\label{ImPi}
\end{align}
where
\begin{align}
\Pi^{(2)}(s)\equiv  \left. \frac{1}{2}\frac{\partial^2}{\partial q^2} \Pi(s,q) \right|_{q=0}
\,.
\end{align}
Here, we have used $\tilde{f}_a|_{q=0} =f_a$ and $\tilde{M}_a^2|_{q=0} =M^2_a$. In this case, the poles of $\Pi^{(2)}(s)$ correspond to $s-M_a^2=0$ with $q=0$, and the imaginary part of $\Pi^{(2)}$ is still non-negative. We also see the asymptotic behaviour
\begin{align}
\lim_{|s|\to \infty} \Pi^{(2)}  \to 0
\label{Pi_asymptotic}
\end{align}
regarding $s$ as a complex variable. This asymptotic behaviour could be expected because the Lorentz-violating correction to the dispersion relation \eqref{dispersion_gen} may vanish in the high-energy limit $|s|\to \infty$ while keeping $q$ finite. As a  result, we find
\begin{align}
\Pi^{(2)}(s) &= \oint_C \frac{ds'}{2\pi i} \frac{ \Pi^{(2)}(s')}{s'-s}  \nn
&= \int_{-\infty}^{\infty} \frac{ds'}{2\pi i} \frac{\Pi^{(2)}(s'+i\epsilon )}{s'-s+i\epsilon} + \int_{\infty}^{-\infty} \frac{ds'}{2\pi i} \frac{\Pi^{(2)}(s'-i\epsilon )}{s'-s-i\epsilon}
+\int_{\mathcal{C}^{\pm}_{\infty}} \frac{ds'}{2\pi i} \frac{ \Pi^{(2)}(s')}{s'-s} 
\nn
&= \int_{-\infty}^{\infty} \frac{ds'}{\pi } \frac{{\rm Im}\, \Pi^{(2)}(s')}{s'-s} 
\end{align}
where we use Cauchy's integral formula in the first equality and then we deform the integration contour to the contour along the real axis and the infinitely large semi-circles $\mathcal{C}^{\pm}_{\infty}$ by using the fact that the poles $s=M_a^2$ exist on the real axis. We finally use Schwarz reflection principle $\Pi^{(2)}(s'-i\epsilon)={\Pi^{(2)}}^*(s'+i\epsilon)$ and \eqref{Pi_asymptotic} to get the last expression. This looks a deliberately complicated derivation of the relation \eqref{Pi=poles} from \eqref{ImPi} and \eqref{Pi_asymptotic}. However, this kind of deformation of the equation is the basis of the derivation of the positivity bounds in the Lorentz-invariant system.

In fact, the following three properties are sufficient to conclude $P_{XX}>0$: (i) the singularities of $\Pi^{(2)}(s)$ are on the positive real axis; (ii) the imaginary part is non-negative; and (iii) the asymptotic behaviour $\lim_{|s|\to \infty} \Pi^{(2)}  \to 0$. We then find
\begin{align}
\Pi^{(2)}(s)=\int^{\infty}_0 \frac{ds'}{\pi } \frac{{\rm Im}\, \Pi^{(2)}(s')}{s'-s} 
\implies
 P_{XX}=\Pi^{(2)}|_{s=0}= \int^{\infty}_0 \frac{ds'}{\pi } \frac{{\rm Im}\, \Pi^{(2)}(s')}{s'} > 0 
\end{align}
where ${\rm Im}\, \Pi^{(2)}(s)$ is assumed to take a non-zero value at some $s>0$. Note that the domain of the integration is $s>0$ since we have assumed that all the singularities exist on the positive real axis here. These three assumptions are similar to the assumptions in the Lorentz-invariant positivity bounds.

However, as we have seen, an eigenvalue of $M^2_a$ can be negative around the timelike background, which means that there can exist a pole on the negative real axis. Then, the positivity of $P_{XX}$ does not necessarily hold. The fact that such a ``tachyonic'' pole is physically allowed is one of the major differences from the standard argument of the positivity bounds.

In this subsection, we have been identifying the poles with the on-shell heavy states by taking the limit $q\to 0$. Part of our consistency conditions for the EFT reduction arise from the stability conditions of the heavy modes. Therefore, if the limit $q \to 0$ could be properly taken,
tachyonic poles could not be identified as physical states, since they would have to represent unstable heavy state(s). The 
limit would make sense around the spacelike background $C_{\mu}=(0,C_i)$, since $q=0$ only corresponds to a particular configuration of the momentum, $p^{\mu}=(\omega, k^i_{\perp})$. Indeed, we have found that the positivity of $P_{XX}$ has to be satisfied in the spacelike case.

On the other hand, $q \to 0$ would result in $p^{\mu}=(0,k^i)$ around the timelike background $C_{\mu}=(C_0,\bm{0})$, implying that the poles in the limit $q \to 0$ could not be identified with on-shell heavy states. Hence, there would be no reason to forbid a ``tachyonic'' pole in the timelike case. In fact, we have found that non-zero $C_0$ stabilizes the heavy modes as seen in \eqref{Omega_timelike} and that the ``tachyonic'' pole is allowed around the timelike background at least in our models.

The purpose of this subsection is not to derive the positivity $P_{XX}>0$ (and the negativity $-P_{XX} > P_X/2X>0$) from general properties such as unitarity. We have just illustrated similarities and differences from the standard argument of the positivity bounds around the Lorentz-invariant background based on our concrete UV setup. In addition, we should carefully think of particles either when gravity is turned on or the shift symmetry is not exact, since the background should depend on time and/or space, i.e.~the translation invariance does not hold.


\subsection{Non-minimal coupling}

Throughout the present work, we have assumed that scalar fields $\Phi^A$ minimally couple to gravity. However, one can introduce a non-minimal coupling, say $\chi^2 R$ which is not forbidden by any symmetries and should be added in the spirit of EFT. Nonetheless, we can perform a field redefinition to eliminate such a non-minimal coupling. Then, the effects of the non-minimal coupling are embedded in the way of the matter coupling,
\begin{align}
S_{\rm m}=S_{\rm m}[\psi, g^J_{\mu\nu} ]
\,,
\end{align}
where $g^J_{\mu\nu}$ is the Jordan frame metric which is not the same as $g_{\mu\nu}$, in general. When the Jordan frame metric depends on $\chi$, e.g.~$g^J_{\mu\nu}=\Omega^2(\chi)g_{\mu\nu}$, the matter fields act as a source of the equation of motion of $\chi$.
Under the condition that the backreaction from the matter fields onto the equation for the heavy field $\chi$ is negligible, then the solution of $\chi$ is given by the same solution $\chi=\chi(\varphi,X)$ as in the cases we have considered in the main body of this paper. The matter action is thus
\begin{align}
S_{\rm m}=S_{\rm m}[\psi, \Omega^2(\varphi,X)g_{\mu\nu}]
\,,
\end{align}
where $\Omega(\varphi,X)=\Omega(\chi)|_{\chi=\chi(\varphi,X)}$. This type of matter coupling yields a subclass of DHOST theories~\cite{Zumalacarregui:2013pma,Achour:2016rkg,Crisostomi:2016czh}. A class of the DHOST theory is obtained as an EFT of the two-field model with the non-minimal coupling under the assumption of small backreaction of matter fields to the heavy field. Our results hold in the DHOST theory in the Einstein frame as long as effects from the matter field are subdominant. The theory deviates from the DHOST theory when the backreaction of the matter to the heavy field is included. It would be intriguing to see how two theories, the EFT of two-field model and the DHOST theory, could differ due to the backreaction of matter.


\subsection{Comparison with previous results}
\label{sec:comparison}
Finally, we compare our top-down bounds with the bounds of bottom-up approaches discussed by~\cite{Adams:2006sv,Baumann:2015nta,Grall:2021xxm} and~\cite{Chandrasekaran:2018qmx}.

\begin{itemize}
\item {\it Lorentz-invariant positivity bounds}~\cite{Adams:2006sv}. As we have mentioned, the negative sign of $P_{XX}$ is allowed only around the timelike backgrounds with the IR instability. The UV consistency around the Lorentz-invariant background requires $P_{XX}>0$ which correctly reproduces the well-established bound~\cite{Adams:2006sv}. We also notice that, although the naively applied positivity bound can be violated, our consistency conditions are essentially the conditions on the sound speed (and the energy condition) even around the Lorentz-violating backgrounds (see Table~\ref{tab:summary} below), agreeing with the claim of~\cite{Adams:2006sv}: the existence of superluminal propagation obstructs a UV completion.

\item {\it Positivity bounds without boost}~\cite{Baumann:2015nta,Grall:2021xxm}. The papers~\cite{Baumann:2015nta,Grall:2021xxm} argued that 2-to-2 scattering around a Lorentz-violating background also provides a consistency condition on EFT. It is fair to say that our results are not inconsistent with their results, but we do not find any top-down guarantee for their bounds. As explicitly written in Appendix A of~\cite{Davis:2021oce}, their bounds involve $P_{XXX}$ and $P_{XXXX}$ while we do not find any general bound on the values and signs of $P^{(n)}$ with $n\geq 3$ since $P^{(n)}~(n\geq 3)$ is linearly related to $\mathcal{V}^{(n)}$, the coefficient of the $n$-point interaction of the UV theory. However, our UV theory is just a partial UV completion with a general field space and a general potential. A restriction of the UV theory leads to additional conditions on the resultant EFT. For instance, Appendix B of Reference~\cite{Baumann:2015nta} considered the UV theory having the interactions $f_{\chi}\neq 0, V_{\chi\chi\chi}\neq 0$ only. In such a case, there is a condition on 2-to-2 scattering since $P_{XXXX}$ is determined by the factorized diagram as shown in \eqref{diagram_PX4}. If there are hidden UV constraints on the field space and the potential from the fundamental theory, we may find non-trivial bounds involving $P^{(n)}$.

\item {\it Higher-point positivity bounds}~\cite{Chandrasekaran:2018qmx}. The paper~\cite{Chandrasekaran:2018qmx} discussed $n$-to-$n$ scattering around the Lorentz-invariant background to explore positivity bounds on the higher-point interaction, $P^{(n)}$. They concluded that the $n$th-order theory, $P(X)=X+\sum_{i=n}^{\infty} c_i X^i/\Lambda_2^{4(i-1)}=X+c_n X^n /\Lambda_2^{4(n-1)}+ c_{n+1}X^{n+1}/\Lambda_2^{4n}+\cdots$, requires $c_n>0$ around the Lorentz-invariant background. In our partial UV models, as explicitly computed in \eqref{direct_PX3} and \eqref{direct_PX4}, $P^{(n)}\neq 0~(n\geq 3)$ inevitably requires $f_{\chi}\neq 0$, concluding $P_{XX} \neq 0$.\footnote{Reference~\cite{Chandrasekaran:2018qmx} mentioned that loops generate the $X^2$ term even in the $n$th-order theory, which can be ignored in a sufficiently weakly coupled theory. Here, we claim that the $X^2$ term has to appear already at the tree-level if a tree-level process generates $P^{(n)}~(n\geq 3)$ at least when its (partial) UV completion is the nonlinear sigma model. The difference is that our (partial) UV theory only has the derivative coupling of the heavy fields to $(\partial \varphi)^2$ while their (partial) UV theory is supposed to have a coupling to the higher-order term, $(\partial \varphi)^n$, without the coupling to the lower-order terms, $(\partial \varphi)^j~(j<n)$.} Therefore, only the second-order theory, $P(X)=X+\sum_{i=2}^{\infty} c_i X^i/\Lambda_2^{4(i-1)}$ with $c_2>0$, is UV completable at least by the nonlinear sigma model around the Lorentz-invariant background. In addition, Reference~\cite{Chandrasekaran:2018qmx} argued the subtleties of the bounds on higher-point interactions in the general k-essence theory, e.g.~the bound on $c_3$ in the second-order theory $P(X)=X+c_2 X^2/\Lambda_2^{4}+c_3 X^3/\Lambda_2^8$. Although our result provides a bound on higher-order terms $c_i~(i\geq 3)$ around the Lorentz-violating backgrounds (for instance, the UV consistency around the spacelike background with $|X| \gg |c_2/c_3| \Lambda_2^4$ requires $c_3<0$), the consistency condition is background dependent. Around the Lorentz-invariant background, we have found no consistency conditions on the higher-order term, $c_i~(i \geq 3)$.
\end{itemize}

\section{Conclusion}
\label{sec:conclusion}

\begin{table}[t]
\begin{tabular}{| c || c | c | c | c | c | c | c | }
\hline \\ [-2.4ex]
& $~ M^2 ~$ & mass square & $~ P_X ~$ & $~ P_{XX} ~$ & ~ sound speed square ~ & ~ UV consistency ~ & ~ EFT predictivity ~ \\ [0.5ex]
\hline\hline \\ [-2.4ex]
\multirow{3}{*}{\begin{tabular}{c} ~ Spacelike:  \\ $X<0$ \end{tabular} } & $+$ & $M^2>0$ & $+$ & $+$ & $ -\mathcal{O}(1)<c_{\parallel}^2 < 1 $ & \checkmark & \checkmark \\ [0.5ex]
    & $+$ & $M^2>0$ & $+$ & $+$ & $c_{\parallel}^2 < -\mathcal{O}(1)$ & \checkmark & $\times$ \\ [0.5ex]
    & $-$ & $M^2<0$ & $+$ & $-$ & $c_{\parallel}^2 > 1$ & $\times$ & \\ [0.5ex] 
\hline \\ [-2.4ex]
\multirow{4}{*}{ \begin{tabular}{c} ~ Timelike:  \\$X>0$ \end{tabular} } & $+$ & $~ M^2+\delta M^2 > 0 ~$ & $+$ & $+$ & $0< c_s^2 <1$ & \checkmark & \checkmark \\ [0.5ex]
    & $-$ & $~ M^2+\delta M^2>0 ~$ & $+$ & $-$ & $ -\mathcal{O}(1)< c_s^2<0$ & \checkmark & \checkmark  \\ [0.5ex]
    & $-$ & $~ M^2+\delta M^2>0 ~$ & $+$ & $-$ & $ c_s^2<-\mathcal{O}(1)$ & \checkmark & $\times$ \\ [0.5ex]
    & $-$ & $~ M^2+\delta M^2<0 ~$ & $+$ & $-$ & $c_s^2>1$ & $\times$ &  \\ [0.5ex]
\hline
\end{tabular}
\caption{Classification of the EFT reduction from the multi-field models with curved field space studied in this work where, in the last two columns, $\checkmark$ and $\times$ mean that the conditions are satisfied and violated, respectively. The parameter $c_{\parallel}$ is the sound speed for the modes of which momenta are parallel to the background, $k_i \propto \partial_i \bar{\varphi}$, around spacelike backgrounds $X<0$,
while $c_s^2$ corresponds to the sound speed of the perturbation around timelike backgrounds $X>0$. When $c_{\parallel}^2<0$ or $c_s^2<0$, a local higher-derivative operator (for $X<0$) or a non-local one (for $X>0$) is needed to cure the IR instability. The absolute values of the imaginary sound speed are bounded by order- or under-unity numbers, e.g.~$-0.2 \lesssim c_s^2 < 1$ for the two-field model. }
\label{tab:summary}
\end{table}
In the present paper, we have clarified consistency conditions, the UV consistency conditions and the EFT predictivity conditions, of the single-field effective field theory (EFT) with spontaneously broken Lorentz symmetry in the Einstein frame by studying explicit EFT reductions from multi-scalar theories. All the classifications of the EFT reduction we consider are summarized in Table~\ref{tab:summary}, and the conclusions of the present paper are listed as follows:
\begin{enumerate}
\item The EFT preserves the null energy condition (NEC);
\item The sound speed of perturbations can be either subluminal or imaginary;
\item The case with imaginary sound speed can accommodate a violation of the naively applied positivity bounds, $P_{XX}>0$;
\item Both ghost and gradient instabilities at IR can be as harmless as Jeans instability;
\item The effective coupling of the fifth force at a finite distance should be larger than that at infinity; and
\item These properties are inferred from the two-point function, not four-point functions, in Lorentz-violating EFTs.
\end{enumerate}
A few comments are in order. First of all, consistency conditions arise from stability conditions of the UV modes which are integrated out. In particular, the UV consistency conditions are the conditions to avoid the ghost and the tachyon in UV while the EFT predictivity conditions are imposed to resolve the IR instability, if exists, within the regime of validity of the EFT. The NEC preservation in IR is a direct consequence of the lack of the source of NEC violation (ghost-free condition at UV) in the particular class of our partial UV completion, and the subluminal propagation is anticipated by the standard positivity bounds~\cite{Adams:2006sv}. We emphasize, however, that stability of IR degrees of freedom is not mandatory in considering either the UV consistency or the EFT predictivity, since our EFTs do not necessarily describe the true vacuum. This is one of the differences from the existing arguments about the Lorentz-invariant positivity bounds, and, indeed, the order-unity violation of the naively applied positivity bound is found when a system exhibits the IR instability. Some EFTs are defined only around a Lorentz-violating background, even if the UV theory has a stable Lorentz-invariant background, in which there is no need to respect the Lorentz-invariant positivity bounds. In Appendix~\ref{app:U(1)}, we provide concrete examples which explain how EFT properties are related to properties of the UV theory.  Physically, the IR instability present in the EFT is analogous to the Jeans instability and is automatically cured at a relatively low scale either by higher-derivative terms that are also present in the EFT or by the multi-field dynamics of the assumed (partial) UV completion. It is interesting that the naively applied positivity bound $P_{XX}>0$ does not necessarily hold around a timelike background, whereas it does hold around a spacelike background. The consistency condition $P_{XX}>0$ for the spacelike ($X<0$) case concludes that there should be no screening mechanism by the nonlinear kinetic terms, if we assume the particular partial UV completion that we studied in the present paper. Finally, we emphasize that these conclusions are obtained by studying the quadratic action, rather than four-point functions, which are typically used to obtain the positivity bounds on Lorentz-invariant EFTs.


\section*{Acknowledgments}
K.A. would like to thank Toshifumi Noumi for insightful discussions, and R.N. is grateful to Masahito Yamazaki for inspiring comments on related topics during the period of ``Workshop on Gravity and Cosmology by Young Researchers 2021'' (YITP-W-20-11) at the Yukawa Institute for Theoretical Physics, Kyoto University.
The work of K.A. was supported in part by Grants-in-Aid from the Scientific Research Fund of the Japan Society for the Promotion of Science, No.~19J00895 and No.~20K14468.
The work of S.M was supported in part by Japan Society for the Promotion of Science Grants-in-Aid for Scientific Research No.~17H02890, No.~17H06359, and by World Premier International Research Center Initiative, MEXT, Japan.


\appendix

\section{General multi-field UV models with DBI-type kinetic terms}
\label{subsec:DBI}

Ref.~\cite{Mukohyama:2016ipl} has shown that DBI-type kinetic terms lead to another class of models without caustic singularities. In this sense, this class can also be regarded as a candidate of partially UV-complete models. While ref.~\cite{Mukohyama:2016ipl} focuses on the case of a single light field, we can extend it to a general multi-field case with a curved field space.
The Lagrangian of this class thus reads
\begin{equation}
    {\cal L}_{\rm DBI} = - \sqrt{1 + \gamma_{AB}(\Phi) \, \dd\Phi^A \dd\Phi^B} - V(\Phi) \; ,
    \label{Lag_DBI}
\end{equation}
where the $N$ fields $\Phi^A$ contains one light field degree $\varphi$ that respects a (approximate) shift symmetry. Note that the caustic free condition does not exclude the DBI model with the opposite sign, $\mathcal{L}=+ \sqrt{1 - \gamma_{AB}(\Phi) \, \dd\Phi^A \dd\Phi^B} - V(\Phi)$; nonetheless, this model admits the Minkowski vacuum and then the standard positivity bounds exclude this model. In addition, although the cuscuton theory~\cite{Afshordi:2006ad,Afshordi:2007yx,Afshordi:2009tt} is free from caustic singularities, the cuscuton is non-local in the sense that there exists an instantaneous/shadowy mode~\cite{DeFelice:2018ewo} and will not be discussed here in a sprite of the positivity bounds. We only consider \eqref{Lag_DBI} as a partial UV completion of the single-field EFT.

In this appendix, we only focus on the EFT at the leading order in $\tilde{\epsilon}$,
\begin{align}
    {\cal L}_{\rm DBI}&=Q\left( \chi , \varphi, X \right) +\mathcal{O}(\tilde{\epsilon}) \; , 
    \label{LUV_DBI_approx}
    \\
    Q\left( \chi , \varphi, X \right) &\equiv  - \sqrt{1 + f(\chi,\varphi) \left( \partial\varphi \right)^2} - V(\chi,\varphi) 
\end{align}
where $Q$ is defined as a function of $\chi^a,\varphi$ and $X = - \left( \partial\varphi \right)^2 / 2$ for a compact notation.
The kinetic terms for $\chi^a$ are absent in \eqref{LUV_DBI_approx} at the leading-order expansion.
The ``constraint'' equations, which are the same as the equations of motion for $\chi^a$, are obtained by variations of \eqref{LUV_DBI_approx} with respect to $\chi^a$, i.e.
\begin{equation}
    \partial_a Q = 0 \; .
    \label{constraint_DBI}
\end{equation}
Then the resultant single-field EFT must obey the following relations:
\begin{align}
{\cal L}_{\rm IR} = P(\varphi, X)+\mathcal{O}(\tilde{\epsilon})&=Q \, \big|_{\chi^a=\chi_0^a(\varphi, X)} +\mathcal{O}(\tilde{\epsilon}) \; ,
\end{align}
with
\begin{align}
P_X&=\left. Q_X+ \partial_a Q \, \frac{\partial \chi^a}{\partial X} \right\vert_{\chi^a=\chi^a_0(\varphi, X)}= Q_X \, \big\vert_{\chi^a=\chi^a_0(\varphi, X)}
\,, \\
P_{XX}&=\left. Q_{XX}+ \partial_a Q_X \, \frac{\partial \chi^a}{\partial X} \right\vert_{\chi^a=\chi^a_0(\varphi, X)}
= \left. Q_{XX} + \partial_a Q_X \left( M^{-2} \right)^{ab} \partial_b Q_X \, \right\vert_{\chi^a = \chi_0^a(\varphi, X)} \; ,
 \label{PXX_DBI}
\end{align}
where $\chi_0^a(X)$ are the solutions to \eqref{constraint_DBI} and $M^2_{ab} \equiv - \partial_a \partial_b Q = \partial_a \partial_b V + \cdots$. Note that the DBI-type kinetic term yields
\begin{align}
Q_X&=\frac{f}{\sqrt{1+ \gamma_{AB}\, \dd\Phi^A \dd\Phi^B} } >0
\,, \\
Q_{XX}&=\frac{f^2}{ ( 1+ \gamma_{AB}\, \dd\Phi^A \dd\Phi^B)^{3/2} } >0
\end{align}
where the components of the field space metric is written as \eqref{multi_fieldmetric}. In particular, the positive definiteness of the field space metric implies $P_X>0$.

We then investigate the stability conditions of the UV modes to discuss the sign of $P_{XX}$. Following the main text, we assume the exact shift symmetry and consider the perturbations around the constant background,
\begin{align}
 \bar{\chi}^a={\rm constant}\,, \quad C_{\mu}\equiv \partial_{\mu}\bar{\varphi}={\rm constant}\,.
\end{align}
We perform a field redefinition to eliminate the kinetic mixing between $\varphi$ and $\chi^a$, as we did in Sec.~\ref{sec:multi_pert}.
Then, the quadratic Lagrangian of the DBI-type partial UV model is
\begin{align}
\mathcal{L}_{\rm DBI}^{(2)} &= -\frac{1}{2}\Gamma_{ab} (\partial \delta \chi^a \cdot \partial \delta \chi^b)-\frac{1}{2}M^2_{ab} \delta \chi^a \delta \chi^b - H_{[ab]}  (C \cdot \partial \delta \chi^a) \delta \chi^b  
-\frac{1}{2} F (\partial \pi)^2 - F_a \delta \chi^a (C\cdot \partial \pi)
+\frac{1}{2}Q_{XX}(C\cdot \partial \pi)^2
\,,
\end{align}
where the coefficients are evaluated at the background and
\begin{align}
\Gamma_{ab} &\equiv \frac{\gamma_{ab}}{\sqrt{1-2f X}}
\,, \quad
F \equiv \frac{f}{\sqrt{1-2fX }}
\,, \quad
F_a \equiv \partial_a F\,, 
\quad 
H_{[ab]}\equiv \frac{h_{[ab]} }{\sqrt{1-2fX }}
\,.
\end{align}
Note that $F_a=\partial_a Q_X$ holds when evaluated at the background.
The no-tachyon condition of the heavy mode is immediately found as $M^2_{ab}>0$ around the spacelike background. Since the DBI theory satisfies $Q_{XX}>0$, the positive definiteness of $M^2_{ab}$ concludes $P_{XX}>0$.

We study the Hamiltonian to obtain the stability conditions around the timelike background $C_\mu = (C_0 , \bm{0})$. The conjugate momenta are
\begin{align}
p_a&=\Gamma_{ab}\delta \dot{\chi}^b+H_{[ab]}C_0\delta \chi^b
\,,\\
p_{\pi}&=(F+2X Q_{XX})\dot{\pi}+F_a C_0 \delta \chi^a
\,.
\end{align}
and the Hamiltonian is
\begin{align}
\mathcal{H}^{(2)}_{\rm DBI}&=\frac{1}{2}\Gamma_{ab} \delta \dot{\chi}^a{}^{\dagger}  \delta \dot{\chi}^b+\frac{1}{2}(F+2X Q_{XX}) |\dot{\pi}|^2+ \frac{1}{2}(k^2 \gamma_{ab}+M^2_{ab})\delta \chi^a{}^{\dagger} \delta \chi^b+ \frac{1}{2}k^2 F |\pi|^2
\,,
\end{align}
in the momentum space where $\delta \dot{\chi}^a$ and $\dot{\pi}$ are understood as the functions of the conjugate variables. We can repeat the same analysis as in Sec.~\ref{sec:multi_pert}. We impose the boundedness of the 
Hamiltonian under $k=0$ and $p_{\pi}=0$, leading to
\begin{align}
M^2_{ab}+\frac{2X F_a F_b}{F+2X Q_{XX}}>0
\; ,
\label{no-tachyon_DBI0}
\end{align}
at least under the situation we are considering. 
We then use the freedom of field redefinitions to diagonalize $M^2_{ab}={\rm diag}[M^2_1,M^2_2,\cdots ]$ and set $F_a=(F_1,0,0,\cdots )$. In this coordinate choice, the conditions \eqref{no-tachyon_DBI0} are
\begin{align}
M_1^2+ \frac{2X F_1^2}{F+2XQ_{XX}}>0 \,, \quad M^2_{a'}>0 \; ,
\label{no-tachyon_DBI}
\end{align}
where the primed index $a'$ denotes the fields other than $\delta \chi^1$ in this field basis.
Hence the relation to the k-essence function is
\begin{align}
P_{XX}=Q_{XX} + \frac{F_1^2}{M_1^2}
\,.
\end{align}
The case $M_1^2>0$, i.e.~$M^2_{ab}>0$, concludes $P_{XX}>0$. On the other hand, the case $M_1^2<0$ does not immediately suggest the negative sign of $P_{XX}$. Nonetheless, the former one of the condition \eqref{no-tachyon_DBI} is given by
\begin{align}
F+2X\left( Q_{XX}+\frac{F_1^2}{M_1^2} \right)<0
\implies P_X+2XP_{XX}<0
\,,
\end{align}
where $M_1^2<0$ and $F+2XQ_{XX}>0$ are used, the latter is required by the no ghost for $\pi$. Since we have $P_X>0$ and $X>0$, $P_{XX}$ has to be negative if there is a negative eigenvalue of $M^2_{ab}$. As a result, the UV consistency conditions for the DBI-type partial UV completion are the same as those in the general multi-field models shown in Sec.~\ref{sec:multi_pert}.

\section{EFT from $U(1)$ scalar field}
\label{app:U(1)}

As a concrete example, we consider a $U(1)$ scalar field described by the Lagrangian
\begin{align}
\mathcal{L}_{U(1)}=-\frac{1}{2}\partial_{\mu}\Phi^\dagger  \partial^{\mu} \Phi - \sigma \frac{M_{\Phi}^2}{2}|\Phi|^2 - \frac{\lambda}{4}|\Phi|^4
\,, \label{U(1)_Lag}
\end{align}
with $\sigma = \pm 1$. We introduce the real variables $\chi$ and $\varphi$ via
\begin{align}
\Phi = \cos \theta \chi^{1+i \tan \theta} e^{i \varphi/ \cos \theta}
\,,
\end{align}
with $\chi>0$ where $\theta $ is a constant parameter which determines the kinetic mixing between $\chi$ and $\varphi$. In this appendix, the variable $\varphi$ is dimensionless since it is essentially the angular coordinate of the field space. The $U(1)$ symmetry of $\Phi$ recasts the shift symmetry of $\varphi$. The Lagrangian is written as
\begin{align}
\mathcal{L}_{U(1)}=-\frac{1}{2}\tilde{\epsilon}^2 (\partial \chi)^2 +\tilde{\epsilon} \sin \theta \chi \nabla_{\mu} \chi \nabla^{\mu} \varphi - \frac{1}{2} \chi^2 (\partial \varphi)^2  - \sigma \frac{M_{\Phi}^2}{2}\chi^2 - \frac{\lambda}{4}\chi^4 \,,
\end{align}
reading
\begin{align}
f=\chi^2\,, \quad V=\sigma \frac{M_{\Phi}^2}{2}\chi^2 + \frac{\lambda}{4}\chi^4 \,,
\end{align}
where $\tilde{\epsilon}$ is introduced which will be set to be $\tilde{\epsilon}=1$ at the end of the calculation (see Sec.~\ref{sec:epsilon}). In this example, the constant background, $\partial_{\mu}\bar{\varphi},\bar{\chi}={\rm constant}~(\bar{\chi}\neq 0)$, is determined by
\begin{align}
-(\partial \bar{\varphi})^2 = \sigma M^2_{\Phi}+\lambda \bar{\chi}^2
\,.
\label{background_U(1)}
\end{align}
In particular, as for the timelike background, $\bar{\chi}=$ constant and $\bar{\varphi}\propto t$, the background motion is a uniform circular rotation in the field space which is realized by balancing the potential force $V_{\chi}$ and the centrifugal force $-f_{\chi} X$.

We consider the solution up to $\mathcal{O}(\tilde{\epsilon}^2)$. Writing $\chi=\chi_0+\tilde{\epsilon} \chi_1 + \tilde{\epsilon}^2 \chi_2 + \mathcal{O}(\tilde{\epsilon}^3)$, we find
\begin{align}
\chi_0&= \sqrt{ \frac{2X-\sigma M_{\Phi}^2}{\lambda} }
\,, \label{chi0_ex} \\
\chi_1&=-\frac{\sin \theta }{2\sqrt{\lambda( 2X-\sigma M_{\Phi}^2) } } \Box \varphi
\,, \\
\chi_2&=\frac{1}{2\sqrt{ \lambda (2X -\sigma M_{\Phi}^2)^3}}\left( \Box X - \frac{1}{4}\sin^2 \theta (\Box \varphi)^2 - \frac{\nabla_{\mu} X \nabla^{\mu}X }{2X-\sigma M_{\Phi}^2} \right)
\,, \label{chi2_ex}
\end{align}
where $(2X-\sigma M_{\Phi}^2)/\lambda>0$ has to be satisfied so that $\chi$ is real. In this example, the positivity of $M^2$ is related to the boundedness of the potential $\lambda>0$ because
\begin{align}
M^2 \equiv V_{\chi\chi}-f_{\chi\chi} X=-2\sigma M_{\chi}^2+ 4X  =2 \lambda \chi_0^2(X)\,,
\end{align}
while $\Meff^2$, which is the effective mass around the timelike background, is given by
\begin{align}
\Meff^2\equiv M^2+ 2X \frac{f_{\chi}^2}{f}=-2\sigma M_{\chi}^2+12 X
\,.
\end{align}
Following the analysis in Sec.~\ref{sec_twofield}, the EFT Lagrangian is
\begin{align}
\mathcal{L}_{\rm IR}=\frac{1}{\lambda} \left[ (X-\sigma M_{\Phi}^2/2)^2 - \tilde{\epsilon} \sin \theta X \Box \varphi + \tilde{\epsilon}^2 \left( -\frac{(\nabla X)^2}{M^2} +\frac{1}{4}\sin^2 \theta (\Box \varphi)^2 \right) + \mathcal{O}(\tilde{\epsilon}^3) \right]
\,.
\end{align}
The k-essence part is
\begin{align}
P(X)=\frac{1}{\lambda} (X-\sigma M_{\Phi}^2/2)^2
\,,
\label{kessence_U(1)}
\end{align}
which yields
\begin{align}
P_X=\frac{2X-\sigma M_{\Phi}^2}{\lambda} \,, \quad
P_{XX}=\frac{2}{\lambda}
\,.
\end{align}
One can explicitly confirm the general relations $P_{X}=f$ and $P_{XX}=f_{\chi}^2/M^2$.

We first discuss the positive $\lambda$ where the potential is bounded from below.
The sign of $\sigma$ determines the sign of the linear kinetic term $P(X)=-\sigma M_{\Phi}^2 X/\lambda + \cdots$. From the UV point of view, $\sigma$ determines whether $\chi=0$ is the true vacuum or the false one.
Let us set $\theta=0$ to have an intuition. Then, the variables $\chi$ and $\varphi$ are nothing but the radial and the angular coordinates in the field space. The motion $\chi=$ constant and $\varphi \propto t$ is a uniform circular motion. When $\sigma = -1$, the fields settle down the true vacuum $\chi=M_{\Phi}/\sqrt{\lambda}$ when $\varphi$ stops where there exists the massless direction $\varphi$. Hence, the EFT is still valid at $X=0$ which can be seen as the positive sign of the linear kinetic term of $P(X)$. On the other hand, the true vacuum is at $\chi=0$ for $\sigma=+1$ where the EFT reduction is no longer consistent since there is no massless mode. From the EFT perspective, the invalidity of EFT reduction can be seen as the opposite sign of the linear kinetic term $P(X)=-M_{\Phi}^2 X /\lambda +\cdots$. However, the massless direction appears when the field $\chi$ goes away from the origin $\chi=0$ thanks to the non-zero angular momentum $\varphi \propto t$. As a result, the EFT reduction becomes consistent around such a Lorentz-violating background. This consistency condition is captured by our conditions $P_X>0$ and $P_{XX}>0$, which conclude $X>M_{\Phi}^2/2$ for $\sigma=+1$. This EFT is valid if and only if $\varphi$ has a non-vanishing timelike gradient.

When we have a kinetic mixing at UV $(\Leftrightarrow \theta \neq 0)$, the EFT has the operator $(\Box \varphi)^2$. It seems that the operator $(\Box \varphi)^2$ pushes up the validity of the EFT under the limit $P_X \to 0$ like the ghost condensate. However, this is not the case. The positive definiteness of the kinetic matrix at UV requires that the coefficient of $(\Box \varphi)^2$ is not arbitrary large, i.e.~$|\sin \theta| < 1$. The operator $(\nabla X)^2$ eventually dominates over $(\Box \varphi)^2$ in the limit $P_X \propto M^2 \to 0$ which spoils the scenario of the ghost condensate. In fact, changing $\theta$ is just a change of the variable at UV which does not change physics.

We then consider the non-relativistic limit of the $U(1)$ scalar which corresponds to the limit $M^2 \to 0$. As discussed in \ref{sec:extendEFT}, the validity of EFT can be extended around the timelike background. Then $M^2$ is no longer the cutoff of the single-field EFT. Note that the EFT breaks down when we take the limit $\chi_0 \to 0$. Therefore, we should take the limit $\lambda \to 0$ while keeping $\chi_0$ finite to consistently take the limit $M^2 \to 0$. In terms of the k-essence $P(X)$, this limit yields $P_X \to$ finite and $P_{XX}\to \infty$, implying that $\varphi$ is not a good variable.
We instead use the following parametrization
\begin{align}
\Phi=\psi e^{-iM_{\Phi}t}
\,,
\label{phi_ansatz}
\end{align}
where $\psi$ is a complex variable. We suppose that the variable $\psi$ slowly varies with respect to $M_{\Phi}$ under $\sigma=+1$ and $\lambda \to 0$. The choice \eqref{U(1)_Lag} with $\sigma=+1$ and $\lambda \to 0$ is a free massive $U(1)$ scalar, meaning that the general solution is given by a superposition of the negative frequency mode solutions $\Phi \propto e^{i(\omega t - k_i x^i)}$ as well as the positive frequency mode solutions $\Phi \propto e^{-i(\omega t - k_i x^i)}$ with $\omega = \sqrt{M_{\Phi}^2+k^2}$. The assumption of the slow variation of $\psi$ is thus to ignore, namely to integrate out, the negative frequency modes as well as the positive frequency modes with $k\gtrsim M_{\Phi}$. We only consider the dynamics of the positive frequency modes with $k \ll M_{\Phi}$. The Lagrangian of the free massive $U(1)$ scalar in terms of $\psi$ is
\begin{align}
\mathcal{L}_{\lambda \to 0}=\frac{1}{2}|\dot{\psi}|^2 + \frac{1}{2}iM_{\Phi}(\psi^\dagger  \dot{\psi}-\dot{\psi}^\dagger  \psi) -\frac{1}{2}\partial_i \psi^\dagger  \partial^i \psi
\end{align}
where the dot is the time derivative and $\partial_i$ is the spatial derivative, respectively. Since we have assumed the slowly varying $\psi$, the first term can be ignored compared with the second term. We then obtain the Lagrangian for the free Schr\"{o}dinger field,
\begin{align}
\mathcal{L}_{\rm Sch}= \frac{1}{2}iM_{\Phi}(\psi^\dagger  \dot{\psi}-\dot{\psi}^\dagger  \psi) -\frac{1}{2}\partial_i \psi^\dagger  \partial^i \psi
\,,
\end{align}
of which equation of motion is the well-known Schr\"{o}dinger equation,
\begin{align}
i\dot{\psi}=-\frac{1}{2M_{\Phi}}\partial^i \partial_i \psi
\,.
\end{align}
This confirms that $\lambda \to 0 ~(\Rightarrow M^2 \to 0)$ consistently provides the effective theory with the non-relativistic dispersion relation. The physical meaning of the effective mass,
\begin{align}
\Meff^2 = (2M_{\Phi})^2 \quad {\rm as}\quad \lambda \to 0\,,
\end{align}
is the energy gap between the positive and negative frequency modes. The relation to the polar coordinates $\Phi=\chi e^{i\varphi}$ with $\chi=\bar{\chi}+\delta \chi$ and $\varphi=-M_{\Phi} t + \pi$ is
\begin{align}
{\rm Re}\, \psi = \bar{\chi}+\delta \chi + \cdots \,, \quad
{\rm Im}\, \psi = \bar{\chi} \pi + \cdots
\,,
\end{align}
up to linear in the perturbations. One can consider the quadratic action in terms of $\delta \chi$ and $\pi$ instead of the complex variable $\psi$. Integrating out the imaginary part $\delta \chi$, one then finds a non-local action $S_{\Lambda}[\pi]$ which agrees with \eqref{L_Meff2} in the limit $M^2 \to 0$.

We discuss the unstable theory $\lambda<0$. Since one can add a higher-order polynomial of $|\Phi|$ to bound the potential at a sufficiently large $\chi$, this theory can be still regarded as a partially UV complete theory. The inequalities $M^2<0$ and $\Meff^2>0$ are satisfied when
\begin{align}
2X<M_{\Phi}^2<6X
\label{ineq_U(1)}
\end{align}
with $\sigma=+1$. The simplest example of $\lambda<0$ is found by taking $\lambda \to -0$ and $2X\to M^2_{\Phi}$. We again use the parametrization \eqref{phi_ansatz} and include the leading order effect of $\lambda$. Now, $\psi$ contains not only inhomogeneous oscillations but also a homogeneous oscillation since the background equation \eqref{background_U(1)} implies $\Phi \propto e^{iC_0 t}$ with $C_0=-M_{\Phi}+\mathcal{O}(\lambda)$. Nonetheless, we have $|\dot{\psi}|^2=\mathcal{O}(\lambda^2)$ from the homogeneous oscillation which can be ignored at the leading order in $\lambda$. As a result, the effective Lagrangian is given by the Gross–Pitaevskii form (see e.g.~\cite{Dalfovo:1999zz}),
\begin{align}
\mathcal{L}_{\rm GP}= \frac{1}{2}iM_{\Phi}(\psi^\dagger  \dot{\psi}-\dot{\psi}^\dagger  \psi) -\frac{1}{2}\partial_i \psi^\dagger  \partial^i \psi -\frac{\lambda}{4}|\psi|^4
\,,
\end{align}
for a small $\lambda$.
The negative sign $\lambda<0$ corresponds to the attractive interaction, leading to the simple physical interpretation of the IR instability discussed in Sec.~\ref{sec:extendEFT}: the IR instability is the consequence of the attractive force, similarly to the Jeans instability.

Finally, we argue how an inconsistency arises if an unstable mode is integrated out. Let us consider the case, $\lambda<0,~\sigma=+1$ and $M_{\Phi}^2>6X$, where we have $M^2<0$ and $\Meff^2<0$. The background is unstable in both timelike and spacelike cases. However, the $P(X)$ theory satisfies
\begin{align}
P_X>0 \,, \quad P_X+2XP_{XX}>0
\,,
\end{align}
with $P_{XX}<0$, meaning that the EFT does not see any instability. Therefore, if one uses the EFT without knowing its UV completion \eqref{U(1)_Lag}, one concludes that the system is stable for a long time, which is inconsistent with the prediction made by \eqref{U(1)_Lag}: the system develops the instability on the short timescale of either $M^2$ (the spacelike case) or $\Meff^2$ (the timelike case).

\section{Comparison with ghost condensate}
\label{sec:ghostcondensate}

In order to examine the claim after \eqref{quadratic_GC}, that is the non-equivalence between the $\Lambda$-EFT and the ghost condensate in the presence of gravity, we now introduce the gravity sector into the two-field model we have been discussing. In fact, this is rather a natural consideration, since Lorentz-violating backgrounds are most relevant for gravitating systems.
We consider coupling the two-field model \eqref{UV_two-field} to gravity in the Einstein frame as a minimal setup. The (partial) UV Lagrangian is
\begin{align}
\mathcal{L}_{\rm UV}=\frac{\Mpl^2}{2} \, R-\frac{1}{2} \left( \nabla \chi \right)^2-\frac{1}{2} \, f(\chi) \left( \nabla \varphi \right)^2-V(\chi)
\; ,
\label{UV_minimal}
\end{align}
where $\Mpl$ is the reduced Planck mass, and $R$ and $\nabla_\mu$ the Ricci scalar and the covariant derivative, respectively, associated with the spacetime metric $g_{\mu\nu}$.
When the gravitational interactions are turned on, the background of $\chi$ and $\varphi$ gravitates, and then the constant solution \eqref{background} is no longer a solution of the system. For instance, one can consider the Friedmann-Lema\^{i}tre-Robertson-Walker (FLRW) metric for the background,
\begin{align}
\dd s^2=g_{\mu\nu} \, \dd x^{\mu} \dd x^{\nu}=-\bar{N}^2(t) \, \dd t^2 + a^2(t) \, \delta_{ij} \, \dd x^i \dd x^j
\,,
\label{FLRW_metric}
\end{align}
where $\bar{N}$ and $a$ are the background lapse and the scale factor, respectively.
We can then search for a time-dependent homogeneous solution $\bar\chi = \bar\chi(t)$ and $\bar\varphi = C_0 t$, where $C_0$ is a constant and $\bar\chi$ is now promoted to a time-dependent function instead of a constant.%
\footnote{In fact, this form of the solution can always be achieved, at the cost of the choice of $\bar{N}$. Since the general homogeneous solution for $\bar\varphi$ is $\partial_t \bar\varphi = C_0 / ( \bar{N} a^3 f)$ with an integration constant $C_0$, we can choose the (background) gauge by $\bar{N} = (a^3 f t)^{-1}$ to achieve the aforementioned form of the background solution. This way all the complication of time dependence is taken care of by $\bar\chi$.}
A tiny variation of the background is allowed to integrate out the UV modes, and the backreaction of gravity does not change the main part of the previous analysis as long as the effect of gravity is small compared to the mass scale of the decoupling.

In order to make comparison, we now prepare the candidate effective Lagrangian of ghost condensate,
\begin{align}
\mathcal{L}_{\rm GC}=\frac{\Mpl^2}{2} \, R+P(\tilde{X})- \frac{1}{2\tilde{\mathcal{M}}^2} (\Box \tilde{\varphi})^2 +\cdots \; ,
\label{low?_gra}
\end{align}
with a scalar field $\tilde{\varphi}=\bar{\varphi}+\tilde{\pi}$. As we have discussed in the main text, the linear dynamics of the ghost condensate is the same as the EFT of the two-field model if gravity is absent.

We study linear perturbations around the homogeneous and isotropic background to clarify the non-equivalence in the presence of gravity. We should compare two theories in the same gauge choice. Here, we adopt the unitary gauge in which the quadratic Lagrangian is given by a simple form.
Looking into the sector of scalar modes, the lapse $N$, the shift $N^i$, and the spatial metric $\gamma_{ij}$ are given by
\begin{align}
N=\bar{N}(t)(1+\alpha )\,,\quad
N^i=\bar{N}(t)\delta^{ij}\partial_j \beta\,, \quad
\gamma_{ij}=a^2 e^{2\zeta} \delta_{ij}
\,.
\end{align}
The scalar fields are perturbed as
\begin{align}
\chi=\bar{\chi}(t)+\delta \chi \,, \quad
\varphi=C_0 t
\,,
\end{align}
as for the UV two-field model \eqref{UV_minimal}, whereas the scalar field for the ghost condensate \eqref{low?_gra} is
\begin{align}
\tilde{\varphi}=\tilde{C}_0 t
\,.
\end{align}
As we mentioned after \eqref{L_Meff2}, the $\Lambda$-EFT can be derived by treating the time derivative of $\chi$ as perturbations. At the leading order, the kinetic term of $\chi$ in \eqref{UV_minimal} is approximated by
\begin{align}
(\nabla \chi)^2\simeq \gamma^{ij}\partial_i \chi \partial_j \chi
\,.
\end{align}
On the other hand, we only keep the highest spatial derivative term for $\tilde\varphi$ in \eqref{low?_gra}, i.e.,
\begin{align}
(\Box \tilde{\varphi})^2 \simeq 2\tilde{X} (\partial_i \partial^i \beta)^2
\, .
\end{align}
The quadratic Lagrangian for the curvature perturbation $\zeta$ of \eqref{UV_minimal} after integrating out $\delta \chi$ and that of \eqref{low?_gra} are
\begin{align}
\mathcal{L}^{(2)}_{\Lambda} & \simeq \left( \frac{f X}{H^2} \right)\left[ \frac{\Meff^2}{k^2/a^2+M^2} \, \frac{\vert \dot{\zeta} \vert^2}{\bar{N}^2}- \frac{k^2}{a^2} \left\vert \zeta \right\vert^2 \right]
\nn
&=\frac{M^2}{k^2/a^2 +M^2} \left( \frac{X}{H^2} \right)\left[ (P_X+2XP_{XX}) \frac{\vert \dot{\zeta} \vert^2}{\bar{N}^2}- \left(P_X + P_X \, \frac{k^2}{a^2 M^2} \right) \frac{k^2}{a^2} \, \vert \zeta \vert^2 \right]
,
\label{zeta_low}
\\
\mathcal{L}^{(2)}_{\rm GC} & \simeq \left( \frac{ \tilde{X} }{H^2} \right)\left[ ( \tilde{P}_X+2\tilde{X} \tilde{P}_{XX})\frac{\vert \dot{\zeta} \vert^2}{\bar{N}^2}-\left( \tilde{P}_X + \frac{k^2}{a^2 \tilde{\mathcal{M}}^2} \right) \frac{k^2}{a^2} \, \vert \zeta \vert^2 \right],
\label{zeta_low2}
\end{align}
in the momentum space, respectively, where we have used the relations $f=P_X$ and $f_{\chi}^2/M^2=P_{XX}$. It is clear that two theories \eqref{zeta_low} and \eqref{zeta_low2} describe the same dynamics only for the modes $k^2 / a^2 \ll |M^2|$ due to the prefactor $\frac{M^2}{k^2/a^2+M^2}$. Although two theories describe the same dynamics when the perturbations are decoupled from the background, the background motion of $\frac{M^2}{k^2/a^2+M^2}$ provides the difference. The EFT of \eqref{UV_minimal} and the ghost condensate \eqref{low?_gra} are distinguished by how they gravitate.

\bibliography{ref}
\bibliographystyle{JHEP}

\end{document}